\documentclass[aps,prl,twocolumn,superscriptaddress,noeprint,longbibliography, nofootinbib,nobibnotes]{revtex4-1}

\usepackage{graphicx}
\usepackage{bm}
\usepackage{physics}
\usepackage{xcolor}
\usepackage{enumitem}
\usepackage{amsmath, amssymb}
\usepackage[normalem]{ulem}
\usepackage{natbib}
\usepackage{bbm}
\usepackage{comment}
\usepackage{mathrsfs}
\usepackage{amsmath}
\usepackage{amsthm}
\usepackage{hyperref}
\usepackage{CJKutf8}

\usepackage{pdfpages}
\makeatletter
\AtBeginDocument{\let\LS@rot\@undefined}
\makeatother

\usepackage[dvipsnames]{xcolor}

\newcommand{\br}{\bm{r}}
\newcommand{\zb}{\bar{z}}

\begin{document}
\begin{CJK*}{UTF8}{gbsn}
\title{Skyrmion Fractional Chern Insulator: \\
An Intrinsically Multiband Route to Fractionalization in Rhombohedral Graphene}
\date{\today}
\author{Julian May-Mann}
\thanks{These authors contributed equally.}
\affiliation{Department of Physics, Stanford University, Stanford, CA 94305, USA}
\author{Tixuan Tan (谈体轩)}
\thanks{These authors contributed equally.}
\affiliation{Department of Physics, Stanford University, Stanford, CA 94305, USA}
\author{Patrick J. Ledwith}
\affiliation{Department of Physics, Massachusetts Institute of Technology, Cambridge, MA 02139, USA}
\author{Zhengyan Darius Shi (石铮岩)}
\affiliation{Department of Physics, Stanford University, Stanford, CA 94305, USA}
\author{Trithep Devakul}
\email{tdevakul@stanford.edu}  
\affiliation{Department of Physics, Stanford University, Stanford, CA 94305, USA}
\begin{abstract}
We propose an unconventional microscopic origin for the fractional quantum anomalous Hall (FQAH) effect in rhombohedral graphene moir\'e superlattices: skyrmion fractionalization.
We view the state at filling $\nu<1$ as a metal of skyrmion vacancies, charge $+e$ objects formed by removing layer-pseudospin skyrmions from the interaction-generated skyrmion lattice Chern insulator at $\nu=1$.
These vacancies are intrinsically multiband degrees of freedom, absent in single Chern band-projected studies.
Building on a recently proposed ideal limit, we first develop an effective field theory showing that skyrmion vacancies can themselves fractionalize, thereby inducing charge fractionalization.
Focusing on $\nu=\frac{2}{3}$, we then construct explicit variational trial wavefunctions for the resulting skyrmion fractional Chern insulator and provide numerical evidence, together with general arguments, showing that this process is energetically favored.
Our results establish a realistic route to the FQAH that does not rely on a partially filled Chern band, but instead arises from fractionalization of collective pseudospin textures.
\end{abstract}

\maketitle
\end{CJK*}

\section{Introduction}
Electron fractionalization, as exemplified by the fractional quantum Hall (FQH) effect, is a hallmark of strongly correlated topological quantum matter.
The recent experimental observations of its zero magnetic field counterpart, the fractional quantum anomalous Hall (FQAH) effect~\cite{ju2024fractional,cao2025fractional}, in twisted MoTe$_2$~\cite{cai2023signatures,zeng2023thermodynamic,park2023observation,xu2023observation} and rhombohedral $N$-layer graphene (R$N$G) on hBN~\cite{lu2024fractional,waters2025chern,aronson2025displacement,choi2025superconductivity,xie2025tunable,zheng2025switchable,lu2025extended,chen2020tunable}, have renewed interest in these phases, particularly in the novel aspects of their realization at zero field
and their microscopic origin.

\begin{figure}[t]
\includegraphics[width=1\linewidth]{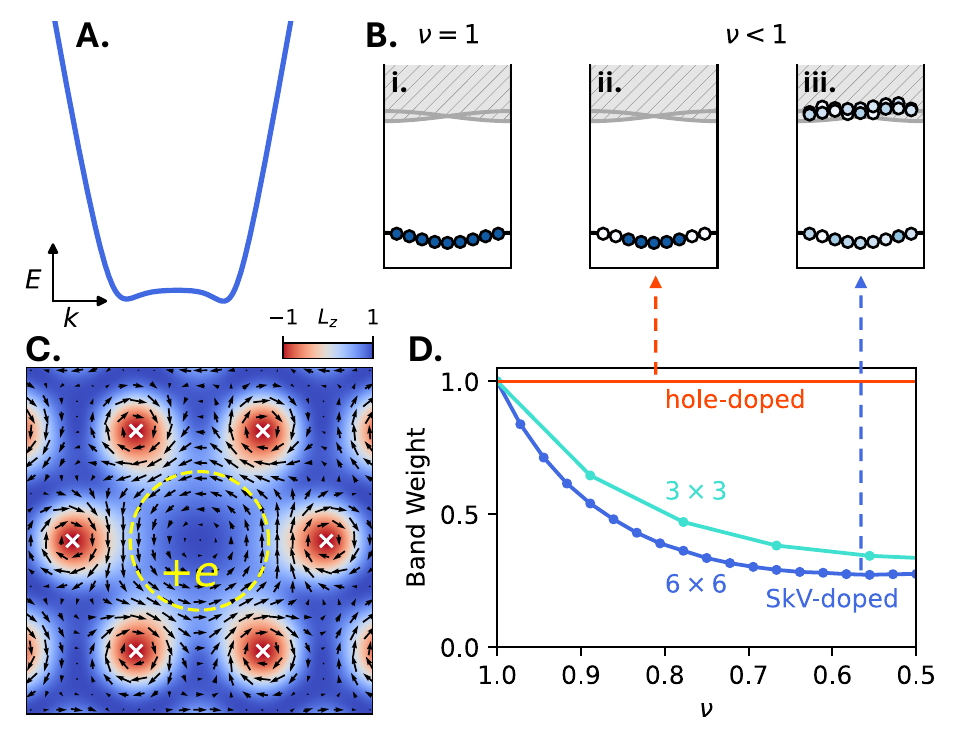}
\caption{\textbf{A.} The realistic R5G dispersion.
\textbf{B.}
(i.)
The state at $\nu=1$ electrons per moir\'e unit cell of R5G/hBN can be understood as a filled interaction-generated $C=1$ band in the folded zone, corresponding to a layer-pseudospin skyrmion lattice in real-space.
At $\nu<1$, two possibilities are  (ii.) holes in the Chern band, or (iii.) skyrmion vacancies (SkVs), which involves mixing into higher bands when viewed in the $\nu=1$ band basis.
\textbf{C.}
Illustration (using R$2$G) of the real-space texture of a SkV, which carries $+e$ charge.
White crosses mark the skyrmion cores, zeros of $\chi_0^{}$.
\textbf{D.} The band weight, defined as the fraction of the density in the lowest $\nu=1$ band, is plotted for $\nu<1$ as holes or SkVs are doped.
For SkVs, the weight is computed numerically for the trial SkV wavefunction (Eq~\ref{eq:chi0skv}) on a $L\times L$ torus, averaged over SkV positions.
}
\label{fig:skv}
\end{figure}

At its core, the conventional picture for the FQAH closely parallels the FQH effect:
strongly interacting electrons partially fill a $C=1$ band and form a fractional Chern insulator (FCI)~\cite{parameswaran2013fractional,bergholtz2013topological,tang2011high,neupert2011fractional,regnault2011fractional,sheng2011fractional,liu2022recent,bernevig2025fractional}.
In tMoTe$_2$~\cite{mak2022semiconductor,li2026quantum,wu2019topological} (and finite-field FCIs~\cite{spantonObservationFractionalChern2018,xieFractionalChernInsulators2021,finneyExtendedFractionalChern2025a,ledwith2020fractional,abouelkomsanParticleHoleDualityEmergent2020a,repellinChernBandsTwisted2020,wilhelmInterplayFractionalChern2021,shefferChiralMagicangleTwisted2021a,parkerFieldtunedZerofieldFractional2021}), an isolated $C=1$ band makes this picture natural~\cite{reddy2023fractional,wang2024fractional,dong2023composite,goldman2023zero,jia2024moire}, and band-projected exact diagonalization (ED) at fractional moir\'e filling factors $0<\nu<1$ yields FCIs.
R5G/hBN is less straightforward because its single-particle spectrum lacks a well-isolated Chern band
(Fig~\ref{fig:skv}A shows the unfolded R5G conduction band).
Instead, Hartree-Fock (HF) theory produces an interaction-generated $C=1$ band at filling $\nu=1$ in the folded moir\'e Brillouin zone~\cite{dong2024anomalous,dong2024theory,zhou2024fractional,guo2024fractional,tan2024parent,kwan2025moire,herzog2024moire,guo2025correlation,huang2024self,huang2025fractional,uchida2026non,xie2024integer,bernevig2025berry,kwan2025moire,tan2025ideal} (Fig~\ref{fig:skv}B.i.),
which is then used as a basis for band-projected ED at $0<\nu<1$~\cite{zhou2024fractional,dong2024theory,guo2024fractional,dong2024anomalous}.
This HF-ED approach works even in the moir\'e-less limit, in which case they spontaneously break continuous translation symmetry realizing  anomalous Hall crystals
~\cite{soejima2024anomalous,valenti2025quantum,hirsbrunner2026topological,zeng2024sublattice,desrochers2025elastic,dong2024stability,tan2025variational,soejima2025topological,dong2025phonons,zeng2025berry,desrochers2026electronic,soejima2025jellium,soejima2024anomalous,crepel2025efficient,song2024intertwined,lu2026generic,may2026composite, desrochers2026energetics, bhattacharjee2026anyon}.
Despite the more roundabout approach, these studies ultimately yield FCIs at partial filling of a single $C=1$ band. 

However, this single-band projection in R5G/hBN is uncontrolled because the $C=1$ band is itself interaction-generated. 
Indeed, calculations that take into account multiple HF bands find that band mixing effects are strong~\cite{yu2025moire,li2025multiband,moirecapacitor}.
The nature of these multiband effects is not fully understood:
whether they
merely dress a single-band FCI or reflect something intrinsic about its microscopic origin remains unclear.

Here, we propose an unconventional, intrinsically multiband, route to fractionalization in rhombohedral graphene that we term the skyrmion FCI (SkFCI).
Our starting point is the perspective~\cite{tan2025ideal} that the $\nu=1$ Chern insulator in R5G/hBN is an interaction-generated layer-pseudospin skyrmion lattice (SkL), pinned by the moir\'e potential, whose texture produces the emergent LLL-like $C=1$ band (Fig~\ref{fig:skv}B.i.).
Upon reducing the filling below $\nu=1$,
the conventional picture, assumed by HF-ED, dopes holes into this $C=1$ band (Fig~\ref{fig:skv}B.ii.).
We instead consider skyrmion vacancies (SkVs), missing skyrmions in the lattice, as the lowest energy charge $+e$ objects.  
Because the SkV involves a modification of the collective layer-pseudospin texture (Fig~\ref{fig:skv}C), it is an intrinsically multiband and multielectron object (Fig~\ref{fig:skv}B.iii.), absent in band-projected studies.
This paints a very different picture for the metallic phase at $\nu<1$ as a metal of charge-carrying texture fluctuations.

Motivated by this, we focus on $\nu=\frac{2}{3}$ and ask whether the parent state of the FCI could be a SkV metal rather than a hole metal.
Do SkVs experience an effective magnetic field, and if so, can they fractionalize?
Can such skyrmion fractionalization account for the observed $\sigma_H=\frac{2}{3}$ FQAH, and is it energetically competitive?
We answer all of these questions in the affirmative.

\section{Ideal limit framework}
Our analysis utilizes the ideal limit framework recently introduced in Ref~\cite{tan2025ideal}.  
We therefore start by reviewing the necessary aspects of this formalism.  
In this limit, the Bloch states of the unfolded conduction band of one spin-valley flavor of R$N$G (shown in Fig~\ref{fig:skv}A) are $\psi_{\bm{k},\ell}(\br)=e^{i\bm{k}\cdot\br}s_{\bm{k},\ell}$ with the layer-spinor 
\begin{equation}
s_{\bm{k},\ell}=N_{\bm{k}} [\gamma (k_x+ik_y)^\ell]; \;\;
N_{\bm{k}}^{-2}={\textstyle \sum_{\ell=0}^{N-1}}|\gamma\bm{k}|^{2\ell}
\label{eq:spinor}
\end{equation}
where $\ell=0,\dots,N-1$ is the layer index and $\gamma\approx1.73$nm.
For microscopic justification that this is a good approximation to the realistic R$N$G spinor, we refer to Refs~\cite{bernevig2025berry,tan2024parent,tan2025ideal,han2025exact,desrochers2026electronic,soejima2024anomalous,desrochers2026energetics}.
We assume full spin-valley polarization throughout, and all the physics to be discussed takes place entirely within this parent R$N$G conduction band.

We work with many-body wavefunctions in real-space, $\Psi_{\{\ell\}}(\{\br\})$, where $\{\ell\}=(\ell_1,\dots,\ell_{N_e})$ and $\{\br\}=(\br_1,\dots,\br_{N_e})$ are the layer indices and 2D position coordinates of the $N_{e}$ electrons.
The spinor structure Eq~\ref{eq:spinor} manifests in real space as the relation
\begin{equation}
\Psi_{\{\ell\}}(\{\br\})=\prod_{i=1}^{N_e}(-2i\gamma\partial_{\bar{z}_i})^{\ell_i}\Psi_{\{0\}}(\{\br\})\label{eq:manybody_recursion}
\end{equation}
where $z=(\bar{z})^*=x+iy$.
This relation implies that all layer components of $\Psi_{\{\ell\}}$  are uniquely determined by $\Psi_{\{0\}}$, the component with all electrons on layer $\ell=0$, which remains unspecified.

In Ref~\cite{tan2025ideal}, a class of wavefunctions were derived of the form
\begin{equation}
\Psi_{\{\ell\}}(\{\br\})=\Phi^{}(\{\br\})X_{\{\ell\}}(\{\br\}).
\label{eq:parton}
\end{equation}
Here,
\begin{equation}
\Phi^{}(\{\br\})=F(\{z\})e^{-\sum_i|z_i|^2/4l_B^2}
\end{equation}
with $F$ an antisymmetric analytic function, describes a fermionic quantum Hall wavefunction in the lowest Landau level (LLL) of an effective magnetic field $B=1/l_B^2$.
The $X$ describes a bosonic wavefunction in the opposite field $-B$ (though not necessarily in the LLL).
Eq~\ref{eq:manybody_recursion} implies that 
\begin{equation}
X_{\{\ell\}}(\{\br\})={\textstyle\prod_{i=1}^{N_e}}(-2i\gamma[\partial_{\zb_i}-z_i/4l_B^2])^{\ell_i}X_{\{0\}}(\{\br\})
\end{equation}
thus, all $X_{\{\ell\}}$ are uniquely determined by $X_{\{0\}}$.
Because the $\Phi$ and $X$ transform under magnetic translations with opposite fields, their product $\Psi=\Phi X$ transforms under ordinary translations.

Eq~\ref{eq:parton} defines a special class of wavefunctions that are exact zero-energy eigenstates of contact interactions $V_0\sum_{i<j}\delta(\br_i-\br_j)$.
This follows since $\Psi_{\{\ell\}}\rightarrow 0$  whenever any $|\br_i-\br_j|\rightarrow 0$ (for all combinations of $\ell_i$ and $\ell_j$), due to the antisymmetric $\Phi^{}$ factor (see Refs~\cite{tan2025ideal,desrochers2026energetics}).
Within this framework, natural trial wavefunctions for integer and fractional Chern insulators have been proposed~\cite{tan2025ideal,desrochers2026energetics}, which we briefly review now.  

The Chern insulators rely on $X=X^{\text{SkL}}$ forming a skyrmion lattice, with one skyrmion per moir\'e unit cell.
This corresponds to the $\ell=0$ bosons condensing into an Abrikosov antivortex lattice, $X^{\text{SkL}}_{\{0\}}(\{\br\})=\prod_i\chi_0^{\text{SkL}}(\br_i)$,
\begin{equation}
\chi^{\text{SkL}}_{0}(\br)=e^{-\frac{\pi}{2A}|z|^2}\sigma(\bar{z})\sim e^{-\frac{\pi}{2A}|z|^2}\prod_{\bm{R}\in\Lambda}(\bar{z}-\bar{Z})
\label{eq:chi0SkL}
\end{equation}
where $\sigma(z)$ is the modified Weierstrass sigma~\cite{haldane2018modular,supp} function satisfying $\sigma(z+a_i)=-e^{\frac{\pi}{A}\bar{a}_i(z+a_i/2)}\sigma(z)$, and $a_{i}\in\{a_1,a_2\}$ are moir\'e unit vectors and $A=|\bm{a}_1\times\bm{a}_2|$. The final expression on the right hand side of Eq~\ref{eq:chi0SkL} is only meant to capture the structure of the zeros of $\sigma$, where $\Lambda=\{n_1 \bm{a}_1+n_2\bm{a}_2\}$ defines the moir\'e lattice vectors and $\bar{Z}={R}_x-i{R}_y$.
The effective magnetic field is at one flux per unit cell, $B=2\pi/A$.
Each antivortex core of $\chi_0$ corresponds to a skyrmion core in the full spinor $\chi_\ell$,
thus it describes a skyrmion lattice~\cite{tan2025ideal}.
The skyrmion lattice $\chi^{\text{SkL}}$ \emph{defines} an electronic $C=1$ band in the folded Brillouin zone, spanned by the single-particle Bloch states $\psi_{\bm{k}}^{\text{SkL}}=\varphi_{\bm{k}}\chi_{\ell}^{\text{SkL}}$, where $\varphi_{\bm{k}}$ are the LLL magnetic Bloch states~\cite{tan2025ideal,desrochers2026energetics,tan2025variational}.

The $C=1$ Chern insulator is obtained by fully filling this Chern band, i.e. using the fully filled LLL for $\Phi$, 
\begin{equation}
\Phi^{\text{IQH}}(\{\br\})=\textstyle{\prod_{i<j}}(z_i-z_j)\prod_ie^{-\frac{\pi}{2A}|z_i|^2}
\end{equation}
combined with $X^{\text{SkL}}$.
The FCIs are similarly obtained with $\Phi=\Phi^{\text{FQH}}$, the corresponding fractional quantum Hall states (e.g. Laughlin states~\cite{laughlin1983anomalous}).
The $C=1$ state in this construction can be understood as a Slater determinant of the fully filled ideal~\cite{ledwith2020fractional,wang2021exact,ledwith2023vortexability} $C=1$ band defined by $\chi^{\text{SkL}}$, and the FCIs as Laughlin states at partial filling of that same band.
We refer to this FCI wavefunction, 
\begin{equation}
\Psi^{\text{FCI}}_{\{\ell\}}(\{\br\})=\Phi^{\text{FQH}}(\{\br\})X^{\text{SkL}}_{\{\ell\}}(\{\br\}),
\end{equation} as ``conventional'', since it describes a state at partial filling of a single $C=1$ band.
Thus, $\Psi^{\text{FCI}}$ represents a model wavefunction for what one might find from HF-ED.
We will later also consider a generalization of $\Psi^{\text{FCI}}$ to include a variational parameter.

\section{Skyrmion vacancies}
Within the ideal limit formalism, holes doped into the $C=1$ band can be described as holes in the $\Phi^{\text{IQH}}$ factor.
The SkVs, however, cannot.  
They correspond to \emph{missing zeros} in $\chi_0^{\text{SkL}}$.
Schematically, a state with SkVs at positions $\{\bm{\eta}\}\subset\Lambda$ is described by $X^{\text{SkV}}=\prod \chi^{\text{SkV}}$ with
\begin{equation}
\begin{split}
\chi_0^{\text{SkV}}(\br;\{\bm{\eta}\})&\sim e^{-\frac{\pi}{2A}|z|^2}{\textstyle\prod_{\bm{R}\in \Lambda\backslash\{\bm{\eta}\}}}(\bar{z}-\bar{Z})\\
&\sim \chi_0^{\text{SkL}}(\br)\textstyle{\prod_{k=1}^{N_v}}(\bar{z}-\bar{\eta}_k)^{-1}
\label{eq:chi0skv}
\end{split}
\end{equation}
where, in the second line, the $(\bar{z}-\bar{\eta})^{-1}$ factors do not result in singularities because they cancel with zeros present in $\chi_0^{\text{SkL}}$.
Each such factor results in a topological change of the skyrmion texture, reducing the total skyrmion number by one.
Since removal of zeros leads to a smoother wavefunction overall, this also motivates why SkVs can be kinetically favored over holes.
As this cannot be accomplished by any LLL factor, SkVs are intrinsically multiband objects that cannot be faithfully represented within the original SkL band.

Fig~\ref{fig:skv}C illustrates a skyrmion lattice with an SkV at the origin, shown for R$2$G for visualization purposes, representing the layer pseudospin $\bm{L}(\bm{r})$ on the Bloch sphere~\cite{tan2025ideal}.  
In Fig~\ref{fig:skv}D, we quantify the amount of ``band mixing'' necessary to describe the SkVs.
It shows the band weight, defined as $\langle N_{\text{SkL}}\rangle/N_e$, where $N_{\text{SkL}}$ is the total density of electrons in the $C=1$ band defined by $\chi^{\text{SkL}}$, evaluated for $\Psi=\Phi^{\text{IQH}}X^{\text{SkV}}$  generalized to the torus~\cite{supp}.
Fig~\ref{fig:skv}D shows that for reasonable densities of SkVs, this weight becomes small, indicating strong band mixing.  
As the electron density approaches $\nu\sim \frac{2}{3}$, the majority of the weight is already outside of the SkL band. 
SkVs are therefore highly multiband objects.

\section{Effective field theory}
We first use an effective field theory approach to determine the universal long-wavelength properties of SkVs, and address the possibility of their fractionalization.
Our starting point is the observation that Eq~\ref{eq:parton} takes the form of a parton wavefunction, motivating a parton effective field theory in which the electron is decomposed as $c_{\ell}=fz_{\ell}$ into fermionic $f$ and bosonic $z_{\ell}$ partons~\cite{coleman1984_parton,baskaran1988_parton,jain1989incompressible, jain1990theory} (see SM~\cite{supp} for extended discussion).
This decomposition introduces an emergent $U(1)$ gauge field $a$.
The large R$N$G band gap imposing Eq~\ref{eq:manybody_recursion} implies that all $z_{\ell>0}$ can be integrated out, leaving only $z_0$.  
After coarse-graining with respect to the moir\'e scale, the remaining $f$ and $z_0$ degrees of freedom are described by a continuum Lagrangian of the form
\begin{equation}
\mathcal{L}=\mathcal{L}_f[f,A+a]+\mathcal{L}_z[z_0,-a]
\end{equation}
where $A$ is a probe physical electromagnetic gauge field.

SkVs correspond to vortices of $z_0$, measured relative to the antivortex lattice configuration (Eq~\ref{eq:chi0SkL}) at $\nu=1$.
To describe them, we invoke a particle-vortex duality on $z_0$\cite{peskin1978mandelstam, Lee1990_bosonvortexskyrmion, zhang1992chern},
\begin{equation}
\mathcal{L}_{z}[z_0,-a]\rightarrow\mathcal{L}_v[v,-b-A^{\text{Sk}}]-\frac{1}{2\pi}bda+\frac{1}{2\pi}(b+A^{\text{Sk}})d\omega^{\text{SkL}}
\end{equation}
where $v$ is the field describing SkVs, $b$ is a new emergent gauge field introduced by the particle-vortex duality, $\omega^{\text{SkL}}$ is a non-dynamic gauge field encoding the background skyrmion density 
$\bm{\nabla}\times \bm{\omega}^{\text{SkL}}/2\pi=1$, and we have abbreviated $bda=b\wedge da$.
Throughout this section, we work in units where the moir\'e unit cell area is $1$.
We have also introduced a new probe gauge field $A^{\text{Sk}}$ to keep track of skyrmion density.

The equations of motion for $a$ and $b$ specify
\begin{equation}
\begin{split}
\frac{\bm{\nabla}\times\bm{a}}{2\pi}=1-\rho_v;\;\;\;\;
\frac{\bm{\nabla}\times\bm{b}}{2\pi}=\rho_f\\
\end{split}
\label{eq:eoms}
\end{equation}
where $\rho_{v},\rho_f$ are the densities of $v,f$.
These equations confirm that the magnetic field experienced by $f$ is equal to the skyrmion density, and furthermore, reveal that SkVs also experience a field equal to the density of $f$. 

To proceed, we assume that the $f$ fermions form a gapped IQH state at magnetic filling factor $\nu_{\text{mag}}=1$. 
Integrating out $f$ and $a$, and shifting $b\rightarrow b+A$ leads to the following SkV Lagrangian 
\begin{equation}\begin{split}
\mathcal{L}=&\mathcal{L}_v[v,-b - A -A^{\text{Sk}}] -\frac{1}{4\pi}bdb\\ &+\frac{1}{2\pi}(b + A +A^{\text{Sk}})d\omega^{\text{SkL}} + \frac{1}{4\pi}AdA
\label{eq:SkVEffectiveLag}\end{split}\end{equation}
which confirms that the SkVs are indeed electrically charged due to the Hall response of the $f$ fermions.

Now, let us consider SkV doping $\rho_v = \frac{1}{3}$, corresponding to the electron density of $\nu = \rho_f =\frac{2}{3}$. 
According to Eq~\ref{eq:eoms}, the SkVs are at \emph{magnetic} filling factor $\nu_{\text{mag}}=-\rho_v/\rho_f=-\frac{1}{2}$, and 
can therefore be placed into a bosonic Laughlin FQH state~\cite{ read1990excitation, cooper2008rapidly}. 
Since $v$ is gapped in this phase, it can be integrated out, leaving
\begin{equation}
\begin{split}
\mathcal{L}_v[v,-b - A -A^{\text{Sk}}] \rightarrow \frac{2}{4\pi}\alpha d\alpha-\frac{1}{2\pi}\alpha d\left(b + A + A^{\text{Sk}}\right),
\label{eq:BosonicHalf}\end{split}
\end{equation}
where $\alpha$ is the emergent dynamical gauge field for the $\nu_{\text{mag}}=-\frac{1}{2}$ SkV FQH state. Combining Eq~\ref{eq:SkVEffectiveLag} and~\ref{eq:BosonicHalf}, and integrating out $b$, we find that the full topological field theory takes the form
\begin{equation}
\begin{split}
\mathcal{L}_{\text{eff}}=&\frac{3}{4\pi}\alpha d\alpha-\frac{1}{2\pi}\alpha d\left(A+A^{\text{Sk}}+\omega^{\text{SkL}}\right)\\
&+\frac{1}{4\pi}AdA + \frac{1}{2\pi} \left(A+A^{\text{Sk}}\right)d\omega^{\text{SkL}}.\\
\end{split}
\end{equation}
If $A^{\text{Sk}}$ is ignored, this describes the same topological order as the conventional $\nu=\frac{2}{3}$ FQH state~\cite{blok1990effective}, consistent with the experimental $\sigma_{H}=\frac{2}{3}$ FQAH.
The presence of $A^{\text{Sk}}$, however, implies that the anyons carry both fractional electric  and skyrmion charge, $q^{e}_{\text{anyon}}=q^{\text{Sk}}_{\text{anyon}}=\frac{1}{3}$.
{This skyrmion fractionalization manifests as a topologically quantized skyrmion-electric Hall response $\sigma_{\text{Sk-}e}=-\frac{1}{3}$, where an electric field drives a perpendicular skyrmion current, as well as a related skyrmion-skyrmion Hall response, $\sigma_{\text{Sk-}\text{Sk}}=-\frac{1}{3}$.}
Our findings are summarized in Fig~\ref{fig:energetics}A.

We note that this analysis implies that the conventional and skyrmion FCIs are sharply distinct only when skyrmion number is conserved, but belong to the same topological order otherwise.
Since skyrmion number is not a true microscopic conserved quantity, the FCI and SkFCI in R5G/hBN should therefore not be regarded as distinct topological phases of matter, but rather as two distinct microscopic realizations of the same topological order.

\section{Microscopic energetics}
Our field theory analysis has identified a route to realizing the FQAH through skyrmion fractionalization.
The remainder of this paper is devoted to analyzing the microscopic energetics of this process, focusing on the comparison to the conventional FCI.  We do this by constructing trial wavefunctions and computing their variational energy with realistic terms in the Hamiltonian,
specializing to $\nu=\frac{2}{3}$.

For notational simplicity, we will work on the disc geometry in the following.  However, all our numerical results are computed for the torus geometry~\cite{supp}. 
For convenience, we define the shorthand
\begin{equation}
\begin{split}
J_{zz}=G_{z}^{\frac{2}{3}}\prod_{i<j}^{N_e}(z_i-z_j); &\;\; J_{z\eta}=G_z^{\frac{1}{3}}G_{\eta}^{\frac{2}{3}}\prod_{i=1}^{N_e}\prod_{k=1}^{N_v}(z_i-\eta_k); \;\\
J_{\eta\eta}=G_{\eta}^{\frac{2}{3}}\prod_{k<l}^{N_{v}}(\eta_k-\eta_l); & \;\; J_{z\Lambda}=G_{z}\prod_{i=1}^{N_e}\sigma(z_i)
\end{split}
\end{equation}
where $N_e$ and $N_v$ are the numbers of electrons and SkVs, satisfying $N_e=2N_v$.
Appropriate Gaussian factors
$G_z=\prod_{i}^{}e^{-\frac{\pi}{2A}|z_i|^2}$ and $G_\eta=\prod_{k}^{}e^{-\frac{\pi}{2A}|\eta_k|^2}$
have been included to ensure each corresponds to the correct uniform average density of $z$ and $\eta$ within the disc.

\begin{figure}[t]
\includegraphics[width=1\linewidth]{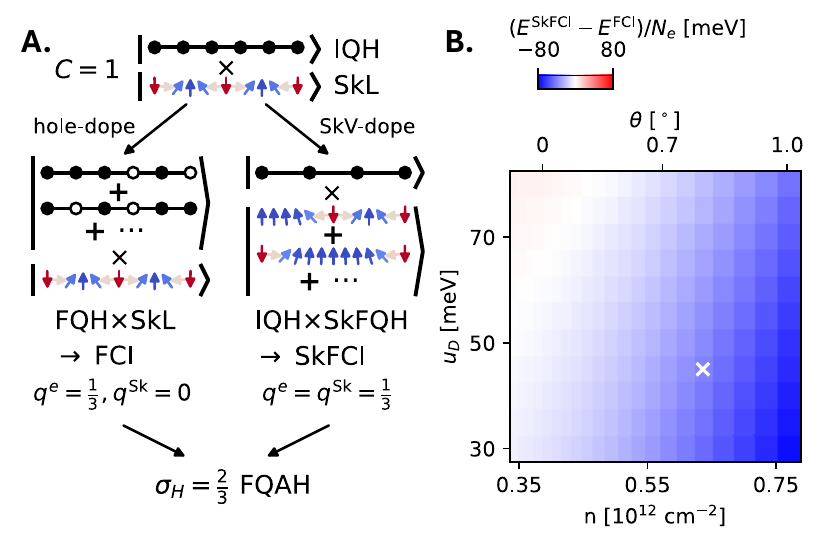}
\caption{\textbf{A.} Two routes to realizing the $\nu=\frac{2}{3}$ FQAH, starting from the $C=1$ Chern insulator at $\nu=1$.
In the conventional route, holes are doped in to the $C=1$ band.
In the skyrmion route, charged SkVs are instead doped into the layer-pseudospin texture.
\textbf{B.}
Energy difference between the SkFCI and conventional FCI trial wavefunctions with realistic dispersion and interactions, as a function of displacement field $u_D$ and density $n$ or, equivalently, hBN twist angle $\theta$. 
White cross marks the parameter value explored in Fig~\ref{fig:variational}.
}
\label{fig:energetics}
\end{figure}

Motivated by the field theory analysis, we place the bosonic partons into a correlated state in which SkVs form a $\nu_{\text{mag}}=-\frac{1}{2}$ bosonic Laughlin state. 
Note that this is \emph{not} the same as putting the bosonic $X$ partons themselves into the Laughlin state, but rather, the SkVs.
Similar to the hierarchy construction of FQH states~\cite{haldane1983fractional,halperin1984statistics,read1990excitation}, this is accomplished by taking the expression for a particular configuration of SkVs, $X^{\text{SkV}}_{\{0\}}(\{\br\},\{\bm{\eta}\})=\bar{J}_{z\Lambda}\bar{J}_{z\eta}^{-1}$(Eq~\ref{eq:chi0skv}), and summing over all lattice configurations $\{\bm{\eta}\}$ with the coefficients $\bar{J}_{\eta\eta}^2$, 
\begin{equation}
X_{\{0\}}^{\text{SkFQH}}(\{\br\})=\bar{J}_{z\Lambda}\textstyle{\sum_{\{\bm{\eta}\}}} \bar{J}_{z\eta}^{-1} \bar{J}_{\eta\eta}^2.
\end{equation}
The full wavefunction is then obtained by combining this with the fermionic parton in the IQH state $J_{zz}$,
\begin{equation}
\Psi_{\{0\}}^{\text{SkFCI}}(\{\br\})=J_{zz}\bar{J}_{z\Lambda}\textstyle{\sum_{\{\bm{\eta}\}}}\bar{J}_{z\eta}^{-1}\bar{J}_{\eta\eta}^2
\label{eq:skfci}
\end{equation}
with all higher-layer components obtained via Eq~\ref{eq:manybody_recursion}.
This wavefunction, illustrated in Fig~\ref{fig:energetics}A, is one of the main results of this paper.

Evaluating expectation values of $\Psi^{\text{SkFCI}}$ is challenging
because of the sum over configurations $\{\bm{\eta}\}$.
We are not aware of an efficient algorithm for doing so in the thermodynamic limit.
We therefore take a brute-force approach here. 
We use the fact that each term in the sum of Eq~\ref{eq:skfci}, $|\{\bm{\eta}\}\rangle\sim J_{zz}\bar{J}_{z\Lambda}\bar{J}_{z\eta}^{-1}$ is, by itself, an electron Slater determinant.
In the SM, we obtain a formula for each $|\{\bm{\eta}\}\rangle$ as a determinant of electron orbitals expressed in the plane-wave basis,
 allowing for a direct calculation of its variational energy.
In practice, this calculation is very computationally costly and we are restricted to a torus of size $3\times 3$, but this will be sufficient to identify general trends in energetics.

We first confirm in the SM~\cite{supp} that $\Psi^{\text{SkFCI}}$ has the correct many-body Chern number for a Hall conductivity of $\sigma_H=\frac{2}{3}$, through flux threading on the torus.
We also confirm~\cite{supp}, by looking at the eigenvalues of $P_{\bm{g}\bm{g}'}(\bm{k})=\langle c_{\bm{k}+\bm{g}'}^\dagger c_{\bm{k}+\bm{g}}\rangle$ where $\bm{g}$ are reciprocal lattice vectors, that $\Psi^{\text{SkFCI}}$ is truly multiband: it cannot be viewed as a state at partial filling of any single band.

We next, in the spirit of Ref~\cite{desrochers2026energetics}, consider the energy of $\Psi^{\text{SkFCI}}$ and $\Psi^{\text{FCI}}$ under realistic dispersion and interaction.  
For $\Psi^{\text{FCI}}$, we take $\Phi^{\text{FQH}}$ to be the particle-hole conjugate of the $\nu_{\text{mag}}=\frac{1}{3}$ Laughlin state.
We obtain the R$5$G dispersion $\mathcal{E}(\bm{k})$ (Fig~\ref{fig:skv}A) by diagonalizing a realistic R5G model~\cite{supp}, and take the gate-screened Coulomb interactions $V(\bm{q})=\frac{e^2\tanh(|\bm{q}|d)}{2\epsilon_r\epsilon_0|\bm{q}|}$ with $\epsilon_r=5$ and $d_{}=25$nm. 
Taken together, the only approximation of our model relative to standard R5G models is to replace the Bloch spinors with Eq~\ref{eq:spinor}.

Fig~\ref{fig:energetics}B shows the total (kinetic plus interaction) energy difference between our two trial wavefunctions, across the parameter space of displacement field $u_D$ and density $n$. 
For each $n$, we assume a moir\'e periodicity such that the electron filling is at $\nu=\frac{2}{3}$ per unit cell.
We will return to discuss the moir\'e potential later.
As we are only concerned with the competition between $\Psi^{\text{SkFCI}}$ and $\Psi^{\text{FCI}}$, we do not consider any other competing states.
Strikingly, $E^{\text{SkFCI}}<E^{\text{FCI}}$ across the majority of the phase diagram by a substantial margin. 

\begin{figure}[t]
\includegraphics[width=1\linewidth]{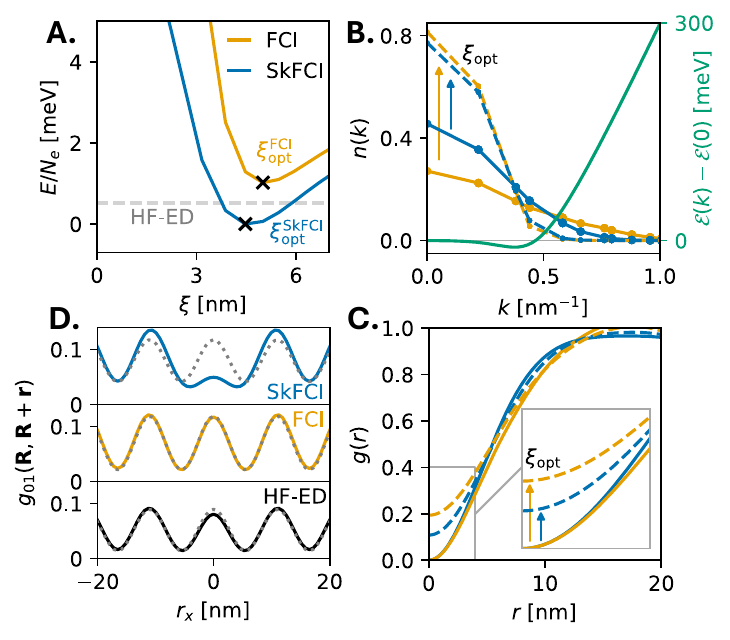}
\caption{\textbf{A.} Energy of the FCI and SkFCI trial wavefunctions (Eq~\ref{eq:variational}) as functions of the variational parameter $\xi$.  
Dashed line is the result of single-band HF-ED for the exact same system.
Energies are shown relative to the minimum.
\textbf{B.} Parent band momentum distribution $n(\bm{k})$ for the unoptimized (solid) and $\xi$-optimized (dashed) trial wavefunctions.
The angle-averaged R5G dispersion $\mathcal{E}(k)$ is also shown.
\textbf{C.} 
The angle-averaged pair correlation function $g(r)$ is shown for the trial wavefunctions.
Inset shows a zoomed-in view of the short-range correlations.
\textbf{D.}
Solid lines show the interlayer pair correlation functions (defined in Eq~\ref{eq:gll}), for the $\xi$-optimized trial states and HF-ED.
Dashed line shows the uncorrelated expectation.
Only the SkFCI contains a well-developed correlation hole.
}
\label{fig:variational}
\end{figure}

To understand why the SkFCI is so energetically favored, Fig~\ref{fig:variational}B (solid lines) shows the parent band momentum distribution function $n(\bm{k}) = \langle c_{\bm{k}}^\dagger c_{\bm{k}}\rangle$ for the two states, and Fig~\ref{fig:variational}C (solid lines) shows their real-space pair correlation function defined as $g(\br)=(1/n N_e)\int d\br'\langle\colon\rho(\bm{r}')\rho(\br'+\br)\colon\rangle$, where $\rho(\bm{r})=\sum_{\ell}\rho_{\ell}(\br)$ is the density operator summed over all layers.
These directly determine the energies via $E_{\text{kin}}=\sum_{\bm{k}}\mathcal{E}(\bm{k})n(\bm{k})$ and $E_{\text{int}}=(nN_e/2)\int d\br V(\bm{r}) g(\br)$, where $V(\bm{r})$ is the Fourier transform of $V(\bm{q})$ (periodized on the torus).
It reveals that the SkFCI has a considerably more compact $n(\bm{k})$, while achieving comparable short-range correlations in $g(\br)$ as the conventional FCI.
This can be understood from their momentum-space envelopes, which scales as $\sim e^{-\frac{1}{2}l_B^2|\bm{k}|^2}$~\cite{tan2025ideal,supp}.
Because $\Psi^{\text{SkFCI}}$ contains fewer skyrmions than $\Psi^{\text{FCI}}$, it has a larger $l_B$, and thus a more compact $n(\bm{k})$.

At this level, the large energy difference comes from the tail of the $n(\bm{k})$ distribution, which has weight into the highly costly region of $\mathcal{E}(\bm{k})$ (shown in Fig~\ref{fig:variational}B).

To further improve our trial wavefunctions,
we now introduce a variational parameter $\xi$. 
Following Ref~\cite{tan2025variational}, natural generalizations of $\Psi^{\text{SkFCI}}$ and $\Psi^{\text{FCI}}$ are 
\begin{equation}
\begin{split}
\Psi^{(\xi,\text{SkFCI})}_{\{\ell\}}(\{\br\})&=e^{\frac{1}{4}\xi^2\sum_{i}\vec{\nabla}_i^2}\Psi^{\text{SkFCI}}_{\{\ell\}}(\{\br\})\\
\Psi^{(\xi,\text{FCI})}_{\{\ell\}}(\{\br\})&=e^{\frac{1}{4}\xi^2\sum_{i}\vec{\nabla}_i^2}\Psi^{\text{FCI}}_{\{\ell\}}(\{\br\}).
\end{split}
\label{eq:variational}
\end{equation}
For $\xi>0$, these states are no longer contact interaction zero modes, and $g(\br\rightarrow0)>0$.
In momentum space, this transformation acts as a Gaussian factor $\sim e^{-\frac{1}{4}\xi^2|\bm{k}|^2}$. 
Thus, it trades short-range correlations for a more compact $n(\bm{k})$. 
Importantly, $\Psi^{(\xi,\text{FCI})}$ remains an FCI within a single (now non-ideal) $C=1$ band~\cite{tan2025variational}, whereas $\Psi^{(\xi,\text{SkFCI})}$ remains multiband.

Fig~\ref{fig:variational}A shows the energies $E^{\text{SkFCI}}(\xi)$ and $E^{\text{FCI}}(\xi)$.
Incorporating $\xi>0$ allows both trial wavefunctions to dramatically lower their energy, yet still the optimized $E^{\text{SkFCI}}_{\text{opt}}<E^{\text{FCI}}_{\text{opt}}$.
To understand why, 
Fig~\ref{fig:variational}B (dashed lines) shows that the optimized $n(\bm{k})$ have both been squeezed just enough to avoid the highly costly portion of $\mathcal{E}(\bm{k})$, so they both have comparable $E_{\text{kin}}$.
However, Fig~\ref{fig:variational}C (dashed lines) shows that the short-range correlations in $g(\br)$ are worse for the FCI than for the SkFCI, resulting in higher $E_{\text{int}}$.
We shall return to this shortly.

We now discuss the role of the moir\'e potential in this competition.
We treat the moir\'e phenomenologically as a scalar potential $U(\bm{r})=2U_0\sum_{i=1}^{3}\cos(\br\cdot\bm{g}_i)$ where $\{\pm\bm{g}_{i}\}$ are the first shell moir\'e reciprocal lattice vectors~\cite{desrochers2026energetics,moirecapacitor}.
While the FCI and SkFCI have similar charge distributions, the SkFCI has slightly weaker modulation because removing skyrmions weakens the charge minima associated with their cores~\cite{tan2025ideal}.
In the SM~\cite{supp}, we evaluate the energies of the $\xi$-optimized states and find a crossover at $U_{0}\sim 15$meV, above which the conventional FCI is favored.
We note that recent estimates give $U_0\sim8$meV~\cite{moirecapacitor}, placing realistic R5G/hBN in regime where the SkFCI is favored.

Next, for comparison, we also repeat the single-band HF-ED analysis of Refs~\cite{zhou2024fractional,dong2024theory,guo2024fractional,dong2024anomalous} for the same system, thus enabling a direct comparison of variational energies.
Fig~\ref{fig:variational}A shows the resulting HF-ED energy.   
Note that we chose $u_D=45$meV for Fig~\ref{fig:variational} based on where the HF-ED FCI gap is most well-formed for this system size.
Incredibly, both $\xi$-optimized trial wavefunctions are competitive with this fully numerical HF-ED approach, to within 1meV per particle. 
In fact, the SkFCI energy dips \emph{even lower} than HF-ED, despite being a fully analytic wavefunction with only one variational parameter.
We take this as evidence that $\Psi^{\text{SkFCI}}$ contains the correct kind of multiband effects that are energetically favored in this system.

To gain more insight into how these multiband effects enhance the SkFCI's short-range correlations, we examine the layer-resolved pair correlation function.
We define
\begin{equation}
\begin{split}
g_{\ell\ell'}(\br,\br') &= \frac{1}{n^2}\langle\colon \rho_\ell(\br)\rho_{\ell'}(\br')\colon\rangle\\
g^{\text{uncorr}}_{\ell\ell'}(\br,\br') &= \frac{N_e-1}{N_e}\frac{1}{n^2}\langle \rho_\ell(\br)\rangle\langle\rho_{\ell'}(\br')\rangle.
\label{eq:gll}
\end{split}
\end{equation}
The interlayer components $\ell\neq \ell'$ of $g_{\ell\ell'}$ dominate at short distances because they need not vanish by Pauli exclusion.
Fig~\ref{fig:variational}D shows $g_{01}(\bm{R},\bm{R}+\br)$, where $\bm{R}\in\Lambda$ is a lattice site, for both $\xi$-optimized trial states and the HF-ED state;
the dashed line shows the completely uncorrelated expectation $g_{01}^{\text{uncorr}}$.
While the single-band FCIs show little correlation, the SkFCI has developed a pronounced ``correlation hole'', where $g_{01}\ll g_{01}^{\text{uncorr}}$. 

The origin of the correlation hole in the SkFCI can be understood as follows. Acting with $\rho_0(\bm{R})$ projects the wavefunction onto configurations with an SkV at $\bm{R}$, in which the local spinor is pointing ``up'' (as in Fig~\ref{fig:skv}C). 
Subsequently acting with $\rho_1(\bm{R})$ then attempts to project the spinor to ``down'', thereby annihilating the state.
In this way, $g_{01}(\bm{R},\bm{R})$ can be small even while $g_{01}^{\text{uncorr}}(\bm{R},\bm{R})$ is not.
Thus, the SkFCI contains precisely the multiband effects, absent in the single-band states, needed to develop such a correlation hole.

This implies that the interlayer correlation hole in Fig~\ref{fig:variational}D can be used as a numerical signature of SkVs in future numerical studies.
We can quantify it by
\begin{equation}
G=1-\frac{g_{01}(\bm{R},\bm{R})}{g_{01}^{\text{uncorr}}(\bm{R},\bm{R})}
\end{equation}
which measures the relative magnitude of this interlayer correlation hole.
It satisfies $G\sim 1$ for the ($\xi=0$) SkFCI, and $G\sim 0$ for the single-band FCIs.
Evaluating it for the states in Fig~\ref{fig:variational}D, we find
\begin{equation}
G^{\text{SkFCI}}\approx0.58;\;\;
G^{\text{FCI}}\approx0.01;\;\;
G^{\text{HF-ED}}\approx0.10.
\end{equation}
Thus, a large $G$ can serve as an easily-computable numerical signature of doped SkVs.

\section{Conclusions}
In this work, we have established an intrinsically multiband route to the FQAH driven by skyrmion fractionalization and shown that such a scenario is energetically competitive in R5G/hBN.
Our theory provides an explanation for the strong band mixing effects found in multiband studies and suggests a qualitatively unconventional microscopic origin of the observed phenomenology in R5G/hBN.
More broadly, our results demonstrate that the FQAH need not originate from a partially filled Chern band, but can instead emerge from fractionalization of collective pseudospin textures.

\section{Acknowledgments}
TD acknowledges useful discussion with Yves Kwan, Steve Kivelson, Mike Zaletel, Felix Desrochers, Ashvin Vishwanath. T.T. is supported by the Department of Energy, Laboratory Directed Research and Development program at SLAC National Accelerator Laboratory, under contract DE-AC02-76SF00515.
ZDS was supported by a Leinweber Institute for Theoretical Physics postdoctoral fellowship at Stanford University and in part by the Gordon and Betty Moore Foundation EPiQS initiative, Grant GBMF8686.01. 
This material is based upon work supported by the Air Force Office of Scientific Research under award number FA9550-25-1-0343. 

\bibliography{main_refs}

\clearpage
\onecolumngrid


\section*{Supplementary material}

# Supplemental Material for "Skyrmion Fractional Chern Insulator: An Intrinsically Multiband Route to Fractionalization in Rhombohedral Graphene

Julian May-Mann,[1,*] Tixuan Tan (谈体轩),[1,*] Patrick J. Ledwith,[2] Zhengyan Darius Shi (石铮岩),[1] and Trithep Devakul[1,†]

[1]*Department of Physics, Stanford University, Stanford, CA 94305, USA*
[2]*Department of Physics, Massachusetts Institute of Technology, Cambridge, MA 02139, USA*

## CONTENTS



## I. EXTENDED DISCUSSION OF THE PARTON EFFECTIVE FIELD THEORY

In this section, we provide additional details of the parton effective field theory. In Sec. I A we discuss the parton decomposition, gauge redundancy, and particle-vortex duality used in the main text. In Sec. I B we discuss the parton

* These authors contributed equally.
† tdevakul@stanford.edu

effective field theory for the $C = 1$ skyrmion lattice state at $\nu = 1$. In Sec. I C we discuss the parton effective field theory for the conventional fractional Chern insulator (FCI).

### A. Parton decomposition and particle-vortex duality

The starting point for our parton effective theory is the decomposition of the layer-dependent fermionic operator $c_\ell = f z_\ell$, where $f$ is a fermionic parton, and $z_\ell$ is a multi-component bosonic parton that carries the layer index. In this decomposition, there is an emergent U(1) gauge ambiguity, $f \to f e^{i\theta}$, $z_\ell \to z_\ell e^{-i\theta}$ that leaves the physical $c_\ell$ invariant. To enforce invariance under such gauge transformations, we include an emergent U(1) gauge field $a$. In the coarse-grained continuum limit, the effective Lagrangian for $f$ and $z_\ell$ can be written as

$$\mathcal{L} = \mathcal{L}_f[f, a + A] + \sum_\ell \mathcal{L}_{z_\ell}[z_\ell, -a] \tag{1}$$

where we have included a probe physical electromagnetic gauge field $A$ to keep track of the physical charge. Without loss of generality, we have chosen to assign the charge to $f$. We could have just as easily assigned the charge to $z_\ell$, which amounts to a shift $a \to a - A$.

For the ideal limit of RMG, there is a recursive relationship that relates the different $z_\ell$'s to each other in the low-energy subspace (see Sec. IV A). Working in this low-energy subspace, we can integrate out the $\ell \geq 1$ component of $z_\ell$ to arrive at a parton effective theory for just $f$ and $z = z_0$,

$$\mathcal{L} = \mathcal{L}_f[f, a + A] + \mathcal{L}_z[z, -a]. \tag{2}$$

where $\mathcal{L}_z$ implicitly accounts for renormalization arising from integrating out the $\ell \geq 1$ components of the bosonic parton.

Let us now discuss the details of the particle-vortex duality used in the main text. For our purposes, we are interested in the skyrmion vacancies, which are defined relative to the $C = 1$ skyrmion lattice configuration that occurs at $\nu = 1$. We will discuss this state in more detail in Sec. I B, but, for the time being, we will use the fact that the skyrmion lattice corresponds to a fixed antivortex lattice configuration of $z$ (the recursive relationship of the bosonic parton implies that antivortices of $z_0$ correspond to skyrmions of $z_\ell$), and skyrmion vacancies are vortices of $z$ defined *relative* to the antivortex lattice configuration.

With this in mind, let us define the classical skyrmion lattice configuration as $z_0^{\mathrm{SkL}} \equiv |z_0^{\mathrm{SkL}}| e^{-i\theta_{\mathrm{SkL}}}$. Since there is one skyrmion per unit cell, $\theta_{\mathrm{SkL}}$ will wind by $2\pi$ per unit cell. Note that $\theta_{\mathrm{SkL}}$ is singular at vortex cores, but this will not be important in the coarse-grained limit we are considering. Using $\theta_{\mathrm{SkL}}$ we can define a phase-rotated bosonic parton, $\tilde{z} = z e^{i\theta_{\mathrm{SkL}}}$, such that the phase of $\tilde{z}$ is constant in the skyrmion lattice configuration, and skyrmion vacancies correspond to the vortices of $\tilde{z}$. Introducing the static field $\omega_\mu^{\mathrm{SkL}} = \partial_\mu \theta_{\mathrm{SkL}}$ ($\mu = t, x, y$), the Lagrangian in Eq. 2 can be rewritten in terms of $\tilde{z}$ as

$$\mathcal{L} = \mathcal{L}_f[f, a + A] + \mathcal{L}_z[\tilde{z}, -a + \omega^{\mathrm{SkL}}]. \tag{3}$$

Since the skyrmion lattice has one skyrmion per unit cell, there is one flux of $\omega^{\mathrm{SkL}}$ per unit cell, and one can view $\boldsymbol{\nabla} \times \boldsymbol{\omega}^{\mathrm{SkL}}/2\pi$ as the background skyrmion density.

Since $\tilde{z}$ is uniform in the skyrmion lattice state, and skyrmion vacancies correspond to simple vortices of $\tilde{z}$, we can implement a particle-vortex duality on $\tilde{z}$ [1–3]

$$\mathcal{L}_z[z, -a] = \mathcal{L}_z[\tilde{z}, -a + \omega^{\mathrm{SkL}}] \to \mathcal{L}_v[v, -b - A^{\mathrm{Sk}}] - \frac{1}{2\pi} b da + \frac{1}{2\pi}(b + A^{\mathrm{Sk}}) d\omega^{\mathrm{SkL}} \tag{4}$$

where $v$ is the vortex dual of $\tilde{z}$, $b$ is the emergent gauge field associated with the particle-vortex duality, and we have included a new probe gauge field $A^{\mathrm{Sk}}$ to keep track of the skyrmion number. We have also abbreviated $bda = b \wedge da$.

For the remaining subsections, we will need the following equations of motion for the $a$ and $b$ gauge fields,

$$\frac{\boldsymbol{\nabla} \times \boldsymbol{a}}{2\pi} = 1 - \rho_v, \qquad \frac{\boldsymbol{\nabla} \times \boldsymbol{b}}{2\pi} = \rho_f \tag{5}$$

where $\rho_v$ and $\rho_f$ are the densities of the vortex field $v$ and fermionic field $f$ respectively.

### B. Skyrmion lattice state

In the skyrmion lattice at $\nu = \rho_f = 1$, there are no skyrmion vacancies. In the effective field theory, this amounts to placing $v$ in a trivial insulating phase. In this case, $v$ can be trivially integrated out

$$\mathcal{L}_v[v, -b - A^{\text{Sk}}] \to 0, \tag{6}$$

where we are ignoring any non-topological contributions, which can, in general, be non-zero when $v$ is in a trivial insulating phase. After integrating out $v$, the gauge field $b$ acts as a Lagrange multiplier fixing $a = \omega^{SkL}$. The effective Lagrangian is then simply

$$\mathcal{L} = \mathcal{L}_f[f, \omega^{\text{SkL}} + A] + \frac{1}{2\pi} A^{\text{Sk}} d\omega^{\text{SkL}} \tag{7}$$

The $f$ fermions therefore feel one magnetic flux per unit cell, and since they are also at $\rho_f = 1$, they can enter an integer quantum Hall state. After integrating out the gapped $f$ fermion, the effective Lagrangian is

$$\mathcal{L}_{\text{eff}} = \frac{1}{4\pi}\left(\omega^{\text{SkL}} + A\right) d \left(\omega^{\text{SkL}} + A\right) + \frac{1}{2\pi} A^{\text{Sk}} d\omega^{\text{SkL}} \tag{8}$$

Taking functional derivatives with respect to $A$ and $A^{\text{Sk}}$ we find that this state has all the quantum numbers expected for the $C = 1$ skyrmion lattice at $\nu = 1$: there is a single charge and skyrmion per unit cell, and there is a $C = 1$ quantized Hall response.

### C. Conventional FCI

Let us now consider a conventional $\frac{2}{3}$ FCI that is formed by doping conventional quasi-particles of the skyrmion lattice, instead of skyrmion vacancies. This amounts to keeping $\rho_v = 0$, but doping $f$ such that $\rho_f = 2/3$. Again, since there are no skyrmion vacancies $v$ remains in a trivial insulating phase, and we again arrive at eq. 7.

Here, since the $f$ fermions are at magnetic filling 2/3 they can be placed into a 2/3 fractional quantum Hall state. Integrating out the $f$ fermions, we arrive at the effective Lagrangian

$$\mathcal{L}_{\text{eff}} = \frac{3}{4\pi} cdc - \frac{1}{2\pi} cd\left(\omega^{\text{SkL}} + A\right) + \frac{1}{4\pi}\left(\omega^{\text{SkL}} + A\right) d\left(\omega^{\text{SkL}} + A\right) + \frac{1}{2\pi} A^{\text{Sk}} d\omega^{\text{SkL}} \tag{9}$$

where $c$ is a new gauge field that reflects the non-trivial topology of the $-1/3$ state of holes. Note that here the anyons carry charge 1/3, but do not carry any fractional skyrmion number. The skyrmion number is fixed to be equal to $\boldsymbol{\nabla} \times \boldsymbol{\omega}^{\text{SkL}}/2\pi$, i.e., one skyrmion per unit cell. The Hall conductivity and charge density of this state are both 2/3 in relative units. Note that there is no skyrmion-electric Hall effect or skyrmion-skyrmion Hall effect here.

It is worth noting that adding a non-dynamic skyrmion to the conventional FCI (which is a flux of $\omega^{\text{SkL}}$ in Eq. 9) binds a net charge 2/3, as the skyrmions act as the background magnetic field of an effective fractional quantum Hall state. In contrast, for the skyrmion FCI discussed in the main text, such a background skyrmion instead binds a charge 1.

## II. REVIEW OF THE $\nu = \frac{2}{3}$ FRACTIONAL QUANTUM HALL WAVEFUNCTION ON THE TORUS

In this section, we review the explicit constructions of fractional quantum Hall wavefunctions on a torus at filling factors $\nu = \frac{1}{3}$ and $\nu = \frac{2}{3}$. The results developed here will be used later to construct skyrmion-vacancy states in the ideal limit of rhombohedral multilayer graphene.

We begin in Sec. II A by reviewing the properties of single-particle magnetic Bloch states in the lowest Landau level (LLL). In Sec. II B, we explicitly construct the $\nu = \frac{1}{3}$ Laughlin state on a torus. We then discuss two constructions of the $\nu = \frac{2}{3}$ fractional quantum Hall state: direct particle-hole transformation of the $\nu = \frac{1}{3}$ Laughlin state in Sec. II C, and the Haldane–Halperin hierarchy construction in Sec. II D. In Sec. IV, we show that these two constructions motivate two corresponding routes for constructing fractionalized skyrmion-vacancy wavefunctions. Throughout this section, we present the relevant expressions in forms suitable for numerical evaluation.

### A. Magnetic Bloch States

When studying the LLL on a torus, it is useful to consider the magnetic Bloch-state basis, $\phi_{\boldsymbol{k}}^{\text{LLL}}(\boldsymbol{r})$. The normalized magnetic Bloch states can be represented using the modified Weierstrass sigma function:

$$\phi_{\boldsymbol{k}}^{\text{LLL}}(\boldsymbol{r}) = \mathcal{N}^{\text{LLL}} \exp\left(\frac{i}{2}(\bar{k} - \frac{\bar{b}_1 + \bar{b}_2}{2})z\right)\sigma(z/l_B + il_B(k - \frac{b_1 + b_2}{2})|\{\frac{a_\mu}{l_B}\})\exp\left(-\frac{l_B^2 k\bar{k}}{4} - \frac{z\bar{z}}{4l_B^2}\right)\exp\left(\frac{l_B^2 k(\bar{b}_1 + \bar{b}_2)}{4}\right). \tag{10}$$

Here, $z \equiv x + iy$ and $\bar{z} \equiv x - iy$ for a two-dimensional real vector $\boldsymbol{z} = (x, y)$. In the following, we will frequently switch between the two-dimensional vector and complex-scalar representations. $\mathcal{N}^{\text{LLL}}$ is the normalization constant, the explicit expression for which will be given below. $\boldsymbol{a}_1$ and $\boldsymbol{a}_2$ are the real-space basis vectors spanning the magnetic lattice on which the magnetic Bloch states are defined. $\boldsymbol{b}_\mu$ are the corresponding reciprocal lattice basis vectors, satisfying $\boldsymbol{a}_\mu \cdot \boldsymbol{b}_\nu = 2\pi\delta_{\mu\nu}$. $l_B$ is the magnetic length, defined via $|\boldsymbol{a}_1 \times \boldsymbol{a}_2| = 2\pi l_B^2$. $\sigma(z|\{a_\mu\})$ is the modified Weierstrass sigma function. In Mathematica, these modified sigma functions are implemented as

$$\sigma(z|\{a_\mu\}) = \exp\big(-\alpha z^2\big)\text{WeierstrassSigma}[z, \text{WeierstrassInvariants}\{a_1/2, a_2/2\}], \tag{11}$$

where, $\alpha$ is further defined through the Weierstrass zeta function $\eta_\mu$:

$$\alpha \equiv \frac{\eta_1}{a_1} - \frac{\bar{a}_1}{4a_1} = \frac{\eta_2}{a_2} - \frac{\bar{a}_2}{4a_2}. \tag{12}$$

In Mathematica, $\eta_\mu$ can be computed using

$$\eta_\mu = \text{WeierstrassZeta}[\frac{a_\mu}{2}, \text{WeierstrassInvariants}[\{\frac{a_1}{2}, \frac{a_2}{2}\}]]. \tag{13}$$

The magnetic Bloch states obey the following magnetic translation symmetry,

$$\phi_{\boldsymbol{k}}^{\text{LLL}}(\boldsymbol{r} + \boldsymbol{a}_\mu) = \phi_{\boldsymbol{k}}^{\text{LLL}}(\boldsymbol{r})\exp(i\boldsymbol{k} \cdot \boldsymbol{a}_\mu)\exp\left(-i\frac{\boldsymbol{r} \times \boldsymbol{a}_\mu}{2l_B^2}\right). \tag{14}$$

This can be confirmed using the quasi-periodicity of the modified sigma function,

$$\sigma(z + a_\mu|\{a_\mu\}) = -\exp\left(\frac{\bar{a}_\mu}{2}(z + \frac{a_\mu}{2})\right)\sigma(z|\{a_\mu\}). \tag{15}$$

Note that the modified sigma function only has one simple zero at $z = 0$ per fundamental domain (which is the parallelogram spanned by $\boldsymbol{a}_1$ and $\boldsymbol{a}_2$). In addition, the magnetic Bloch states (Eq. 10) obey the momentum-space boundary conditions,

$$\frac{\phi_{\boldsymbol{k}+\boldsymbol{b}_\mu}^{\text{LLL}}(\boldsymbol{r})}{\phi_{\boldsymbol{k}}^{\text{LLL}}(\boldsymbol{r})} = \exp\left(-\frac{il_B^2\boldsymbol{k} \times \boldsymbol{b}_\mu}{2}\right). \tag{16}$$

The final aspect of Eq. 10 to discuss is the normalization constant $\mathcal{N}^{\text{LLL}}$. As we shall now prove, $\mathcal{N}^{\text{LLL}}$ is independent of $\boldsymbol{k}$. For a system defined on a $N_\phi$-flux torus (i.e., the torus contains $N_\phi$ magnetic unit cells) the normalization constant is defined such that:

$$\int_{\text{torus}} d\boldsymbol{r}|\phi_{\boldsymbol{k}}^{\text{LLL}}(\boldsymbol{r})|^2 = 1, \tag{17}$$

Substituting Eq. 10 into the normalization condition gives:

$$(\mathcal{N}^{\text{LLL}})^2\int_{\text{torus}} d\boldsymbol{r}\exp\left(\frac{l_B^2|b_1 + b_2|^2}{8}\right)|\sigma(\frac{z}{l_B} + il_B(k - \frac{b_1 + b_2}{2})|\{\frac{a_\mu}{l_B}\})|^2\exp\left(-\frac{|\frac{z}{l_B} + il_B(k - \frac{b_1+b_2}{2})|^2}{2}\right) = 1. \tag{18}$$

The integrand is periodic under the shift $z \to z + a_\mu$. The momentum $k$ only occurs as a shift of the integration variable, which does not change the integration result. The normalization constant is therefore independent of $\boldsymbol{k}$ and is given explicitly by

$$\mathcal{N}^{\text{LLL}} = \left(\frac{N_\phi l_B^2}{\sqrt{2}(\text{Im}(\tau))^{\frac{3}{2}}|\eta_D(\tau)|^6}\exp\left(\frac{l_B^2|b_1 + b_2|^2}{8}\right)\right)^{-\frac{1}{2}}. \tag{19}$$

Here, $\tau \equiv \frac{a_2}{a_1}$, and $\eta_D(\tau)$ is the Dedekind eta function, which is implemented in Mathematica as DedekindEta[$\tau$].

Finally, for later reference, we provide the expression for magnetic form factors,

$$F_{\boldsymbol{k}_1,\boldsymbol{k}_2}(\boldsymbol{q}) \equiv \int_{\text{torus}} d\boldsymbol{r}\phi^{\text{LLL}*}_{\boldsymbol{k}_1}(\boldsymbol{r})\exp(-i\boldsymbol{q}\cdot\boldsymbol{r})\phi^{\text{LLL}}_{\boldsymbol{k}_2}(\boldsymbol{r})$$
$$= \exp\left(-\frac{l_B^2|\boldsymbol{q}|^2}{4}\right)\exp\left(\frac{il_B^2\boldsymbol{k}_2\times\boldsymbol{q}}{2}\right)(-1)^{n_1n_2}\exp\left(\frac{il_B^2\boldsymbol{k}_1\times\boldsymbol{g}}{2}\right)\delta^{\text{mBZ}}_{\boldsymbol{k}_1,\boldsymbol{k}_2-\boldsymbol{q}}, \quad \boldsymbol{g}\equiv\boldsymbol{k}_1+\boldsymbol{q}-\boldsymbol{k}_2 = n_1\boldsymbol{b}_1+n_2\boldsymbol{b}_2, \tag{20}$$

where, $\delta^{\text{mBZ}}_{\boldsymbol{k}_1,\boldsymbol{k}_2}$ is non-zero only when the two wavevectors are the same modulo the magnetic reciprocal lattice basis vectors $\boldsymbol{b}_\mu$.This expression is derived using the magnetic algebra, as reviewed in the supplementary material of Ref. [4], and is useful for exact diagonalization calculations in the LLL.

### B. $\nu = \frac{1}{3}$ Laughlin state on the torus

We now consider the quantum Hall system on an $N_\phi$-flux torus. We take the torus to be spanned by vectors $\boldsymbol{L}_\mu$ ($\mu = 1, 2$), satisfying

$$\boldsymbol{L}_\mu = N_\mu\boldsymbol{a}_\mu, \quad N_\mu\in\mathbb{Z}^+, \quad L_B^2 \equiv \frac{|\boldsymbol{L}_1\times\boldsymbol{L}_2|}{2\pi} = N_\phi l_B^2, \quad N_\phi\equiv N_1N_2. \tag{21}$$

The corresponding reciprocal lattice basis vectors $\boldsymbol{T}_1$ and $\boldsymbol{T}_2$ satisfy $\boldsymbol{T}_\mu\cdot\boldsymbol{L}_\nu = 2\pi\delta_{\nu\mu}$. $L_B$ is an auxiliary length scale associated with one flux quantum through the torus, whereas $l_B$ is the physical magnetic length. The fully filled Landau level wavefunction is a $N_\phi$-electron Slater determinant of the magnetic Bloch states:

$$\Psi^{\nu=1,\text{LLL}}(\{\boldsymbol{r}\};\boldsymbol{\Delta\theta}) \equiv \det_{\boldsymbol{k}\in\text{BZ1},p}[\phi^{\text{LLL}}_{\boldsymbol{k}-\boldsymbol{\Delta\theta}}(\boldsymbol{r}_p)], \quad \boldsymbol{k}\in\text{BZ1}\Leftrightarrow\boldsymbol{k}=n_1\boldsymbol{T}_1+n_2\boldsymbol{T}_2, \quad n_\mu = 0,1..N_\mu-1. \tag{22}$$

We use BZ1 to denote the discrete set of $N_\phi$-element momenta generated by $\boldsymbol{T}_\mu$ within the magnetic Brillouin zone spanned by $\boldsymbol{b}_\mu$. The momenta $\boldsymbol{k}\in\text{BZ1}$ satisfy $\exp(i\boldsymbol{k}\cdot\boldsymbol{L}_\mu)=1$. The vector $\boldsymbol{\Delta\theta}$ parametrizes the twisted boundary condition of $\Psi^{\nu=1,\text{LLL}}(\{\boldsymbol{r}\};\boldsymbol{\Delta\theta})$:

$$\frac{\Psi^{\nu=1,\text{LLL}}(\boldsymbol{r}_1...,\boldsymbol{r}_p+\boldsymbol{L}_\mu,\boldsymbol{r}_{p+1},...;\boldsymbol{\Delta\theta})}{\Psi^{\nu=1,\text{LLL}}(\boldsymbol{r}_1...,\boldsymbol{r}_p,\boldsymbol{r}_{p+1},...;\boldsymbol{\Delta\theta})} = \exp\left(-\frac{i\boldsymbol{r}_p\times\boldsymbol{L}_\mu}{2l_B^2}\right)\exp(-i\boldsymbol{\Delta\theta}\cdot\boldsymbol{L}_\mu). \tag{23}$$

We will later study the evolution of the wavefunction under flux threading by varying $\boldsymbol{\Delta\theta}$.

The fully filled LLL wavefunction has a particularly useful form, which we now derive. When two particles approach the same position, the Slater determinant vanishes. Moreover, up to the overall Gaussian factor, $\Psi^{\nu=1,\text{LLL}}(\{\boldsymbol{r}\};\boldsymbol{\Delta\theta})$ is holomorphic in each coordinate $z_p$ and antisymmetric under the exchange of any two electron coordinates. It must therefore contain the pairwise factor $\prod_{p<q}\sigma\left(\frac{z_p-z_q}{L_B}\middle|\left\{\frac{L_\mu}{L_B}\right\}\right)$. We consequently write

$$\Psi^{\nu=1,\text{LLL}}(\{\boldsymbol{r}\};\boldsymbol{\Delta\theta}) = \prod_p\exp\left(-\frac{z_p\bar{z}_p}{4l_B^2}\right)\prod_{p<q}\sigma(\frac{z_p-z_q}{L_B}|\{\frac{L_\mu}{L_B}\})\times\underbrace{\text{Something holomorphic in all z}}_{F(\{z\};\boldsymbol{\Delta\theta})} \tag{24}$$

The quasi-periodicity lattice of $\sigma\left(\frac{z_i-z_j}{L_B}\middle|\left\{\frac{L_\mu}{L_B}\right\}\right)$ is generated by $L_1$ and $L_2$, so its fundamental domain coincides with the entire torus. This function has one simple zero in each fundamental domain. We will refer to sigma functions with these periods as torus sigma functions, in order to distinguish them from sigma functions associated with smaller fundamental domains. To match the boundary condition of the two sides of the equation above, we find that $F(\{z\})$ must obey the boundary condition,

$$\frac{F(\{z_1...z_p+L_\mu,z_{p+1}...\};\boldsymbol{\Delta\theta})}{F(\{z\};\boldsymbol{\Delta\theta})} = \exp\left(\frac{L_\mu\bar{L}_\mu}{4L_B^2}\right)\exp\left(\frac{\bar{L}_\mu Z}{2L_B^2}\right)(-1)^{N_1N_2-1}\exp(-i\boldsymbol{\Delta\theta}\cdot\boldsymbol{L}_\mu), \quad Z\equiv\sum_{p=1}^{N_\phi}z_p. \tag{25}$$

Here, we have introduced the center of mass (COM) coordinate $Z$. The equation above has a unique solution, up to

an overall normalization constant [1]:

$$F(\{z\},\boldsymbol{\Delta\theta}) \propto F^{\nu=1,\mathrm{LLL}}(Z;\boldsymbol{\Delta\theta}) = \sigma(\frac{Z}{L_B} + i\frac{L_B N_1 N_2}{2}(T_1+T_2) - iL_B\Delta\theta|\{\frac{L_\mu}{L_B}\})\exp\Big((\frac{iN_1N_2(\bar{T}_2+\bar{T}_1)}{4} - i\frac{\Delta\bar{\theta}}{2})Z\Big), \tag{26}$$

such that

$$\Psi^{\nu=1,\mathrm{LLL}}(\{\boldsymbol{r}\};\boldsymbol{\Delta\theta}) \propto \prod_{p<q}\sigma(\frac{z_p-z_q}{L_B}|\{\frac{L_\mu}{L_B}\})\prod_p \exp\Big(-\frac{z_p\bar{z}_p}{4l_B^2}\Big)F^{\nu=1,\mathrm{LLL}}(Z;\boldsymbol{\Delta\theta}). \tag{27}$$

Here, the proportionality constant is independent of $\{\boldsymbol{r}\}$, but can be dependent on other factors, such as the boundary condition twist $\boldsymbol{\Delta\theta}$.

To construct the Laughlin state at $\nu=\frac{1}{3}$, we replace the first power of the pairwise sigma-function factor in Eq. 27 by its third power, reduce the electron number from $N_\phi$ to $\frac{N_\phi}{3}$, and replace the COM factor $F^{\nu=1,\mathrm{LLL}}$ with one of three new COM factors, $F^{(s)}$ for $s=0,1,2$:

$$\Psi_{(s)}^{\nu=\frac{1}{3},\mathrm{Laughlin}}(\{\boldsymbol{r}\};\boldsymbol{\Delta\theta}) = \prod_{i<j}\sigma^3(\frac{z_i-z_j}{L_B}|\{\frac{L_\mu}{L_B}\})\prod_i\exp\Big(-\frac{z_i\bar{z}_i}{4l_B^2}\Big)F^{(s)}(Z;\boldsymbol{\Delta\theta}), \quad Z=\sum_{p=1}^{N_\phi/3} z_p. \tag{28}$$

Here, $s$ labels the three topological sectors. Requiring the Laughlin state to obey the twisted boundary condition,

$$\frac{\Psi_{(s)}^{\nu=\frac{1}{3},\mathrm{Laughlin}}(\boldsymbol{r}_1...,\boldsymbol{r}_p+\boldsymbol{L}_\mu,\boldsymbol{r}_{p+1},...;\boldsymbol{\Delta\theta})}{\Psi_{(s)}^{\nu=\frac{1}{3},\mathrm{Laughlin}}(\boldsymbol{r}_1...,\boldsymbol{r}_p,\boldsymbol{r}_{p+1},...;\boldsymbol{\Delta\theta})} = \exp\Big(-\frac{i\boldsymbol{r}_p\times\boldsymbol{L}_\mu}{2l_B^2}\Big)\exp(-i\boldsymbol{\Delta\theta}\cdot\boldsymbol{L}_\mu), \tag{29}$$

gives the following boundary condition for the COM factor,

$$\frac{F^{(s)}(Z+L_\mu;\boldsymbol{\Delta\theta})}{F^{(s)}(Z;\boldsymbol{\Delta\theta})} = (-1)^{3(N_e-1)}\exp\Big(\frac{3\bar{L}_\mu Z}{2L_B^2}\Big)\exp\Big(\frac{3L_\mu\bar{L}_\mu}{4L_B^2}\Big)\exp(-i\boldsymbol{L}_\mu\cdot\boldsymbol{\Delta\theta}), \quad N_e=\frac{N_\phi}{3}. \tag{30}$$

This equation has three linearly independent solutions, labeled by $s$. Since $N_\phi$ is divisible by 3, we will, without loss of generality, assume $N_1$ is divisible by 3 and use the following ansatz, which is parameterized by $\alpha^{(s)}$ and $\bar{\beta}^{(s)}$,

$$F^{(s)}(Z;\boldsymbol{\Delta\theta}) = \sigma(\frac{Z+i\alpha^{(s)}}{L_{Bf}}|\{\frac{A_\mu}{L_{Bf}}\})\exp\Big(i\bar{\beta}^{(s)}Z\Big), \tag{31}$$

where the auxiliary lattice is defined by

$$\boldsymbol{A}_1 = \frac{1}{3}\boldsymbol{L}_1, \quad \boldsymbol{A}_2=\boldsymbol{L}_2, \quad |\boldsymbol{A}_1\times\boldsymbol{A}_2| = 2\pi L_{Bf}^2. \tag{32}$$

Matching this ansatz to the boundary condition in Eq. 30 gives

$$(-1)\exp\Big(\frac{3\bar{L}_\mu(Z+i\alpha^{(s)})}{2L_B^2}\Big)\exp\Big(i\bar{\beta}^{(s)}L_\mu\Big) = (-1)^{3(N_e-1)}\exp\Big(\frac{3\bar{L}_\mu Z}{2L_B^2}\Big)\exp(-i\boldsymbol{L}_\mu\cdot\boldsymbol{\Delta\theta}), \tag{33}$$

and the resulting solutions are

$$\beta^{(s)} = \frac{3\alpha^{(s)}}{2L_B^2} \tag{34}$$

$$\alpha^{(s)} = -\frac{L_B^2\Delta\theta}{3} + \frac{N_e L_B^2}{3}(\frac{T_1+T_2}{2}) + \frac{sL_B^2T_1}{3}, \quad s=0,1,2. \tag{35}$$

When one flux quantum is threaded in the $\boldsymbol{L}_1$ direction, $\Delta\theta\to\Delta\theta+T_1$, and consequently $\alpha^{(s)}\to\alpha^{(s-1)}$, where the sector index is understood modulo 3. Thus, threading one flux quantum permutes the three states, while each

[1] The exact proof of this statement is beyond the scope of this work

individual state returns to itself after threading three flux quanta. This is the expected spectral flow of the threefold topologically degenerate Laughlin ground-state manifold. The zero of $F^{(s)}(Z)$ is located at $Z = -i\alpha^{(s)}$. Since the three zeros occur at distinct positions modulo the fundamental domain spanned by $\boldsymbol{A}_\mu$, the corresponding solutions are linearly independent.

Eq. 31 can also be written in terms of torus sigma functions by matching the positions of zeros inside the torus spanned by $\boldsymbol{L}_\mu$:

$$F^{(s)}(Z, \boldsymbol{\Delta\theta}) \propto \exp\Big(i\bar{\beta}^{(s)}Z\Big)\exp\Big(-\frac{\bar{L}_1 Z}{2L_B^2}\Big)\prod_{j=0}^{2}\sigma\Big(\frac{Z + i\alpha^{(s)} + j\frac{L_1}{3}}{L_B}\Big|\{\frac{L_\mu}{L_B}\}\Big). \tag{36}$$

Using the explicit expression of $F^{(s)}(Z; \boldsymbol{\Delta\theta})$, we can derive that these three solutions have definite total Bloch momentum,

$$\frac{\Psi_{(s)}^{\nu=\frac{1}{3},\text{Laughlin}}(\{\boldsymbol{r} + \boldsymbol{a}_\mu\}; \boldsymbol{\Delta\theta})}{\Psi_{(s)}^{\nu=\frac{1}{3},\text{Laughlin}}(\{\boldsymbol{r}\}; \boldsymbol{\Delta\theta})} = \exp\Big(-\frac{i(\sum_p \boldsymbol{r}_p) \times \boldsymbol{a}_\mu}{2l_B^2}\Big)\exp\Big(i\boldsymbol{a}_\mu \cdot \Big(\boldsymbol{k}_{(s)}^{\text{tot}} - N_e\boldsymbol{\Delta\theta}\Big)\Big), \tag{37}$$

where

$$\boldsymbol{k}_{(s)}^{\text{tot}} = (\frac{N_e^2}{2} + \frac{3N_e}{2})(\boldsymbol{T}_1 + \boldsymbol{T}_2) + sN_e\boldsymbol{T}_1. \tag{38}$$

The construction above assumes that $N_1$ is divisible by 3. If instead $N_2$ is divisible by 3 while $N_1$ is not, the ansatz based on $\boldsymbol{A}_1 = \boldsymbol{L}_1/3$ does not produce states with definite total Bloch momentum. In this case, the more natural choice is $\boldsymbol{A}_1 = \boldsymbol{L}_1$ and $\boldsymbol{A}_2 = \boldsymbol{L}_2/3$.

### C. Particle-hole construction of the $\nu = \frac{2}{3}$ state

In this subsection, we construct the $\nu = \frac{2}{3}$ fractional quantum Hall state by applying a particle-hole transformation to $\Psi_{(s)}^{\nu=\frac{1}{3},\text{Laughlin}}(\{\boldsymbol{r}\}; \boldsymbol{\Delta\theta})$. We first present the construction in the second-quantized language and then derive its first-quantized form.

Since the magnetic Bloch states form a complete single-particle basis of the LLL, we can expand the $\nu = \frac{1}{3}$ Laughlin state in the basis of Slater determinants of magnetic Bloch states,

$$\Big|\Psi_{(s)}^{\nu=\frac{1}{3},\text{Laughlin}}(\boldsymbol{\Delta\theta})\Big\rangle = \sum_{\{\boldsymbol{k}\},\sum \boldsymbol{k}=\boldsymbol{k}_{(s)}^{\text{tot}}} C_{\{\boldsymbol{k}\}}^{(s)}(\boldsymbol{\Delta\theta})\prod_{\substack{\boldsymbol{k}_i\in\{\boldsymbol{k}\}\\ I_1<I_2\dots}} c_{\boldsymbol{k}_i-\boldsymbol{\Delta\theta}}^\dagger |0\rangle, \quad \{\boldsymbol{k}\} \subset \text{BZ1}. \tag{39}$$

Here, $|0\rangle$ is the vacuum, and $c_{\boldsymbol{k}}^\dagger$ creates an electron in the magnetic Bloch state $\phi_{\boldsymbol{k}}^{\text{LLL}}(\boldsymbol{r})$. We use $I_i = 1, \dots, N_\phi$ to denote the canonical ordering index of $\boldsymbol{k}_i$ within BZ1. Each momentum set $\{\boldsymbol{k}\}$ contains $\frac{N_\phi}{3}$ elements. Since the Laughlin state has a definite total magnetic Bloch momentum, only momentum sets $\boldsymbol{k}$ with the corresponding total momentum, $\boldsymbol{k}_{(s)}^{\text{tot}}$ (modulo the magnetic reciprocal lattice basis vectors $\boldsymbol{b}_\mu$), contribute to the expansion. The many-body ket is related to the first-quantized wavefunction through

$$\langle\{\boldsymbol{r}\}|\Psi_{(s)}^{\nu=\frac{1}{3},\text{Laughlin}}(\boldsymbol{\Delta\theta})\rangle \equiv \Psi_{(s)}^{\nu=\frac{1}{3},\text{Laughlin}}(\{\boldsymbol{r}\}; \boldsymbol{\Delta\theta}), \quad \langle\{\boldsymbol{r}\}|\prod_{\substack{\boldsymbol{k}_i\in\{\boldsymbol{k}\}\\ I_1<I_2\dots}} c_{\boldsymbol{k}_i-\boldsymbol{\Delta\theta}}^\dagger|0\rangle = \frac{1}{\sqrt{(\frac{N_\phi}{3})!}}\det_{\substack{\boldsymbol{k}_i\in\{\boldsymbol{k}\},p\\ I_1<I_2\dots}}[\phi_{\boldsymbol{k}_i-\boldsymbol{\Delta\theta}}^{\text{LLL}}(\boldsymbol{r}_p)] \tag{40}$$

We define an anti-unitary operator $\mathcal{P}$ as,

$$\mathcal{P}i\mathcal{P}^{-1} = -i, \quad \mathcal{P}c_{\boldsymbol{k}-\boldsymbol{\Delta\theta}}^\dagger\mathcal{P}^{-1} = c_{\boldsymbol{k}-\boldsymbol{\Delta\theta}}, \quad \mathcal{P}|0\rangle = \prod_{\substack{\boldsymbol{k}_i\in\text{BZ1}\\ I_1<I_2\dots}} c_{\boldsymbol{k}_i-\boldsymbol{\Delta\theta}}^\dagger|0\rangle \tag{41}$$

Using $\mathcal{P}$, the particle-hole construction of the $\nu = \frac{2}{3}$ fractional quantum Hall state is defined as:

$$\Big|\Psi_{(s)}^{\nu=\frac{2}{3},\text{PH}}(\boldsymbol{\Delta\theta})\Big\rangle = \mathcal{P}\Big|\Psi_{(s)}^{\nu=\frac{1}{3},\text{Laughlin}}(\boldsymbol{\Delta\theta})\Big\rangle = \sum_{\{\boldsymbol{k}\},\sum \boldsymbol{k}=\boldsymbol{k}_{(s)}^{\text{tot}}} C_{\{\boldsymbol{k}\}}^{(s)*}(\boldsymbol{\Delta\theta})\text{sgn}(\{\boldsymbol{k}\})\prod_{\boldsymbol{k}_i\notin\{\boldsymbol{k}\},I_1<I_2\dots} c_{\boldsymbol{k}_i-\boldsymbol{\Delta\theta}}^\dagger|0\rangle. \tag{42}$$

Here, $\mathrm{sgn}(\{\boldsymbol{k}\}) = (-1)^{\sum_{\boldsymbol{k}_i\in\{\boldsymbol{k}\}}(I_i-1)}$ is the fermion permutation sign. This is the particle-hole construction of the $\nu=\frac{2}{3}$ state in second-quantization language.

Using the second-quantized formulation, we have the following first-quantization wavefunction:

$$\Psi_{(s)}^{\nu=\frac{2}{3},\mathrm{PH}}(\{\boldsymbol{r}\};\boldsymbol{\Delta\theta}) \equiv \left\langle \{\boldsymbol{r}\} \middle| \Psi_{(s)}^{\nu=\frac{2}{3},\mathrm{PH}}(\boldsymbol{\Delta\theta}) \right\rangle = \frac{1}{\sqrt{(\frac{2N_\phi}{3})!}} \sum_{\{\boldsymbol{k}\},\sum \boldsymbol{k}=\boldsymbol{k}^{\mathrm{tot}}_{(s)}} C^{(s)*}_{\{\boldsymbol{k}\}}(\boldsymbol{\Delta\theta})\mathrm{sgn}(\{\boldsymbol{k}\}) \det_{\substack{\boldsymbol{k}_i\notin\{\boldsymbol{k}\},p\\ I_1<I_2...}}[\phi^{\mathrm{LLL}}_{\boldsymbol{k}_i-\boldsymbol{\Delta\theta}}(\boldsymbol{r}_p)] \tag{43}$$

We can also find this wavefunction using the Laplace expansion of determinants. For the fully filled Landau level, this expansion gives

$$\begin{aligned}\Psi^{\nu=1,\mathrm{LLL}}(\{\boldsymbol{r}\};\boldsymbol{\Delta\theta}) &= \frac{1}{\sqrt{N_\phi!}} \det_{\boldsymbol{k}_i\in\mathrm{BZ1},p}[\phi^{\mathrm{LLL}}_{\boldsymbol{k}_i-\boldsymbol{\Delta\theta}}(\boldsymbol{r}_p)]\\ &\propto \sum_{\{\boldsymbol{k}\}} \mathrm{sgn}(\{\boldsymbol{k}\}) \det_{\substack{\boldsymbol{k}_i\notin\{\boldsymbol{k}\}\\ p\in[1,N_\phi-|\{\boldsymbol{k}\}|]\\ I_1<I_2...}}[\phi^{\mathrm{LLL}}_{\boldsymbol{k}_i-\boldsymbol{\Delta\theta}}(\boldsymbol{r}_p)] \det_{\substack{\boldsymbol{k}_i\in\{\boldsymbol{k}\}\\ p\in[N_\phi-|\{\boldsymbol{k}\}|+1,N_\phi]\\ I_1<I_2...}}[\phi^{\mathrm{LLL}}_{\boldsymbol{k}_i-\boldsymbol{\Delta\theta}}(\boldsymbol{r}_p)].\end{aligned} \tag{44}$$

In the equation above, the summation over $\{\boldsymbol{k}\}$ is over all possible $\frac{N_\phi}{3}$-element subsets of BZ1, not only those subsets with fixed total momentum. Combining this with the orthogonality relation

$$\int_{\mathrm{torus}} \prod_{p=1}^{N} d\boldsymbol{r}_i \det_{\substack{\boldsymbol{k}_i,p\\ I_1<I_2...}}[\phi^{\mathrm{LLL}*}_{\boldsymbol{k}_i-\boldsymbol{\Delta}_{\boldsymbol{\theta}}}(\boldsymbol{r}_p)] \det_{\substack{\boldsymbol{k}'_i,p\\ I'_1<I'_2...}}[\phi^{\mathrm{LLL}}_{\boldsymbol{k}'_i-\boldsymbol{\Delta}_{\boldsymbol{\theta}}}(\boldsymbol{r}_p)] = N!\prod_{p=1}^{N}\delta_{\boldsymbol{k}_p,\boldsymbol{k}'_p}, \tag{45}$$

we obtain the first-quantized expression for the PH construction of the $\nu=\frac{2}{3}$ fractional quantum Hall states:

$$\begin{aligned}&\Psi_{(s)}^{\nu=\frac{2}{3},\mathrm{PH}}(\{\boldsymbol{r}\};\boldsymbol{\Delta\theta})\\ &\propto \int_{\mathrm{torus}} \prod_a d\boldsymbol{\eta}_a \Psi^{\nu=1,\mathrm{LLL}}(\{\boldsymbol{r},\boldsymbol{\eta}\};\boldsymbol{\Delta\theta})\Psi_{(s)}^{\nu=\frac{1}{3},\mathrm{Laughlin}*}(\{\boldsymbol{\eta}\};\boldsymbol{\Delta\theta})\\ &\propto \int_{\mathrm{torus}} \prod_a d\boldsymbol{\eta}_a \prod_{p<q}\sigma(\frac{z_p-z_q}{L_B}|\{\frac{L_\mu}{L_B}\})\prod_{p,a}\sigma(\frac{z_p-\eta_a}{L_B}|\{\frac{L_\mu}{L_B}\})\prod_{a<b}\sigma(\frac{\eta_a-\eta_b}{L_B}|\{\frac{L_\mu}{L_B}\})\prod_{a<b}(\sigma^*(\frac{\eta_a-\eta_b}{L_B})|\{\frac{L_\mu}{L_B}\})^3\\ &\exp\left(-\frac{\sum_p z_p\bar{z}_p + 2\sum_a \eta_a\bar{\eta}_a}{4l_B^2}\right)F^{\nu=1,\mathrm{LLL}}(Z+H;\boldsymbol{\Delta\theta})F^{(s)*}(H,\boldsymbol{\Delta\theta}).\end{aligned} \tag{46}$$

In the equation above, we have defined the COM coordinate for $\eta$:

$$H = \sum_{a=1}^{\frac{N_\phi}{3}} \eta_a. \tag{47}$$

### D. Haldane-Halperin construction of the $\nu = \frac{2}{3}$ state

The Haldane-Halperin construction of the $\nu=\frac{2}{3}$ state is best understood in the first-quantized language [5, 6]. We notice that in Eq. 46, for each configuration of $\{\boldsymbol{\eta}\}$, there is a Jastrow factor $\prod_{a<b}|\sigma(\frac{\eta_a-\eta_b}{L_B}|\{\frac{L_\mu}{L_B}\})|^2$. The Haldane-Halperin construction of the $\nu=\frac{2}{3}$ fractional quantum Hall state is obtained by dividing the integrand of Eq. 46 by $F^{\mathrm{Jastrow}}(\{\boldsymbol{\eta}\})$:

$$F^{\mathrm{Jastrow}}(\boldsymbol{\eta}_1,\boldsymbol{\eta}_2...\boldsymbol{\eta}_{\frac{N_\phi}{3}}) = \prod_{a<b}|\sigma(\frac{\eta_a-\eta_b}{L_B}|\{\frac{L_\mu}{L_B}\})|^2\prod_a \exp\left(-\frac{\eta_a\bar{\eta}_a}{6l_B^2}\right)\exp\left(\frac{H\bar{H}}{2L_B^2}\right) \tag{48}$$

This function is periodic when any of the arguments is translated by the torus spanning vector, or all of its arguments are translated by the same amount:

$$F^{\mathrm{Jastrow}}(\boldsymbol{\eta}_1...,\boldsymbol{\eta}_a+\boldsymbol{L}_\mu,...) = F^{\mathrm{Jastrow}}(\{\boldsymbol{\eta}\}) \tag{49}$$

$$F^{\text{Jastrow}}(\boldsymbol{\eta}_1+\boldsymbol{x},\boldsymbol{\eta}_2+\boldsymbol{x},...)=F^{\text{Jastrow}}(\{\boldsymbol{\eta}\}),\quad \forall \boldsymbol{x}. \tag{50}$$

The full Haldane-Halperin state wavefunction is given by

$$\begin{aligned}\Psi_{(s)}^{\nu=\frac{2}{3},\text{HH}}(\{\boldsymbol{r}\};\boldsymbol{\Delta\theta}) \equiv \int_{\text{torus}} \prod_a d\boldsymbol{\eta}_a \prod_{p<q} \sigma(\frac{z_p-z_q}{L_B}|\{\frac{L_\mu}{L_B}\}) \prod_{p,a} \sigma(\frac{z_p-\eta_a}{L_B}|\{\frac{L_\mu}{L_B}\}) \prod_{a<b} (\sigma^*(\frac{\eta_a-\eta_b}{L_B})|\{\frac{L_\mu}{L_B}\})^2 \\ \exp\left(-\frac{\sum_p z_p\bar{z}_p+\frac{4}{3}\sum_a \eta_a\bar{\eta}_a}{4l_B^2}\right)\exp\left(-\frac{H\bar{H}}{2L_B^2}\right)F^{\nu=1,\text{LLL}}(Z+H;\boldsymbol{\Delta\theta})F^{(s)*}(H,\boldsymbol{\Delta\theta}).\end{aligned} \tag{51}$$

Before concluding this review, we make several comments. Using the periodicity of $F^{\text{Jastrow}}(\{\boldsymbol{\eta}\})$, one can show that $\Psi_{(s)}^{\nu=\frac{2}{3},\text{HH}}(\boldsymbol{r};\boldsymbol{\Delta\theta})$ has the same magnetic Bloch periodicity as $\Psi_{(s)}^{\nu=\frac{2}{3},\text{PH}}(\boldsymbol{r};\boldsymbol{\Delta\theta})$ under the simultaneous translation of all electron coordinates by $\boldsymbol{a}_\mu$. Consequently, the Haldane–Halperin state can also be expanded in the basis of Slater determinants of magnetic Bloch states, similarly to Eq. 39.

However, no comparably simple parent Hamiltonian is known for which the $\nu=\frac{2}{3}$ Haldane–Halperin states are exact ground states. Obtaining their Slater-determinant expansion coefficients is therefore numerically difficult. In contrast, the coefficients for the particle-hole construction are inherited directly from the expansion of the $\nu=\frac{1}{3}$ Laughlin state in Eq. 39. They can be obtained either by exact diagonalization using the Trugman–Kivelson pseudopotential or by numerically matching the two sides of Eq. 39 using Eqs. 28 and 10.

Finally, we compare these two constructions of the $\nu=\frac{2}{3}$ fractional quantum Hall state, as defined in Eq. 46 and Eq. 51. Both constructions involve the same number of auxiliary hole coordinates $|\{\boldsymbol{\eta}\}|=\frac{N_\phi}{3}$ and the same number of electrons $|\{\boldsymbol{r}\}|=\frac{2N_\phi}{3}$. As a function of each $\eta_a$, the integrand of Eq. 46 has $N_\phi$ chiral zeros, including the COM contribution, and therefore describes the holes forming a fermionic Laughlin state in the original $N_\phi$-flux Landau level at filling factor $\frac{1}{3}$. On the other hand, as a function of each $\eta_a$, the integrand of Eq. 51 has $\frac{2N_\phi}{3}+1$ chiral zeros, including the COM contribution. The holes therefore form a bosonic Laughlin state in an effective $(2N_\phi/3+1)$-flux Landau level, whose filling factor approaches $\nu=\frac{1}{2}$ in the thermodynamic limit.

## III. REVIEW OF IDEAL LIMIT

The ideal limit of Rhombohedral Multilayer Graphene (RMG) was introduced in Ref. [7]. In this section, we briefly review the single-particle wavefunction. We will first introduce the realistic tight-binding Hamiltonian of RMG in Sec. III A, and then discuss the ideal limit in Sec. III B.

### A. Tight-binding Hamiltonian for realistic RMG

The real-space lattice of RMG is defined by the basis vectors,

$$\boldsymbol{R}_1=a_{\text{Gr}}(1,0),\quad \boldsymbol{R}_2=a_{\text{Gr}}(\frac{1}{2},\frac{\sqrt{3}}{2}), \tag{52}$$

where $a_{\text{Gr}}=0.246$nm is the graphene lattice constant. The reciprocal lattice of RMG is spanned by the basis vectors,

$$\boldsymbol{G}_1=\frac{4\pi}{\sqrt{3}a_{\text{Gr}}}(\frac{\sqrt{3}}{2},-\frac{1}{2}),\quad \boldsymbol{G}_2=\frac{4\pi}{\sqrt{3}a_{\text{Gr}}}(0,1). \tag{53}$$

Based on realistic modeling, the tight-binding Hamiltonian of RMG is,

$$H_{\text{RMG}}=\sum_{\boldsymbol{k}\in\text{BZ},\sigma,\sigma',l,l'} c^\dagger_{\boldsymbol{k},\sigma,l}[h_{\text{RMG}}(\boldsymbol{k})]_{\sigma,l;\sigma',l'}c_{\boldsymbol{k},\sigma',l'}. \tag{54}$$

In this expression, $c$ and $c^\dagger$ are fermion annihilation and creation operators, respectively. $\sigma=A,B$ labels the sublattice, and $l=0...N_L-1$ labels the layer. BZ denotes the graphene Brillouin zone. The intralayer ($l=l'=L$) Hamiltonian is

$$[h_{\text{RMG}}(\boldsymbol{k})]_{\sigma,L;\sigma',L}=\begin{pmatrix} u_L & -t_0f(\boldsymbol{k}) \\ -t_0f^*(\boldsymbol{k}) & u_L\end{pmatrix}_{\sigma\sigma'}. \tag{55}$$

The interlayer tunneling Hamiltonians are

$$[h_{\rm RMG}(\boldsymbol{k})]_{\sigma,L;\sigma',L+1} = \begin{pmatrix} t_4 f(\boldsymbol{k}) & t_3 f^*(\boldsymbol{k}) \\ -t_1 & t_4 f(\boldsymbol{k}) \end{pmatrix}_{\sigma\sigma'}, \tag{56}$$

$$[h_{\rm RMG}(\boldsymbol{k})]_{\sigma,L;\sigma',L+2} = \begin{pmatrix} 0 & t_2/2 \\ 0 & 0 \end{pmatrix}_{\sigma\sigma'}, \tag{57}$$

$$f(\boldsymbol{k}) = \sum_{i=1}^{3} \exp(i\boldsymbol{k}\cdot\boldsymbol{\delta}_i); \qquad \boldsymbol{\delta}_i = R_{\frac{i2\pi}{3}}(0, \frac{a_{Gr}}{\sqrt{3}})^T. \tag{58}$$

Here, $u_L = (-L + \frac{N_L-1}{2})u_D$ is the onsite potential energy on the $L-$th layer. $(t_0, t_1, t_2, t_3, t_4) = (3100, -380, -21, 290, 141)$meV are the hopping parameters of graphene [8]. $R_\theta$ denotes a counter-clockwise rotation by angle $\theta$.

### B. Tight-binding Hamiltonian for RMG in the dispersive ideal limit

Having presented the realistic tight-binding Hamiltonian for RMG, we now consider its ideal limit. To define the ideal limit, we will consider momentum near $\boldsymbol{K} = (\frac{4\pi}{3a_{Gr}}, 0)$ and write the continuum limit Hamiltonian as

$$H_{\rm ideal} = \sum_{\boldsymbol{k}\in\mathbb{R}^2,\sigma,\sigma',l,l'} c^\dagger_{\boldsymbol{k}+\boldsymbol{K},\sigma,l}[h_{\rm flat}(\boldsymbol{k}) + h_{\rm disp}(\boldsymbol{k})]_{\sigma,l;\sigma',l'} c_{\boldsymbol{k}+\boldsymbol{K},\sigma',l'}. \tag{59}$$

We have separated the ideal tight-binding Hamiltonian into two terms: $h_{\rm flat}(\boldsymbol{k})$, which defines a flat ideal band, and $h_{\rm disp}$, which gives the band a finite dispersion. The intralayer hopping matrix for $h_{\rm flat}(\boldsymbol{k})$ is

$$[h_{\rm flat}(\boldsymbol{k})]_{\sigma,L;\sigma',L} = \begin{pmatrix} 0 & v_F(k_x - ik_y) \\ v_F(k_x + ik_y) & 0 \end{pmatrix}_{\sigma\sigma'} \qquad L = 0...N_L - 2, \tag{60}$$

where the Fermi-velicty, $v_F$ is defined via $f(\boldsymbol{K}+\boldsymbol{q}) \approx -\frac{\sqrt{3}}{2}a_{\rm Gr}(q_x - iq_y) \equiv -\frac{v_F}{t_0}(q_x - iq_y)$. The interlayer hopping matrix is

$$[h_{\rm flat}(\boldsymbol{k})]_{\sigma,L;\sigma',L+1} = \begin{pmatrix} 0 & 0 \\ -t_1 & 0 \end{pmatrix}_{\sigma\sigma'}. \tag{61}$$

The dispersion term, $h_{\rm disp}$, is purely onsite,

$$[h_{\rm disp}(\boldsymbol{k})]_{\sigma,L;\sigma',L} = \begin{pmatrix} \mathcal{E}(\boldsymbol{k}) & 0 \\ 0 & -\mathcal{E}(\boldsymbol{k}) - \Delta_B \end{pmatrix}_{\sigma\sigma'} \qquad L = 0...N_L - 1. \tag{62}$$

where $\mathcal{E}(\boldsymbol{k})$ is the dispersion of the ideal band, and $\Delta_B$ is set to some small positive number, the exact value of which will not affect our calculation. For most of our calculations, we take $\mathcal{E}(\boldsymbol{k})$ to be the dispersion of the first conduction band of $h_{\rm RMG}(\boldsymbol{K}+\boldsymbol{k})$ at a fixed displacement field $u_D$.

$h_{\rm flat}(\boldsymbol{k})$ has a zero-energy A-sublattice-polarized eigenstate,

$$h_{\rm flat}(\boldsymbol{k})\,|s_{\boldsymbol{k}}\rangle = 0. \tag{63}$$

This state is also an eigenstate of $h_{\rm flat}(\boldsymbol{k}) + h_{\rm disp}(\boldsymbol{k})$ with energy $\mathcal{E}(\boldsymbol{k})$. The components of $|s_{\boldsymbol{k}}\rangle$ in the basis of $(A_0, B_0, A_1..)$ are

$$|s_{\boldsymbol{k}}\rangle = N_{\boldsymbol{k}}(1, 0, \gamma(k_x + ik_y), 0, (\gamma(k_x + ik_y))^2, 0...), \quad N_{\boldsymbol{k}} = (\sum_{l=0}^{N_L-1}(\gamma|\boldsymbol{k}|)^{2l})^{-\frac{1}{2}}, \quad \gamma \equiv \frac{v_F}{t_1}. \tag{64}$$

Here, $N_{\boldsymbol{k}}$ is the normalization constant. Since this single-particle wavefunction is completely sublattice-polarized, we will use $s_{\boldsymbol{k},l}$ to refer to the $l-$th layer component of this single-particle wavefunction, and use $c^\dagger_{\boldsymbol{k}}$ to refer to the fermion creation operator that creates an electron in this state.

## IV. ANSATZ WAVEFUNCTION OF FRACTIONAL CHERN INSULATORS

In this section, we construct trial wavefunctions for fractional Chern insulators (FCIs) at filling factor $\nu = \frac{2}{3}$. This will be done using a torus geometry (see Sec. II for relevant background on fractional quantum Hall wavefunctions on a torus). To start, we will review the $\nu = 1$ skyrmion-lattice state in Sec. IV A. Based on this, we will consider two classes of descendant FCI trial wavefunctions: In Sec. IV B, we will construct skyrmion FCI states by fractionalizing skyrmion vacancy configurations. In Sec. IV C, we will construct the conventional FCI state obtained by directly fractionalizing the filled skyrmion-lattice Chern band. The physical distinction between these two constructions is discussed in the main text. In Sec. IV D we discuss adding an additional variational parameter to the skyrmion FCI wavefunction that smooths out its short-distance structure. Relevant extended data is presented in Sec. IV E.

### A. $\nu = 1$ skyrmion-lattice wavefunction

In the ideal limit of RMG, the $\nu = 1$ skyrmion-lattice state is described by the Slater determinant wavefunction [7]

$$\Psi^{\nu=1,\mathrm{SkL}}_{\{\ell\}}(\{\boldsymbol{r}\}) = \det_{\boldsymbol{k}m}[\phi^{\mathrm{LLL}}_{\boldsymbol{k}-\boldsymbol{q}}(\boldsymbol{r}_m)\chi_{\ell_m}(\boldsymbol{r}_m)]. \tag{65}$$

Here, $\{\ell\}$ is the layer index of the multi-component many-body wavefunction. In general, $\chi_0(\boldsymbol{r})$ can be parameterized in terms of a list of complex coefficients, $\{c_n\}$,

$$\chi_0(\boldsymbol{r}) = \sum_n c_n \phi^{\mathrm{nLL}*}_{-\boldsymbol{q}}(\boldsymbol{r}). \tag{66}$$

The momentum $\boldsymbol{q}$ specifies the magnetic-Bloch momentum carried by the common skyrmion factor and corresponds to the position of the skyrmion core. For most of the calculations presented in this paper, the Hamiltonian, including both the kinetic and interaction terms, is invariant under continuous translations, and the choice of $\boldsymbol{q}$, which only shifts the origin of the skyrmion lattice, does not affect the energy. However, when a periodic scalar potential is included, the choice of $\boldsymbol{q}$ becomes physically relevant (Sec. IV E). With this convention, the product $\phi^{\mathrm{LLL}}_{\boldsymbol{k}-\boldsymbol{q}}(\boldsymbol{r})\chi_\ell(\boldsymbol{r})$ has ordinary Bloch momentum $\boldsymbol{k}$. The higher-layer components are fixed by the holomorphic structure of the parent band wavefunction $|s_{\boldsymbol{k}}\rangle$:

$$\phi^{\mathrm{LLL}}_{\boldsymbol{k}-\boldsymbol{q}}(\boldsymbol{r})\chi_\ell(\boldsymbol{r}) = (-2i\gamma\partial_{\bar{z}})^\ell \left[\phi^{\mathrm{LLL}}_{\boldsymbol{k}-\boldsymbol{q}}(\boldsymbol{r})\chi_0(\boldsymbol{r})\right]. \tag{67}$$

Each single-particle orbital constructed in this way lies within the holomorphic parent band Hilbert space defined by Eq. 64. The resulting Slater determinant is an exact zero-energy state of the layer-isotropic density-density contact interaction.

Different choices of $\{c_n\}$ have different kinetic energies. Here, the lowest kinetic energy occurs for [2]

$$c_0 = 1, \qquad c_{n\neq 0} = 0, \tag{68}$$

so that [7]

$$\chi_0(\boldsymbol{r}) = \phi^{\mathrm{LLL}*}_{-\boldsymbol{q}}(\boldsymbol{r}), \qquad \chi_\ell(\boldsymbol{r}) = \sqrt{\ell!}\left(-\frac{\sqrt{2}\gamma}{l_B}\right)^\ell \phi^{\ell\mathrm{LL}*}_{-\boldsymbol{q}}(\boldsymbol{r}). \tag{69}$$

In the remainder of this work, we specialize to the choice of $\{c_n\}$ given in Eq. 68. It is, however, straightforward to generalize this construction to other choices of $\{c_n\}$.

### B. Skyrmion FCI

In this subsection we will focus on FCI wavefunctions based on skyrmion vacancy fractionalization. First, we will go over the construction of a skyrmion vacancy Slater determinant for an arbitrary vacancy configuration in Sec. IV B 1. Based on this, we will construct two microscopically distinct skyrmion FCI wavefunctions at $\nu = \frac{2}{3}$ in Secs. IV B 2 and IV B 3. The difference between these wavefunctions lies in the weighting of the vacancy configurations, but their energies are very similar in the parameter regime relevant to this work (Sec. IV E). In Sec. IV B 4, we discuss the numerical procedure for computing observables from our trial wavefunctions.

---

[2] The caveats are discussed in the Supplementary Material of Ref. [7].

### 1. Ansatz wavefunction for a skyrmion vacancy state

The $\nu = 1$ skyrmion lattice wavefunction in Eq. 65 can be factorized into a fully filled lowest Landau level part and a skyrmion lattice part:

$$\Psi^{\nu=1,\mathrm{SkL}}_{\{\ell\}}(\{\boldsymbol{r}\}) = \det_{\boldsymbol{k}m}[\phi^{\mathrm{LLL}}_{\boldsymbol{k}-\boldsymbol{q}}(\boldsymbol{r}_m)] \prod_p \chi_{\ell_p}(\boldsymbol{r}_p). \tag{70}$$

Each occupied orbital contains the same position-dependent layer spinor $\chi_\ell(\boldsymbol{r})$. We first focus on the $\{\ell = 0\}$ component of the many-body wavefunction, $\chi_0(\boldsymbol{r}) = \phi^{\mathrm{LLL}*}_{-\boldsymbol{q}}(\boldsymbol{r})$. The momentum $\boldsymbol{q}$ will be taken to be fixed in this section.

To consider the skyrmion vacancies, it is useful to rewrite the magnetic Bloch state $\phi^{\mathrm{LLL}}_{\boldsymbol{k}}(\boldsymbol{r})$ in terms of torus sigma functions:

$$\phi^{\mathrm{LLL}}_{\boldsymbol{k}}(\boldsymbol{r}) = \mathcal{N}^{\mathrm{LLL}} \exp\left(\frac{i}{2}(\bar{k} - \frac{\bar{b}_1 + \bar{b}_2}{2})z\right)\sigma(z/l_B + il_B(k - \frac{b_1+b_2}{2})|\{\frac{a_\mu}{l_B}\})\exp\left(-\frac{l_B^2 k\bar{k}}{4} - \frac{z\bar{z}}{4l_B^2}\right)\exp\left(\frac{l_B^2 k(\bar{b}_1+\bar{b}_2)}{4}\right), \tag{71}$$

where $\boldsymbol{a}_\mu$ are the skyrmion-lattice primitive vectors, and $\boldsymbol{b}_\mu$ are the corresponding reciprocal-lattice vectors. In this writing, we identify the magnetic lattice introduced in Sec. II A with the $\nu = 1$ skyrmion lattice in the ideal limit of RMG. Based on this, the torus spanning vectors defined in Eq. 21 are given by $\boldsymbol{L}_\mu = N_\mu \boldsymbol{a}_\mu$. We shall use this identification throughout the rest of the analysis.

To proceed, we will use that

$$\phi^{\mathrm{LLL}}_{\boldsymbol{k}}(\boldsymbol{r}) \propto \exp\left(-\frac{z\bar{z}}{4l_B^2}\right)\exp\left(z\frac{\sum_{mn}\bar{\xi}_{mn}(k - \frac{b_1+b_2}{2})}{2L_B^2}\right)\prod_{m,n=0}^{N_1-1,N_2-1}\sigma(\frac{z - \xi_{mn}(k - \frac{b_1+b_2}{2})}{L_B}|\{L_\mu/L_B\}), \tag{72}$$

where $\xi_{mn}(k) \equiv -ikl_B^2 + ma_1 + na_2$, and the proportionality factor is independent of $\boldsymbol{r}$, but can depend on $\boldsymbol{k}$. To understand Eq. 72, note that both expressions have the same $N_\phi$ zeros within the torus spanned by $\boldsymbol{L}_1$ and $\boldsymbol{L}_2$. They also obey the same magnetic-Bloch boundary condition, Eq. 14, under $\boldsymbol{r} \to \boldsymbol{r} + \boldsymbol{a}_\mu$. After factoring out the common Gaussian, the ratio of their holomorphic parts has no zeros or poles, and is periodic under both lattice translations. Consequently, the ratio must be a constant. In what follows, we suppress the momentum dependence of $\xi_{mn}$ to avoid clutter.

The skyrmion vacancy wavefunction with $N_{\mathrm{vac}}$ skyrmion vacancies is constructed by removing $N_{\mathrm{vac}}$ zeros of the torus sigma function:

$$\phi^{\mathrm{vac}}(\boldsymbol{r}, \{\boldsymbol{\eta}\}) \equiv \frac{\exp\left(-\frac{z\bar{z}}{4l_{Bv}^2}\right)\exp\left(\bar{z}\frac{\sum_{mn}\xi_{mn}}{2L_B^2}\right)\prod_{m,n}\sigma^*(\frac{z-\xi_{mn}}{L_B}|\{L_\mu/L_B\})}{\prod_{a=1}^{N_{\mathrm{vac}}}\sigma^*(\frac{z-\eta_a}{L_B}|\{L_\mu/L_B\})\exp\left(\frac{\eta_a\bar{z}}{2L_B^2}\right)}. \tag{73}$$

In the equation above, we have used an $N_{\mathrm{vac}}$-element set of vacancy positions $\{\boldsymbol{\eta}\}$, and we have also introduced the corresponding auxiliary magnetic length $l_{Bv}$, defined via

$$l_{Bv}^2 \equiv \frac{L_B^2}{N_\phi - N_{\mathrm{vac}}}. \tag{74}$$

The change of the Gaussian factor from $l_B$ to $l_{Bv}$ is required for $\phi^{\mathrm{vac}}$ to remain a Landau-level wavefunction after removing $N_{\mathrm{vac}}$ zeros (the anti-holomorphic sector of the skyrmion vacancy wavefunction has $N_\phi - N_{\mathrm{vac}}$ zeros and therefore carries $N_\phi - N_{\mathrm{vac}}$ flux quanta). Eq. 73 obeys the following boundary condition:

$$\frac{\phi^{\mathrm{vac}}(\boldsymbol{r} + \boldsymbol{L}_\mu, \{\boldsymbol{\eta}\})}{\phi^{\mathrm{vac}}(\boldsymbol{r}, \{\boldsymbol{\eta}\})} = \exp\left(i(2\boldsymbol{A}(\{\boldsymbol{\eta}\}) + (N_\phi - N_{\mathrm{vac}})\frac{\boldsymbol{T}_1 + \boldsymbol{T}_2}{2}) \cdot \boldsymbol{L}_\mu\right)\exp\left(-i\frac{\boldsymbol{L}_\mu \times \boldsymbol{r}}{2l_{Bv}^2}\right), \tag{75}$$

with

$$A(\{\eta\}) \equiv i(\sum_a \frac{\eta_a}{2L_B^2} - \sum_{mn}\frac{\xi_{mn}}{2L_B^2}). \tag{76}$$

In order for all poles in $\phi^{\mathrm{vac}}$ to cancel, the set of vacancy positions $\{\boldsymbol{\eta}\}$ must coincide with a subset of the zero positions $\{\boldsymbol{\xi}\}$ modulo the torus spanning vectors $\boldsymbol{L}_\mu$. Note that since points related by $\eta_a \to \eta_a + L_\mu$ represent the same position

on the torus, different choices of representatives, e.g. $\{\boldsymbol{\eta}\} = \{\boldsymbol{\xi}_{00}, \boldsymbol{\xi}_{01}...\}$ and $\{\boldsymbol{\eta}\} = \{\boldsymbol{\xi}_{00} + \boldsymbol{L}_1, \boldsymbol{\xi}_{01} + \boldsymbol{L}_2...\}$, correspond to the same physical skyrmion vacancy configuration. However, it will be useful to retain different representatives $\{\boldsymbol{\eta}\}$ explicitly, as the relative factors associated with the choice of representative must be treated consistently when we later construct a skyrmion FCI wavefunction as a coherent superposition of different vacancy configurations.

Using $\phi^{\text{vac}}(\boldsymbol{r}, \{\boldsymbol{\eta}\})$, let us now construct the full $\{\ell = 0\}$ component of the (unnormalized) skyrmion vacancy Slater determinant for $(N_\phi - N_{\text{vac}})$-electrons. To do this, let us first note that the $\{\ell = 0\}$ component of the full skyrmion lattice ansatz wavefunction, Eq. 65, is a Slater determinant of $\phi^{\text{LLL}}_{\boldsymbol{k}-\boldsymbol{q}}(\boldsymbol{r})\phi^{\text{LLL}*}_{-\boldsymbol{q}}(\boldsymbol{r})$. For the skyrmion vacancy state, we replace the skyrmion lattice factor, $\phi^{\text{LLL}*}_{-\boldsymbol{q}}(\boldsymbol{r})$, with $\phi^{\text{vac}}(\boldsymbol{r}, \{\boldsymbol{\eta}\})$. To ensure that the full wavefunction has ordinary (non-magnetic) twisted boundary conditions, we also need to replace the $\phi^{\text{LLL}}_{\boldsymbol{k}-\boldsymbol{q}}(\boldsymbol{r})$ factors with the magnetic Bloch states from a non-conjugated $N_\phi - N_{\text{vac}}$-flux LLL. The $\{\ell = 0\}$ component of the (unnormalized) skyrmion vacancy Slater determinant is thus

$$\Psi^{\text{SkV}}_{\{\ell=0\}}(\{\boldsymbol{r}\}, \{\boldsymbol{\eta}\}, \boldsymbol{\Delta\theta}) = \det_{\substack{\boldsymbol{k}\in\text{BZ2} \\ p\in[1,N_\phi-N_{\text{vac}}]}} [\phi^{\text{LLL,vac}}_{\boldsymbol{k}-\boldsymbol{\kappa}^{\text{vac}}}(\boldsymbol{r}_p)\phi^{\text{vac}}(\boldsymbol{r}_p, \{\boldsymbol{\eta}\})] = \det_{\substack{\boldsymbol{k}\in\text{BZ2} \\ p\in[1,N_\phi-N_{\text{vac}}]}} [\phi^{\text{LLL,vac}}_{\boldsymbol{k}-\boldsymbol{\kappa}^{\text{vac}}}(\boldsymbol{r}_p)] \prod_p \phi^{\text{vac}}(\boldsymbol{r}_p, \{\boldsymbol{\eta}\}), \tag{77}$$

where

$$\boldsymbol{\kappa}^{\text{vac}}(\{\boldsymbol{\eta}\}, \boldsymbol{\Delta\theta}) \equiv 2\boldsymbol{A}(\{\boldsymbol{\eta}\}) + (N_\phi - N_{\text{vac}})\frac{\boldsymbol{T}_1 + \boldsymbol{T}_2}{2} + \boldsymbol{\Delta\theta}, \tag{78}$$

accounts for the $\boldsymbol{\Delta\theta}$-twisted boundary conditions, similar to Sec. II. Here, $\phi^{\text{LLL,vac}}$ is the magnetic Bloch state on a torus with $N_\phi - N_{\text{vac}}$ magnetic flux,

$$\begin{aligned}\phi^{\text{LLL,vac}}_{\boldsymbol{k}}(\boldsymbol{r}) =& \mathcal{N}^{\text{LLL}}_v \exp\left(\frac{i}{2}(\bar{k} - \frac{\bar{b}_{1v} + \bar{b}_{2v}}{2})z\right)\sigma(\frac{z}{l_{Bv}} + il_{Bv}(k - \frac{b_{1v} + b_{2v}}{2})|\frac{a_{1v}}{l_{Bv}}, \frac{a_{2v}}{l_{Bv}}) \\ &\times \exp\left(-\frac{l^2_{Bv}k\bar{k}}{4} - \frac{z\bar{z}}{4l^2_{Bv}}\right)\exp\left(\frac{l^2_{Bv}k(\bar{b}_{1v} + \bar{b}_{2v})}{4}\right),\end{aligned} \tag{79}$$

with normalization factor $\mathcal{N}^{\text{LLL}}_v$. Here, we have introduced reciprocal lattice vectors $\boldsymbol{b}_{1v}$ and $\boldsymbol{b}_{2v}$ that span an auxiliary $(N_\phi - N_{\text{vac}})$-flux magnetic Brillouin zone, BZ2,

$$\boldsymbol{k} \in \text{BZ2} \Leftrightarrow \boldsymbol{k} = n_1\boldsymbol{T}_1 + n_2\boldsymbol{T}_2, \quad n_\mu = 0, \ldots, N_{\mu v} - 1 \quad N_{1v}N_{2v} = N_\phi - N_{\text{vac}}. \tag{80}$$

The corresponding auxiliary magnetic lattice has primitive vectors $\boldsymbol{a}_{1v}$ and $\boldsymbol{a}_{2v}$,

$$|\boldsymbol{a}_{1v} \times \boldsymbol{a}_{2v}| = 2\pi l^2_{Bv}, \quad \boldsymbol{L}_1 = N_{1v}\boldsymbol{a}_{1v}, \quad \boldsymbol{L}_2 = N_{2v}\boldsymbol{a}_{2v}. \tag{81}$$

With these definitions, it is straightforward to check that $\Psi^{\text{SkV}}_{\{\ell=0\}}(\{\boldsymbol{r}\}, \{\boldsymbol{\eta}\})$ obeys the twisted boundary condition, parametrized by $\boldsymbol{\Delta\theta}$:

$$\frac{\Psi^{\text{SkV}}_{\{\ell=0\}}(\{\boldsymbol{r}_1, ..., \boldsymbol{r}_p + \boldsymbol{L}_\mu...\}, \{\boldsymbol{\eta}\}, \boldsymbol{\Delta\theta})}{\Psi^{\text{SkV}}_{\{\ell=0\}}(\{\boldsymbol{r}\}, \{\boldsymbol{\eta}\}, \boldsymbol{\Delta\theta})} = \exp(-i\boldsymbol{\Delta\theta} \cdot \boldsymbol{L}_\mu). \tag{82}$$

It is worth noting that for fixed $N_\phi - N_{\text{vac}}$, there can be multiple compatible choices of $N_{1v}$ and $N_{2v}$, which correspond to different magnetic Bloch bases of the same $(N_\phi - N_{\text{vac}})$-flux LLL Hilbert space. Since all states in this auxiliary LLL are filled in Eq. 77, changing the auxiliary lattice changes the Slater determinant only by an overall factor.

To conclude this subsection, we now derive an expression of $\Psi^{\text{SkV}}_{\{\ell\}}(\{\boldsymbol{r}\}, \{\boldsymbol{\eta}\})$ that is convenient for numerical computation. The states $\phi^{\text{LLL,vac}*}_{\boldsymbol{k}-\boldsymbol{\kappa}^{\text{vac}}+\boldsymbol{\Delta\theta}}(\boldsymbol{r})$, with $\boldsymbol{k} \in \text{BZ2}$, form a complete basis of the complex-conjugate LLL with the boundary condition in Eq. 75. Therefore,

$$\phi^{\text{vac}}(\boldsymbol{r}, \{\boldsymbol{\eta}\}) = \sum_{\boldsymbol{k}\in\text{BZ2}} W_{\boldsymbol{k}}(\{\boldsymbol{\eta}\})\phi^{\text{LLL,vac}*}_{\boldsymbol{k}-\boldsymbol{\kappa}^{\text{vac}}+\boldsymbol{\Delta\theta}}(\boldsymbol{r}) \tag{83}$$

These coefficients $W_{\boldsymbol{k}}(\{\boldsymbol{\eta}\})$ [3] can be evaluated numerically using Eq. 73 and Eq. 79. In addition, using the results from the Supplementary Material of Ref. [7], the product of two LLL wavefunctions can be expanded in the plane-wave basis

$$\phi^{\text{LLL,vac}}_{-\boldsymbol{\kappa}^{\text{vac}}+\boldsymbol{k}_1}(\boldsymbol{r})\phi^{\text{LLL,vac}*}_{-\boldsymbol{\kappa}^{\text{vac}}+\boldsymbol{k}_2+\boldsymbol{\Delta\theta}}(\boldsymbol{r}) = \mathcal{M}_{\boldsymbol{k}_1,\boldsymbol{k}_2}(\{\boldsymbol{\eta}\}, \boldsymbol{\Delta\theta}) \sum_{\boldsymbol{g}=n_1\boldsymbol{b}_{1v}+n_2\boldsymbol{b}_{2v}} e^{i\pi n_1 n_2} e^{-\frac{l^2_{Bv}}{4}(g\bar{g}+2(k_1-k_2-\Delta\theta)\bar{g})} e^{i(\boldsymbol{k}_1-\boldsymbol{k}_2-\boldsymbol{\Delta\theta}+\boldsymbol{g})\cdot(\boldsymbol{r}-\boldsymbol{R}_0)}, \tag{84}$$

[3] $W_{\boldsymbol{k}}(\{\boldsymbol{\eta}\})$ is independent of $\boldsymbol{\Delta\theta}$, since $\boldsymbol{\kappa}^{\text{vac}} - \boldsymbol{\Delta\theta}$ is independent of $\boldsymbol{\Delta\theta}$.

where

$$\boldsymbol{R}_0 = (k_{2y} - \kappa_y^{\rm vac} + \Delta\theta_y, -k_{2x} + \kappa_x^{\rm vac} - \Delta\theta_x) l_{Bv}^2, \tag{85}$$

and $\mathcal{M}_{\boldsymbol{k}_1,\boldsymbol{k}_2}$ is an $\boldsymbol{r}$-independent normalization constant, which can be evaluated numerically by matching the two sides of the equation above. Combining these results,

$$\Psi^{\rm SkV}_{\{\ell\}}(\{\boldsymbol{r}\},\{\boldsymbol{\eta}\},\boldsymbol{\Delta\theta}) = [\prod_p (-2i\gamma\partial_{\bar{z}_p})^{\ell_p}]\Psi^{\rm SkV}_{\{\ell=0\}}(\{\boldsymbol{r}\},\{\boldsymbol{\eta}\},\boldsymbol{\Delta\theta}) = \det_{\boldsymbol{k}\in{\rm BZ2},p}[\psi^{\rm SkV}_{\boldsymbol{k},\ell_p}(\boldsymbol{r}_p,\{\boldsymbol{\eta}\},\boldsymbol{\Delta\theta})], \tag{86}$$

with

$$\psi^{\rm SkV}_{\boldsymbol{k},\ell}(\boldsymbol{r},\{\boldsymbol{\eta}\},\boldsymbol{\Delta\theta}) = \sum_{\substack{\boldsymbol{k}_2\in{\rm BZ2}\\ \boldsymbol{g}=n_1\boldsymbol{b}_{1v}+n_2\boldsymbol{b}_{2v}}} W_{\boldsymbol{k}_2}\mathcal{M}_{\boldsymbol{k},\boldsymbol{k}_2} e^{i\pi n_1 n_2} e^{-\frac{l_{Bv}^2}{4}(g\bar{g}+2(k-k_2-\Delta\theta)\bar{g})} e^{i(\boldsymbol{k}-\boldsymbol{k}_2-\boldsymbol{\Delta\theta}+\boldsymbol{g})\cdot(\boldsymbol{r}-\boldsymbol{R}_0)} \frac{s_{\boldsymbol{k}-\boldsymbol{k}_2-\boldsymbol{\Delta\theta}+\boldsymbol{g},\ell}}{N_{\boldsymbol{k}-\boldsymbol{k}_2-\boldsymbol{\Delta\theta}+\boldsymbol{g}}}. \tag{87}$$

Here, $|s_{\boldsymbol{k}}\rangle$ is the holomorphic parent-band spinor defined in Eq. 64, and $N_{\boldsymbol{k}}$ is its normalization factor. It is worth noting that the orbitals $\psi^{\rm SkV}_{\boldsymbol{k},\ell}(\boldsymbol{r},\boldsymbol{\eta},\boldsymbol{\Delta\theta})$ are generally not orthogonal for different $\boldsymbol{k}$, since a fixed vacancy configuration breaks the skyrmion-lattice translation symmetry. For numerical convenience, we orthonormalize the orbitals for each vacancy configuration. The corresponding change in the normalization of the Slater determinant must be retained when different vacancy configurations are subsequently superposed, as discussed in Sec. IV B 4.

We note that, from Eq. 87, the momentum-space occupation $\langle c^\dagger_{\boldsymbol{k}} c_{\boldsymbol{k}}\rangle$ of the skyrmion-vacancy wavefunction decays approximately as $\exp\left(-|\boldsymbol{k}|^2 l_{Bv}^2/2\right)$. In contrast, the corresponding occupation of the $\nu = 1$ skyrmion-lattice wavefunction [7] decays as $\exp\left(-|\boldsymbol{k}|^2 l_B^2/2\right)$. Since $l_{Bv} > l_B$, the momentum-space distribution of the skyrmion-vacancy wavefunction is naturally concentrated within a smaller region of $|\boldsymbol{k}|$. For a dispersion $\mathcal{E}(\boldsymbol{k})$ that increases monotonically with $|\boldsymbol{k}|$, this narrower momentum-space distribution provides a natural kinetic-energy advantage over states in the $\nu = 1$ skyrmion-lattice band.

In Fig. 1(a,b), we show the layer-resolved charge density $\rho_\ell(\boldsymbol{r})$ of the full skyrmion-lattice state $\Psi^{\nu=1,{\rm SkL}}_{\{\ell\}}$ on a 4-by-4 torus. For the full skyrmion lattice, the $\ell = 0$ charge density $\rho_{\ell=0}(\boldsymbol{r})$ has 16 zeros within the torus, while the $\ell = 1$ charge density exhibits peaks at the corresponding positions. At these positions, the common layer spinor $\chi_\ell(\boldsymbol{r})$ of the occupied orbitals is rotated away from the zeroth layer. Each zero/peak pair corresponds to one skyrmion [7], giving a total of 16 skyrmions on the torus. In Fig. 1(c,d), we show the layer-resolved charge density of a skyrmion vacancy state with two vacancies. Compared with the full skyrmion lattice, the $\ell = 0$ component has two missing zeros, while the corresponding two peaks are absent from the $\ell = 1$ component. This directly illustrates the removal of two skyrmions from the torus.

### 2. *Skyrmion FCI wavefunction at $\nu = \frac{2}{3}$*

In this subsection, we construct the skyrmion FCI wavefunction at filling factor $\nu = \frac{2}{3}$, for which

$$N_e = N_\phi - N_{\rm vac} = \frac{2N_\phi}{3}. \tag{88}$$

The skyrmion FCI wavefunction is constructed as a coherent superposition of $\Psi^{\rm SkV}_{\{\ell\}}(\{\boldsymbol{r}\},\{\boldsymbol{\eta}\},\boldsymbol{\Delta\theta})$ over different skyrmion vacancy configurations $\{\boldsymbol{\eta}\}$. As discussed in Sec. IV B 1, for each fixed vacancy configuration, $\Psi^{\rm SkV}_{\{\ell\}}(\{\boldsymbol{r}\},\{\boldsymbol{\eta}\},\boldsymbol{\Delta\theta})$ is a Slater determinant. Combining the results of Sec. II and Sec. IV B 1, the $\{\ell = 0\}$ component of the skyrmion vacancy wavefunction can be written explicitly in terms of torus sigma functions:

$$\begin{aligned}\Psi^{\rm SkV}_{\{\ell=0\}}(\{\boldsymbol{r}\},\{\boldsymbol{\eta}\},\boldsymbol{\Delta\theta}) \propto & \prod_p \exp\left(-\frac{z_p\bar{z}_p}{4l_{Bv}^2}\right)\prod_{p<q}\sigma\left(\frac{z_p-z_q}{L_B}\Big|\left\{\frac{L_\mu}{L_B}\right\}\right)\prod_p \exp\left(-\frac{z_p\bar{z}_p}{4l_{Bv}^2}\right)\prod_{p,m,n}\sigma^*\left(\frac{z_p-\xi_{mn}}{L_B}\Big|\left\{\frac{L_\mu}{L_B}\right\}\right)\\ & \times\left(\prod_{p,a}\sigma^*\left(\frac{z_p-\eta_a}{L_B}\Big|\left\{\frac{L_\mu}{L_B}\right\}\right)\right)^{-1} F_1(Z,-\boldsymbol{\kappa}^{\rm vac})\exp\left(\frac{\sum_{mn}\xi_{mn}-H}{2L_B^2}\bar{Z}\right)\\ \equiv & \Phi(\{\boldsymbol{r}\},\{\boldsymbol{\eta}\},\boldsymbol{\Delta\theta}),\end{aligned} \tag{89}$$

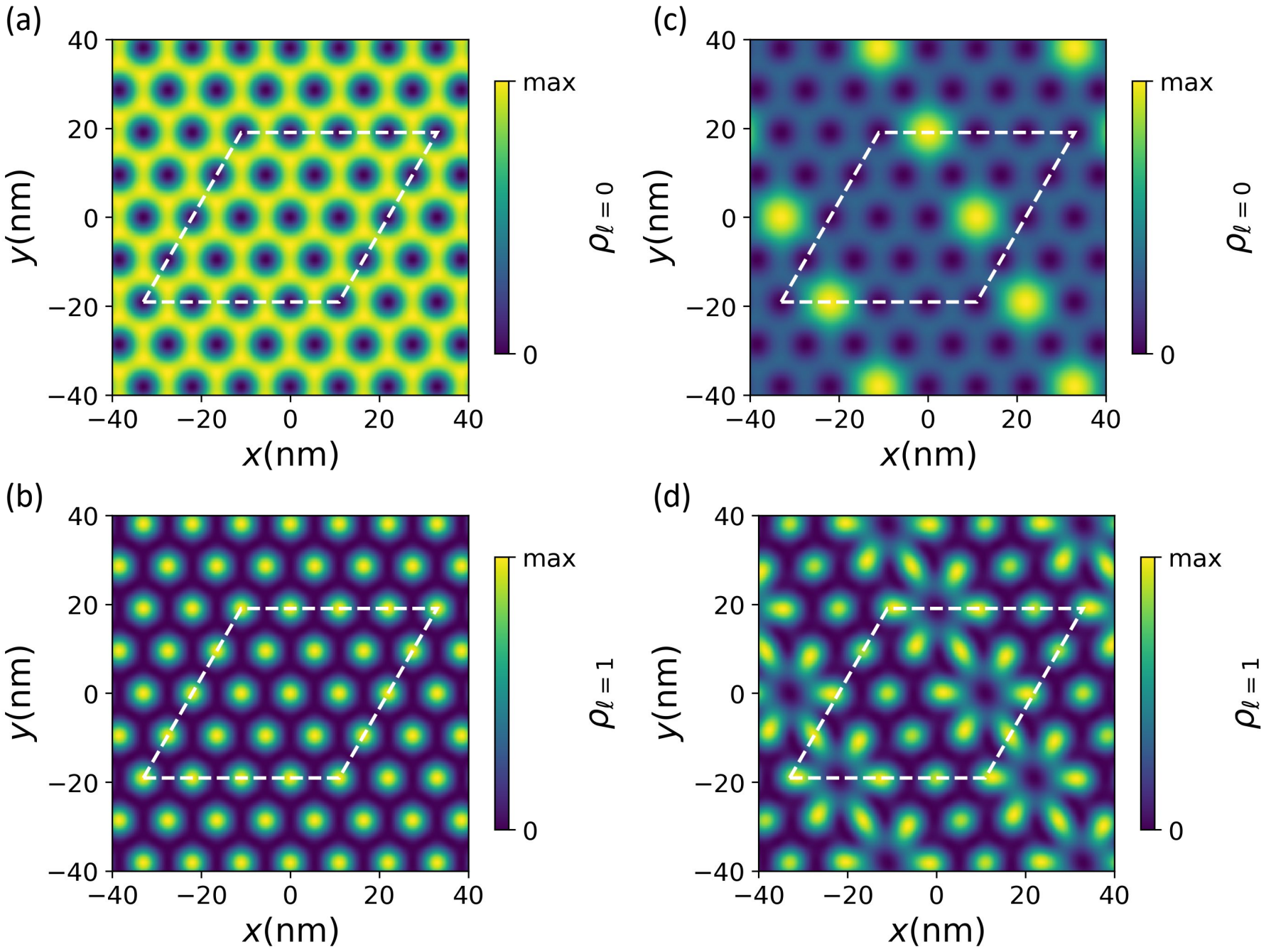


FIG. 1. Layer-resolved charge density of **(a,b)** the full $\nu = 1$ skyrmion-lattice state $\Psi^{\nu=1,\mathrm{SkL}}_{\{\ell\}}$, and **(c,d)** the skyrmion vacancy state $\Psi^{\mathrm{SkV}}_{\{\ell\}}$ for a particular two-vacancy configuration. The system is placed on an $N_1 = N_2 = 4$ torus, whose boundary is indicated by the white dashed line. The wavefunctions are constructed for $N_L = 5$ and skyrmion-lattice period $|\boldsymbol{a}_\mu| = 11\,\mathrm{nm}$.

where we have defined the center-of-mass coordinate of the vacancies as $H \equiv \sum_{a=1}^{N_{\mathrm{vac}}} \eta_a$, and

$$\begin{aligned} F_1(Z, -\boldsymbol{\kappa}^{\mathrm{vac}}) &= \sigma(\frac{Z}{L_B} + 2iL_B(-\frac{\kappa^{\mathrm{vac}}}{2} + \frac{(N_\phi - N_{\mathrm{vac}})(T_1 + T_2)}{4})|\{L_\mu/L_B\}) \exp\left(iZ(-\frac{\bar{\kappa}^{\mathrm{vac}}}{2} + \frac{(N_\phi - N_{\mathrm{vac}})(\bar{T}_1 + \bar{T}_2)}{4})\right) \\ &= \sigma(\frac{Z + H - \sum_{mn}\xi_{mn} - iL_B^2\Delta\theta}{L_B}|\{L_\mu/L_B\}) \exp\left(Z\frac{(-\bar{H} + \sum_{mn}\bar{\xi}_{mn} - iL_B^2\Delta\bar{\theta})}{2L_B^2}\right). \end{aligned} \tag{90}$$

Under a translation $\boldsymbol{\eta}_a \to \boldsymbol{\eta}_a + \boldsymbol{L}_\mu$ for one vacancy coordinate, $\Phi(\{\boldsymbol{r}\}, \{\boldsymbol{\eta}\}, \boldsymbol{\Delta\theta})$ transforms as

$$\begin{aligned} &\frac{\Phi(\{\boldsymbol{r}\}, \{\boldsymbol{\eta}_1, ..., \boldsymbol{\eta}_a + \boldsymbol{L}_\mu, ...\}, \boldsymbol{\Delta\theta})}{\Phi(\{\boldsymbol{r}\}, \{\boldsymbol{\eta}\}, \boldsymbol{\Delta\theta})} \\ &= (-1)^{N_e - 1} \exp\left(\frac{\bar{L}_\mu H}{2L_B^2} - \frac{\bar{L}_\mu \sum_{mn}\xi_{mn}}{2L_B^2}\right) \exp\left(-\frac{L_\mu \bar{L}_\mu (N_e - 1)}{4L_B^2}\right) \exp\left(-\frac{N_e L_\mu \bar{\eta}_a}{2L_B^2}\right) \exp\left(-i\frac{\bar{L}_\mu \Delta\theta}{2}\right) \end{aligned} \tag{91}$$

The skyrmion FCI wavefunction is obtained by placing the vacancies in a $\nu = -\frac{1}{2}$ bosonic Laughlin state, analogous to the Haldane–Halperin construction reviewed in Sec. II D:

$$\Psi^{\text{frac-SkV},(s)}_{\{\ell=0\},1}(\{\boldsymbol{r}\}, \boldsymbol{\Delta\theta}) = \sum_{\{\boldsymbol{\eta}\}} \Phi(\{\boldsymbol{r}\}, \{\boldsymbol{\eta}\}, \boldsymbol{\Delta\theta}) \prod_{a<b} \sigma^{*2}(\frac{\eta_a - \eta_b}{L_B}|\{\frac{L_\mu}{L_B}\}) F_2^{(s)}(H, \bar{H}, \boldsymbol{\Delta\theta}), \tag{92}$$

$$\Psi^{\text{frac-SkV},(s)}_{\{\ell\},1} = \prod_p (-2i\gamma\partial_{\bar{z}_p})^{\ell_p} \Psi^{\text{frac-SkV},(s)}_{\{\ell=0\},1}(\{\boldsymbol{r}\}, \boldsymbol{\Delta\theta}). \tag{93}$$

The subscript 1 distinguishes this state from the microscopically different skyrmion FCI wavefunction that will be introduced in Sec. IV B 3. The superscript $s$ labels the three topological sectors discussed below. The sum over $\{\boldsymbol{\eta}\}$ runs over all $N_{\mathrm{vac}}$-element subsets of the $N_\phi = N_1 N_2$ skyrmion-lattice zero positions, $\{\boldsymbol{\xi}\}$, within a chosen torus fundamental domain. There are a total of $\binom{N_1 N_2}{N_{\mathrm{vac}}}$ distinct vacancy configurations.

We note that each summed term of Eq. 92 must be invariant under $\boldsymbol{\eta}_a \to \boldsymbol{\eta}_a + \boldsymbol{L}_\mu$. Otherwise, the amplitude assigned to a vacancy configuration would depend on the arbitrary choice of its torus representative. This requirement determines the boundary condition of the COM factor $F_2^{(s)}$. This is analogous to the requirement that in the first-quantized construction of the $\nu = \frac{2}{3}$ fractional quantum Hall state, the integrand is periodic when $\boldsymbol{\eta}_a \to \boldsymbol{\eta}_a + \boldsymbol{L}_\mu$ (Eq. 46 and Eq. 51).

$$\frac{F_2^{(s)}(H + L_\mu, \bar{H} + \bar{L}_\mu, \boldsymbol{\Delta\theta})}{F_2^{(s)}(H, \bar{H}, \boldsymbol{\Delta\theta})} = (-1)^{N_e - 1} \exp\left(\frac{L_\mu \bar{L}_\mu}{4L_B^2}\right) \exp\left(-\frac{\bar{L}_\mu H}{2L_B^2}\right) \exp\left(\frac{L_\mu \bar{H}}{L_B^2}\right) \exp\left(\frac{\bar{L}_\mu \sum_{mn} \xi_{mn}}{2L_B^2}\right) \exp\left(i\frac{\bar{L}_\mu \Delta\theta}{2}\right) \tag{94}$$

The COM term $F_2^{(s)}$ takes the ansatz form

$$F_2^{(s)}(H, \bar{H}, \boldsymbol{\Delta\theta}) = f_2^{(s)}(\bar{H}, \boldsymbol{\Delta\theta}) \exp\left(-\frac{H\bar{H}}{2L_B^2}\right) \exp\left(-i\frac{H\Delta\bar{\theta}}{2}\right). \tag{95}$$

This choice of ansatz is motivated by the fact that there must be three topologically degenerate solutions that are related via flux threading, which we will show explicitly below. Eq. 94 implies that $f_2^{(s)}(\bar{H}, \boldsymbol{\Delta\theta})$ obeys the following boundary condition:

$$\frac{f_2^{(s)}(\bar{H} + \bar{L}_\mu)}{f_2^{(s)}(\bar{H})} = (-1)^{N_e - 1} \exp\left(\frac{3L_\mu \bar{L}_\mu}{4L_B^2}\right) \exp\left(\frac{3L_\mu \bar{H}}{2L_B^2}\right) \exp\left(\frac{\bar{L}_\mu \sum_{mn} \xi_{mn}}{2L_B^2}\right) \exp(i\boldsymbol{L}_\mu \cdot \boldsymbol{\Delta\theta}). \tag{96}$$

There are three linearly independent solutions [4]:

$$f_2^{(s)}(\bar{H}) = \exp\left(i\beta^{(s)}\bar{H}\right) \exp\left(-\frac{L_1 \bar{H}}{2L_B^2}\right) \prod_{j=0}^{2} \sigma^*\left(\frac{H - i\alpha^{(s)} + jL_1/3}{L_B} \Big| \left\{\frac{L_\mu}{L_B}\right\}\right), \quad s = 0, 1, 2 \tag{97}$$

$$\beta^{(s)} = -i \sum_{mn} \frac{\xi_{mn}}{2L_B^2} + \frac{3\alpha^{(s)}}{2L_B^2} \tag{98}$$

$$\alpha^{(s)} = \frac{L_B^2}{3}\left(N_e \frac{T_1 + T_2}{2} + sT_1\right) + \frac{L_B^2 \Delta\theta}{3}. \tag{99}$$

Upon the insertion of one flux quantum, $\boldsymbol{\Delta\theta} \to \boldsymbol{\Delta\theta} + \boldsymbol{T}_1$, the three states labeled by $s$ are cyclically permuted. Each state returns to itself only after three flux quanta are inserted. This spectral flow is characteristic of a threefold topological manifold on the torus and is consistent with the expected $\nu = \frac{2}{3}$ FCI state. However, to show that our trial wavefunction indeed carries a fractional Hall response, one must compute the many-body Chern number, discussed in Sec. IV B 4 and Sec. IV E.

Finally, we note that the skyrmion FCI wavefunction has Bloch periodicity when all the particles are translated by $\boldsymbol{a}_\mu$. This can be proved by noticing that [5]

$$\frac{\Phi(\{\boldsymbol{r} + \boldsymbol{a}_\mu\}, \{\boldsymbol{\eta}\}, \boldsymbol{\Delta\theta}) \prod_{a<b} \sigma^{*2}\left(\frac{\eta_a - \eta_b}{L_B} \big| \{\frac{L_\mu}{L_B}\}\right) F_2^{(s)}(H, \bar{H}, \boldsymbol{\Delta\theta})}{\Phi(\{\boldsymbol{r}\}, \{\boldsymbol{\eta'}\}, \boldsymbol{\Delta\theta}) \prod_{a<b} \sigma^{*2}\left(\frac{\eta'_a - \eta'_b}{L_B} \big| \{\frac{L_\mu}{L_B}\}\right) F_2^{(s)}(H', \bar{H}', \boldsymbol{\Delta\theta})} = \exp(-iN_e \boldsymbol{a}_\mu \cdot \boldsymbol{\Delta\theta}) \exp\left(i\boldsymbol{a}_\mu \cdot \left(N_{\text{vac}}(s + \frac{N_e}{2})\boldsymbol{T}_1\right)\right). \tag{100}$$

Here $\boldsymbol{\eta}'_a \equiv \boldsymbol{\eta}_a - \boldsymbol{a}_\mu$. Since the skyrmion FCI wavefunction, Eq. 92, sums over all the skyrmion vacancy configurations, the wavefunction is Bloch periodic with the total momentum determined by the equation above.

### 3. *A second skyrmion FCI wavefunction*

We now consider a second skyrmion FCI wavefunction, which differs from Eq. 92 only in the relative weights assigned to different skyrmion vacancy configurations:

---

[4] The exact proof of this statement is beyond the scope of this work
[5] The proof is straightforward but tedious

$$\Psi^{\text{frac-SkV},(s)}_{\{\ell=0\},2}(\{\boldsymbol{r}\},\boldsymbol{\Delta\theta}) = \sum_{\{\boldsymbol{\eta}\}} \Phi(\{\boldsymbol{r}\},\{\boldsymbol{\eta}\},\boldsymbol{\Delta\theta}) \prod_{a<b} \sigma^{*2}(\frac{\eta_a-\eta_b}{L_B}|\{\frac{L_\mu}{L_B}\}) F_2^{(s)}(H,\bar{H},\boldsymbol{\Delta\theta}) J(\{\boldsymbol{\eta}\}). \tag{101}$$

$$J(\{\boldsymbol{\eta}\}) \equiv \left( \prod_{a<b} |\sigma(\frac{\eta_a-\eta_b}{L_B}|\{\frac{L_\mu}{L_B}\})|^2 \prod_a \exp\left(-\frac{\eta_a\bar{\eta}_a}{4l_{Bv}^2}\right) \exp\left(\frac{H\bar{H}}{2L_B^2}\right) \right) \tag{102}$$

Compared with $\Psi^{\text{frac-SkV},(s)}_{\{\ell=0\},1}(\{\boldsymbol{r}\},\boldsymbol{\Delta\theta})$, this wavefunction contains an additional factor $J(\{\boldsymbol{\eta}\})$ in the weight of each vacancy configuration. This factor is periodic under $\boldsymbol{\eta}_a \to \boldsymbol{\eta}_a + \boldsymbol{L}_\mu$,

$$J(\{\boldsymbol{\eta}_1, ...\boldsymbol{\eta}_a + \boldsymbol{L}_\mu, ...\}) = J(\{\boldsymbol{\eta}_1, ...\boldsymbol{\eta}_a, ...\}). \tag{103}$$

It is also invariant under a simultaneous translation of all vacancy coordinates,

$$J(\{\boldsymbol{\eta}_1 + \boldsymbol{x}, \boldsymbol{\eta}_2 + \boldsymbol{x}, ...\}) = J(\{\boldsymbol{\eta}\}), \quad \forall \boldsymbol{x} \tag{104}$$

Since $J(\{\boldsymbol{\eta}\})$ is invariant under these transformations, it does not modify the periodicity condition of the full summand. Therefore, $F_2^{(s)}(H,\bar{H},\boldsymbol{\Delta\theta})$ obeys the same boundary condition as before, and the corresponding solutions remain those in Eq. 97.

The two skyrmion FCI wavefunctions only differ by the periodic Jastrow factor $J(\{\boldsymbol{\eta}\})$ in the weighting of the vacancy configurations. This is directly analogous to the difference between the particle-hole and Haldane–Halperin hierarchy constructions of the $\nu = \frac{2}{3}$ fractional quantum Hall state reviewed in Sec. II. In this analogy, $\Psi^{\text{frac-SkV}}_{\{\ell\},1}$ is the Haldane–Halperin-like construction, whereas $\Psi^{\text{frac-SkV}}_{\{\ell\},2}$ is the particle-hole-like construction. All our main text calculations are performed using $\Psi^{\text{frac-SkV}}_{\{\ell\},1}$.

#### *4. Numerical calculation*

We now describe the numerical evaluation of the skyrmion FCI wavefunctions constructed above. The key simplification is that the sum over vacancy configurations is finite, and for each fixed configuration $\{\boldsymbol{\eta}\}$ the skyrmion vacancy wavefunction is a Slater determinant. The skyrmion FCI wavefunction can therefore be represented as a finite superposition of Slater determinants.

To obtain the relative coefficients of these Slater determinants, we must retain the configuration-dependent factors that were left implicit in the proportionality relations above. We first define $\mathcal{R}(\{\boldsymbol{\eta}\},\boldsymbol{\Delta\theta})$ through Eq. 89:

$$\prod_p \exp\left(-\frac{z_p\bar{z}_p}{4l_{Bv}^2}\right) \prod_{p<q} \sigma(\frac{z_p-z_q}{L_B}|\{\frac{L_\mu}{L_B}\}) F_1(Z,-\boldsymbol{\kappa}^{\text{vac}}) = \mathcal{R}(\{\boldsymbol{\eta}\},\boldsymbol{\Delta\theta}) \det_{\substack{\boldsymbol{k}\in\text{BZ2} \\ p\in[1,N_\phi-N_{\text{vac}}]}} [\phi^{\text{LLL,vac}}_{\boldsymbol{k}-\boldsymbol{\kappa}^{\text{vac}}}(\boldsymbol{r}_p)]. \tag{105}$$

The values of the $\mathcal{R}$ factors can be numerically evaluated by matching the Eqs. 105 and 79 at generic real-space points. In terms of $\mathcal{R}$, the two skyrmion FCI wavefunctions are:

$$\begin{aligned}\Psi^{\text{frac-SkV},(s)}_{\{\ell\},1}(\{\boldsymbol{r}\},\boldsymbol{\Delta\theta}) = &\sum_{\{\boldsymbol{\eta}\}} \left( \mathcal{R}(\{\boldsymbol{\eta}\},\boldsymbol{\Delta\theta}) \prod_{a<b} \sigma^{*2}(\frac{\eta_a-\eta_b}{L_B}|\{\frac{L_\mu}{L_B}\}) F_2^{(s)}(H,\bar{H},\boldsymbol{\Delta\theta}) \right) \\ &\times \det_{\boldsymbol{k}\in\text{BZ2},p} [\psi^{\text{SkV}}_{\boldsymbol{k},\ell_p}(\boldsymbol{r}_p,\{\boldsymbol{\eta}\},\boldsymbol{\Delta\theta})].\end{aligned} \tag{106}$$

Similarly,

$$\begin{aligned}\Psi^{\text{frac-SkV},(s)}_{\{\ell\},2}(\{\boldsymbol{r}\},\boldsymbol{\Delta\theta}) = &\sum_{\{\boldsymbol{\eta}\}} \left( \mathcal{R}(\{\boldsymbol{\eta}\},\boldsymbol{\Delta\theta}) \prod_{a<b} \sigma^{*2}(\frac{\eta_a-\eta_b}{L_B}|\{\frac{L_\mu}{L_B}\}) F_2^{(s)}(H,\bar{H},\boldsymbol{\Delta\theta}) J(\{\boldsymbol{\eta}\}) \right) \\ &\times \det_{\boldsymbol{k}\in\text{BZ2},p} [\psi^{\text{SkV}}_{\boldsymbol{k},\ell_p}(\boldsymbol{r}_p,\{\boldsymbol{\eta}\},\boldsymbol{\Delta\theta})].\end{aligned} \tag{107}$$

Here, $\psi^{\text{SkV}}_{\boldsymbol{k},\ell}(\boldsymbol{r},\{\boldsymbol{\eta}\},\boldsymbol{\Delta\theta})$ is defined in Eq. 87. Eqs. 106 and 107 express the skyrmion FCI wavefunctions as finite superpositions of Slater determinants.

For numerical calculations, we need to define an orthogonal basis for the single-particle orbitals at fixed vacancy configuration, as the different $\psi^{\rm SkV}_{\boldsymbol{k},\ell}$ are generally not orthogonal for different $\boldsymbol{k}$. Here, we use the basis

$$\psi^{\rm SkV,orth}_{m,\ell}(\boldsymbol{r},\{\boldsymbol{\eta}\},\boldsymbol{\Delta\theta}) = \sum_{\boldsymbol{k}\in {\rm BZ2}} Q_{\boldsymbol{k},m}(\{\boldsymbol{\eta}\},\boldsymbol{\Delta\theta})\psi^{\rm SkV}_{\boldsymbol{k},\ell}(\boldsymbol{r},\{\boldsymbol{\eta}\},\boldsymbol{\Delta\theta}), \tag{108}$$

where the transformation matrix $Q$ is determined from the overlap matrix of the orbitals $\psi^{\rm SkV}_{\boldsymbol{k},\ell}(\boldsymbol{r},\{\boldsymbol{\eta}\},\boldsymbol{\Delta\theta})$, which can be evaluated using Eq. 87. The orthonormalized Slater determinant differs from the original one by a factor of $\det Q$. Therefore,

$$\begin{aligned}\Psi^{\text{frac-SkV},(s)}_{\{\ell\},1}(\{\boldsymbol{r}\},\boldsymbol{\Delta\theta}) &= \sum_{\{\boldsymbol{\eta}\}}\left(\mathcal{R}(\{\boldsymbol{\eta}\},\boldsymbol{\Delta\theta})\prod_{a<b}\sigma^{*2}(\frac{\eta_a-\eta_b}{L_B}|\{\frac{L_\mu}{L_B}\})F_2^{(s)}(H,\bar{H},\boldsymbol{\Delta\theta})\det(Q(\{\boldsymbol{\eta}\},\boldsymbol{\Delta\theta}))^{-1}\right)\\ &\quad\times\det_{i,p}[\psi^{\rm SkV,orth}_{i,\ell_p}(\boldsymbol{r}_p,\{\boldsymbol{\eta}\},\boldsymbol{\Delta\theta})]\\ &\equiv\sum_{\{\boldsymbol{\eta}\}}C^{(s)}(\{\boldsymbol{\eta}\},\boldsymbol{\Delta\theta})\det_{i,p}[\psi^{\rm SkV,orth}_{i,\ell_p}(\boldsymbol{r}_p,\{\boldsymbol{\eta}\},\boldsymbol{\Delta\theta})].\end{aligned}\tag{109}$$

In the equation above, we collect all configuration-dependent prefactors into the coefficients $C^{(s)}(\{\boldsymbol{\eta}\},\boldsymbol{\Delta\theta})$. We explicitly remark that the Slater determinants associated with different vacancy configurations, $\Psi^{\rm SkV}_{\{\ell=0\}}(\{\boldsymbol{r}\},\{\boldsymbol{\eta}\}_1,\boldsymbol{\Delta\theta})$ and $\Psi^{\rm SkV}_{\{\ell=0\}}(\{\boldsymbol{r}\},\{\boldsymbol{\eta}\}_2,\boldsymbol{\Delta\theta})$, are also generally not orthogonal. This will be important when considering overlaps between states later on.

To proceed, it will be useful to switch to second-quantized notation. Here, we will define $d^\dagger_m(\{\boldsymbol{\eta}\},\boldsymbol{\Delta\theta})$ to be the creation operator associated with the orthonormalized orbital $\psi^{\rm SkV,orth}_{m,\ell}(\boldsymbol{r},\{\boldsymbol{\eta}\},\boldsymbol{\Delta\theta})$. Using Eqs. 108 and 87, we can write,

$$d^\dagger_m(\{\boldsymbol{\eta}\},\boldsymbol{\Delta\theta}) = \sum_{\boldsymbol{k}=n_1\boldsymbol{T}_1+n_2\boldsymbol{T}_2} D_{\boldsymbol{k},m}(\{\boldsymbol{\eta}\},\boldsymbol{\Delta\theta})c^\dagger_{\boldsymbol{k}-\boldsymbol{\Delta\theta}}, \tag{110}$$

where $c^\dagger_{\boldsymbol{k}}$ is the creation operator associated with the parent-band state $s_{\boldsymbol{k},\ell}e^{i\boldsymbol{k}\cdot\boldsymbol{r}}$. The expansion coefficients $D_{\boldsymbol{k},m}$ are determined by the transformation matrix $Q$ and Eq. 87, and satisfy,

$$\sum_{\boldsymbol{k}} D^*_{\boldsymbol{k},m}(\{\boldsymbol{\eta}\},\boldsymbol{\Delta\theta})D_{\boldsymbol{k},n}(\{\boldsymbol{\eta}\},\boldsymbol{\Delta\theta}) = \delta_{mn}. \tag{111}$$

In terms of $d^\dagger$, the (unnormalized) skyrmion FCI wavefunction can therefore be written compactly as,

$$\left|\Psi^{\text{frac-SkV},(s)}_1(\boldsymbol{\Delta\theta})\right\rangle = \sum_{\{\boldsymbol{\eta}\}}C^{(s)}(\{\boldsymbol{\eta}\},\boldsymbol{\Delta\theta})\prod_m d^\dagger_m(\{\boldsymbol{\eta}\},\boldsymbol{\Delta\theta})\left|0\right\rangle. \tag{112}$$

We now discuss how various physical quantities are evaluated using the wavefunctions we have constructed here. First, we will summarize how the many-body Chern number of the skyrmion FCI states is calculated. As discussed in Sec. IV B 2, the three states are cyclically permuted under flux threading. The many-body Chern number provides a direct characterization of the corresponding fractional Hall response. We parametrize the boundary-condition twist as,

$$\boldsymbol{\Delta\theta} \equiv \frac{\theta_1}{2\pi}\boldsymbol{T}_1 + \frac{\theta_2}{2\pi}\boldsymbol{T}_2. \tag{113}$$

For a normalized many-body state $|\Psi(\boldsymbol{\Delta\theta})\rangle$, the Berry curvature and Chern number on this twisted-boundary-condition manifold are defined as

$$\mathcal{F}(\theta_1,\theta_2) = -{\rm Im}\sum_{\mu,\nu}\left(\epsilon_{\mu\nu}\langle\partial_{\theta_\mu}[\exp\left(i\sum_p\boldsymbol{r}_p\cdot\boldsymbol{\Delta\theta}\right)\Psi(\boldsymbol{\Delta\theta})]|\partial_{\theta_\nu}[\exp\left(i\sum_p\boldsymbol{r}_p\cdot\boldsymbol{\Delta\theta}\right)\Psi(\boldsymbol{\Delta\theta})]\rangle\right), \tag{114}$$

$$C = \frac{1}{2\pi}\int_{[0,2\pi]\times[0,2\pi]} d\theta_1 d\theta_2\mathcal{F}(\theta_1,\theta_2). \tag{115}$$

Here, $|\Psi(\boldsymbol{\Delta\theta})\rangle$ denotes a many-body wavefunction with twisted boundary condition $\boldsymbol{\Delta\theta}$. The factor $\exp\left(i\sum_p \boldsymbol{r}_p\cdot\boldsymbol{\Delta\theta}\right)$ removes the boundary condition twist, allowing states at different $\boldsymbol{\Delta\theta}$ to be compared in the same periodic Hilbert space. For a $d$-dimensional topological manifold, the Berry curvature must be generalized to its non-Abelian form. If $C$ denotes the total Chern number of the manifold, the corresponding Hall conductivity is $\sigma_{\mathrm{H}} = Ce^2/(dh)$ [9, 10].

We evaluate the Chern number numerically using the discretized non-Abelian formulation of Ref. [11, 12]. We discretize $(\theta_1, \theta_2)$ and construct the three skyrmion FCI states at each value of $\boldsymbol{\Delta\theta}$ using Eq. 92 and Eq. 101. The non-Abelian link variables are then obtained from the $3\times 3$ overlap matrices between the trial wavefunction manifolds at neighboring values of $\boldsymbol{\Delta\theta}$. For reference, the overlap between two unorthogonalized skyrmion FCI states at neighboring twist angles can be evaluated directly from the Slater-determinant expansion,

$$\begin{aligned}&\langle\Psi_1^{\text{frac-SkV},(s_1)}(\boldsymbol{\Delta\theta}_1)|\exp\left(i\sum_p \boldsymbol{r}_p\cdot(\boldsymbol{\Delta\theta}_2-\boldsymbol{\Delta\theta}_1)\right)|\Psi_1^{\text{frac-SkV},(s_2)}(\boldsymbol{\Delta\theta}_2)\rangle\\ &=\sum_{\{\boldsymbol{\eta}\}_1}\sum_{\{\boldsymbol{\eta}\}_2} C^{(s_1)*}(\{\boldsymbol{\eta}\}_1,\boldsymbol{\Delta\theta}_1)C^{(s_2)}(\{\boldsymbol{\eta}\}_2,\boldsymbol{\Delta\theta}_2)\det_{mn}\Big(\sum_{\boldsymbol{k}} D^*_{\boldsymbol{k},m}(\{\boldsymbol{\eta}\}_1,\boldsymbol{\Delta\theta}_1)D_{\boldsymbol{k},n}(\{\boldsymbol{\eta}\}_2,\boldsymbol{\Delta\theta}_2)\langle s_{\boldsymbol{k}-\boldsymbol{\Delta\theta_1}}|s_{\boldsymbol{k}-\boldsymbol{\Delta\theta_2}}\rangle\Big).\end{aligned} \tag{116}$$

For both skyrmion FCI constructions, $\Psi^{\text{frac-SkV},(s)}_{\{\ell\},1}$ and $\Psi^{\text{frac-SkV},(s)}_{\{\ell\},2}$, we obtain a total many-body Chern number $C = 2$ for the three-dimensional manifold (see Sec. IV E and Fig. 2). The corresponding Hall conductivity is therefore $\sigma_{\mathrm{H}} = 2e^2/(3h)$, as expected for a $\nu = \frac{2}{3}$ fractional Chern insulator state.

Next we discuss how we calculate the energy of the state. Other local observables can be evaluated analogously. Since the skyrmion FCI wavefunction is a finite superposition of Slater determinants, it is sufficient to evaluate matrix elements between pairs of Slater determinants. Using the second-quantized notation, we denote the Slater determinant associated with a particular vacancy configuration $\{\boldsymbol{\eta}\}_1$ by,

$$|D_1\rangle \equiv \prod_n d_n^\dagger(\{\boldsymbol{\eta}\}_1,\boldsymbol{\Delta\theta})\,|0\rangle\,. \tag{117}$$

The many-body overlap between two Slater determinants is,

$$\langle D_1|D_2\rangle = \det S, \tag{118}$$

where,

$$S_{ab} \equiv \sum_{\boldsymbol{k}} D^*_{\boldsymbol{k},a}(\{\boldsymbol{\eta}\}_1,\boldsymbol{\Delta\theta})D_{\boldsymbol{k},b}(\{\boldsymbol{\eta}\}_2,\boldsymbol{\Delta\theta}). \tag{119}$$

As mentioned before, the Slater determinants associated with different vacancy configurations are not orthogonal in general.

To find the energy, we will need to consider both one-body and two-body terms in the Hamiltonian. For a general one-body term of the form,

$$H_{\text{single}} = \sum_{\boldsymbol{k}_1,\boldsymbol{k}_2=n_1\boldsymbol{T}_1+n_2\boldsymbol{T}_2} h^{\text{single}}_{\boldsymbol{k}_1,\boldsymbol{k}_2}(\boldsymbol{\Delta\theta})c^\dagger_{\boldsymbol{k}_1-\boldsymbol{\Delta\theta}}c_{\boldsymbol{k}_2-\boldsymbol{\Delta\theta}}, \tag{120}$$

the corresponding one-body energy for a general many-body state $|\Psi(\boldsymbol{\Delta\theta})\rangle$ is

$$E_{\text{single}} = \frac{\sum_{\boldsymbol{k}_1,\boldsymbol{k}_2} h^{\text{single}}_{\boldsymbol{k}_1,\boldsymbol{k}_2}\langle\Psi(\boldsymbol{\Delta\theta})|c^\dagger_{\boldsymbol{k}_1-\boldsymbol{\Delta\theta}}c_{\boldsymbol{k}_2-\boldsymbol{\Delta\theta}}|\Psi(\boldsymbol{\Delta\theta})\rangle}{\langle\Psi(\boldsymbol{\Delta\theta})|\Psi(\boldsymbol{\Delta\theta})\rangle}. \tag{121}$$

For the calculations presented in the main text, the relevant one-body contributions are the kinetic energy. The kinetic Hamiltonian (projected onto the ideal band) is

$$H_{\text{KE}} = \sum_{\boldsymbol{k}} \mathcal{E}(\boldsymbol{k}-\boldsymbol{\Delta\theta})c^\dagger_{\boldsymbol{k}-\boldsymbol{\Delta\theta}}c_{\boldsymbol{k}-\boldsymbol{\Delta\theta}}, \tag{122}$$

where $\mathcal{E}(\boldsymbol{k})$ is the dispersion of the first conduction band of $h_{\text{RMG}}(\boldsymbol{K}+\boldsymbol{k})$ at the corresponding displacement field $u_D$.

To numerically evaluate the energy in Eq. 120, we note that the matrix elements of a one-body operator between the two Slater determinants can be written in terms of the first-order minors of $S$:

$$\langle D_1|c^\dagger_{\boldsymbol{k}_1-\boldsymbol{\Delta\theta}}c_{\boldsymbol{k}_2-\boldsymbol{\Delta\theta}}|D_2\rangle = \sum_{m,n} D^*_{\boldsymbol{k}_1,m}(\{\boldsymbol{\eta}\}_1,\boldsymbol{\Delta\theta})D_{\boldsymbol{k}_2,n}(\{\boldsymbol{\eta}\}_2,\boldsymbol{\Delta\theta})(-1)^{m+n}\det S^{(m|n)}. \tag{123}$$

Here, $S^{(m|n)}$ denotes the matrix obtained by deleting the $m$-th row and the $n$-th column of $S$. We caution the reader that for some vacancy configurations, $S$ can be nearly singular, and we suggest calculating the minors directly rather than relying on the matrix inverse of $S$.

We next consider a translationally invariant two-body interaction $V(\boldsymbol{r})$. Projected onto the ideal band, the interaction Hamiltonian is

$$H_{\text{int}} = \frac{1}{2A}\sum_{\boldsymbol{q},\boldsymbol{k}_1,\boldsymbol{k}_2} V(\boldsymbol{q})c^\dagger_{\boldsymbol{k}_1-\boldsymbol{\Delta\theta}}c^\dagger_{\boldsymbol{k}_2-\boldsymbol{\Delta\theta}}c_{\boldsymbol{k}_2+\boldsymbol{q}-\boldsymbol{\Delta\theta}}c_{\boldsymbol{k}_1-\boldsymbol{q}-\boldsymbol{\Delta\theta}}\langle s_{\boldsymbol{k}_1-\boldsymbol{\Delta\theta}}|s_{\boldsymbol{k}_1-\boldsymbol{q}-\boldsymbol{\Delta\theta}}\rangle\langle s_{\boldsymbol{k}_2-\boldsymbol{\Delta\theta}}|s_{\boldsymbol{k}_2+\boldsymbol{q}-\boldsymbol{\Delta\theta}}\rangle \tag{124}$$

Here, $A$ is the area of the torus, and $V(\boldsymbol{q})$ is the Fourier transform of the interaction,

$$V(\boldsymbol{q}) = \int d\boldsymbol{r}\, V(\boldsymbol{r})e^{-i\boldsymbol{q}\cdot\boldsymbol{r}}, \qquad V(\boldsymbol{r}) = \frac{1}{A}\sum_{\boldsymbol{q}} V(\boldsymbol{q})e^{i\boldsymbol{q}\cdot\boldsymbol{r}}. \tag{125}$$

For a many-body state $|\Psi(\boldsymbol{\Delta\theta})\rangle$, we define the normal-ordered density correlation,

$$g(\boldsymbol{q}) \equiv \frac{\sum_{\boldsymbol{k}_1,\boldsymbol{k}_2}\langle\Psi(\boldsymbol{\Delta\theta})|c^\dagger_{\boldsymbol{k}_1-\boldsymbol{\Delta\theta}}c^\dagger_{\boldsymbol{k}_2-\boldsymbol{\Delta\theta}}c_{\boldsymbol{k}_2+\boldsymbol{q}-\boldsymbol{\Delta\theta}}c_{\boldsymbol{k}_1-\boldsymbol{q}-\boldsymbol{\Delta\theta}}|\Psi(\boldsymbol{\Delta\theta})\rangle\langle s_{\boldsymbol{k}_1-\boldsymbol{\Delta\theta}}|s_{\boldsymbol{k}_1-\boldsymbol{q}-\boldsymbol{\Delta\theta}}\rangle\langle s_{\boldsymbol{k}_2-\boldsymbol{\Delta\theta}}|s_{\boldsymbol{k}_2+\boldsymbol{q}-\boldsymbol{\Delta\theta}}\rangle}{\langle\Psi(\boldsymbol{\Delta\theta})|\Psi(\boldsymbol{\Delta\theta})\rangle}. \tag{126}$$

With this definition, $g(\boldsymbol{0}) = N_e(N_e-1)$. In terms of $g(\boldsymbol{q})$, the interaction energy is

$$E_{\text{int}} = \frac{1}{2A}\sum_{\boldsymbol{q}} V(\boldsymbol{q})g(\boldsymbol{q}). \tag{127}$$

To evaluate the numerator of $g(\boldsymbol{q})$, we use that the matrix element of a two-body operator can be written in terms of the second-order minors of $S$:

$$\begin{aligned}
&\langle D_1|c^\dagger_{\boldsymbol{k}_1-\boldsymbol{\Delta\theta}}c^\dagger_{\boldsymbol{k}_2-\boldsymbol{\Delta\theta}}c_{\boldsymbol{k}_4-\boldsymbol{\Delta\theta}}c_{\boldsymbol{k}_3-\boldsymbol{\Delta\theta}}|D_2\rangle \\
=& \sum_{m<r}\sum_{n<s}(-1)^{m+r+n+s}\det S^{(m,r|n,s)} \\
&\times\left[D^*_{\boldsymbol{k}_1,m}(\{\boldsymbol{\eta}\}_1,\boldsymbol{\Delta\theta})D^*_{\boldsymbol{k}_2,r}(\{\boldsymbol{\eta}\}_1,\boldsymbol{\Delta\theta}) - D^*_{\boldsymbol{k}_1,r}(\{\boldsymbol{\eta}\}_1,\boldsymbol{\Delta\theta})D^*_{\boldsymbol{k}_2,m}(\{\boldsymbol{\eta}\}_1,\boldsymbol{\Delta\theta})\right] \\
&\times\left[D_{\boldsymbol{k}_3,n}(\{\boldsymbol{\eta}\}_2,\boldsymbol{\Delta\theta})D_{\boldsymbol{k}_4,s}(\{\boldsymbol{\eta}\}_2,\boldsymbol{\Delta\theta}) - D_{\boldsymbol{k}_3,s}(\{\boldsymbol{\eta}\}_2,\boldsymbol{\Delta\theta})D_{\boldsymbol{k}_4,n}(\{\boldsymbol{\eta}\}_2,\boldsymbol{\Delta\theta})\right].
\end{aligned} \tag{128}$$

Here, $S^{(m,r|n,s)}$ denotes the matrix obtained by deleting the $m$-th and $r$-th rows and the $n$-th and $s$-th columns of $S$.

For the energy calculations presented in the main text, we use the dual-gate-screened Coulomb interaction,

$$V(\boldsymbol{q}) = \frac{e^2\tanh(|\boldsymbol{q}|D)}{2\epsilon_r\epsilon_0|\boldsymbol{q}|}, \qquad D = 25\,\text{nm}, \qquad \epsilon_r = 5. \tag{129}$$

The expressions above apply to an arbitrary many-body state $|\Psi(\boldsymbol{\Delta\theta})\rangle$. For each skyrmion FCI construction, we obtain three generally nonorthogonal many-body states spanning the three-dimensional topological manifold[6]. To evaluate quantities such as the energy of this manifold, we first construct an orthonormal basis for these three states and then average the expectation value over the three orthonormalized states.

In the main text, we also plot the layer-resolved two-particle correlation function in real space, defined as

$$g_{\ell,\ell'}(\boldsymbol{r}_1,\boldsymbol{r}_2) \equiv \frac{A^2}{N_e^2}\frac{\langle\Psi(\boldsymbol{\Delta\theta})|c^\dagger_\ell(\boldsymbol{r}_1)c^\dagger_{\ell'}(\boldsymbol{r}_2)c_{\ell'}(\boldsymbol{r}_2)c_\ell(\boldsymbol{r}_1)|\Psi(\boldsymbol{\Delta\theta})\rangle}{\langle\Psi(\boldsymbol{\Delta\theta})|\Psi(\boldsymbol{\Delta\theta})\rangle}. \tag{130}$$

---

[6] This nonorthogonality is distinct from the nonorthogonalities discussed in the context of the single-particle orbitals, or the Slater determinants. For the three-by-three geometry we use in this paper, the three many-body skyrmion FCI states, labelled by $s$, are non-orthogonal.

Here, $c^\dagger_\ell(\boldsymbol{r})$ is the field operator projected onto the holomorphic parent band $|s_{\boldsymbol{k}}\rangle$, defined as

$$c^\dagger_\ell(\boldsymbol{r}) = \frac{1}{\sqrt{A}} \sum_{\boldsymbol{k}=n_1\boldsymbol{T}_1+n_2\boldsymbol{T}_2} s^*_{\boldsymbol{k}-\boldsymbol{\Delta\theta},\ell} c^\dagger_{\boldsymbol{k}-\boldsymbol{\Delta\theta}} e^{-i(\boldsymbol{k}-\boldsymbol{\Delta\theta})\cdot\boldsymbol{r}}. \tag{131}$$

Using the equation above, we can prove that the real-space correlation function is related to $g(\boldsymbol{q})$ by

$$\frac{1}{A}\sum_{\boldsymbol{q}} g(\boldsymbol{q}) e^{i\boldsymbol{q}\cdot\boldsymbol{R}} = \frac{N_e^2}{A^2}\sum_{\ell,\ell'}\int d\boldsymbol{r}\, g_{\ell,\ell'}(\boldsymbol{r},\boldsymbol{r}+\boldsymbol{R}). \tag{132}$$

Therefore, the interaction energy $E_{\text{int}}$ can be written as a real space integral:

$$E_{\text{int}} = \frac{N_e^2}{2A^2}\sum_{\ell,\ell'}\int d\boldsymbol{r} d\boldsymbol{R} V(\boldsymbol{R}) g_{\ell\ell'}(\boldsymbol{r},\boldsymbol{r}-\boldsymbol{R}) = \frac{N_e^2}{2A}\int d\boldsymbol{R} V(\boldsymbol{R}) g(\boldsymbol{R}), \tag{133}$$

where we have defined

$$g(\boldsymbol{R}) \equiv \frac{1}{A}\int d\boldsymbol{r}\sum_{\ell,\ell'} g_{\ell\ell'}(\boldsymbol{r},\boldsymbol{r}-\boldsymbol{R}) = \frac{1}{N_e^2}\sum_{\boldsymbol{q}} g(\boldsymbol{q}) e^{-i\boldsymbol{q}\cdot\boldsymbol{R}} \tag{134}$$

$g(\boldsymbol{R})$ is also plotted in the main text. We note that its spatial average is

$$\frac{1}{A}\int d\boldsymbol{R}\, g(\boldsymbol{R}) = \frac{N_e-1}{N_e}. \tag{135}$$

### C. Conventional FCI

We now construct a second class of $\nu = \frac{2}{3}$ trial states, which we refer to as conventional FCI states. Unlike the skyrmion FCI construction, where we first introduce skyrmion vacancies and then form a coherent superposition of vacancy configurations, the conventional FCI construction directly fractionalizes the single-particle states of the filled ideal Chern band. This follows the standard mapping between ideal Chern bands and Landau levels [13]. We begin with the $\nu = 1$ skyrmion-lattice state in the ideal limit. For the $\{\ell = 0\}$ component,

$$\Psi^{\nu=1,\text{SkL}}_{\{\ell=0\}}(\{\boldsymbol{r}\},\boldsymbol{\Delta\theta}) = \det_{\boldsymbol{k}\in\text{BZ1},p}[\phi^{\text{LLL}}_{\boldsymbol{k}-\boldsymbol{\Delta\theta}-\boldsymbol{q}}(\boldsymbol{r}_p)\phi^{\text{LLL}*}_{-\boldsymbol{q}}(\boldsymbol{r}_p)] = \det_{\boldsymbol{k}\in\text{BZ1},p}[\phi^{\text{LLL}}_{\boldsymbol{k}-\boldsymbol{\Delta\theta}-\boldsymbol{q}}(\boldsymbol{r}_p)]\prod_p \phi^{\text{LLL}*}_{-\boldsymbol{q}}(\boldsymbol{r}_p), \tag{136}$$

and,

$$\Psi^{\nu=1,\text{SkL}}_{\{\ell\}}(\{\boldsymbol{r}\},\boldsymbol{\Delta\theta}) = \prod_p(-2i\gamma\partial_{\bar{z}_p})^{\ell_p}\Psi^{\nu=1,\text{SkL}}_{\{\ell=0\}}(\{\boldsymbol{r}\};\boldsymbol{\Delta\theta}). \tag{137}$$

Here, BZ1 is defined in Eq.22. As shown in the Supplementary Material of Ref. [7], the same state can be written as a Slater determinant obtained by filling the entire $\nu = 1$ skyrmion-lattice band:

$$\Psi^{\nu=1,\text{SkL}}_{\{\ell\}}(\{\boldsymbol{r}\};\boldsymbol{\Delta\theta}) \propto \det_{\boldsymbol{k}\in\text{BZ1},p}\left[\psi^{\text{SkL}}_{\boldsymbol{k}-\boldsymbol{\Delta\theta},\ell_p}(\boldsymbol{r}_p)\right], \tag{138}$$

where the normalized Bloch states are,

$$\psi^{\text{SkL}}_{\boldsymbol{k}-\boldsymbol{\Delta\theta},\ell}(\boldsymbol{r}) = \mathcal{N}_{\boldsymbol{k}-\boldsymbol{\Delta\theta}}\sum_{\boldsymbol{g}=n_1\boldsymbol{b}_1+n_2\boldsymbol{b}_2} e^{i\pi n_1 n_2} e^{-\frac{l_B^2}{4}(g\bar{g}+2(k-\Delta\theta)\bar{g})} e^{i(\boldsymbol{k}-\boldsymbol{\Delta\theta}+\boldsymbol{g})\cdot(\boldsymbol{r}-\boldsymbol{R}_1)}\frac{s_{\boldsymbol{k}-\boldsymbol{\Delta\theta}+\boldsymbol{g},\ell}}{N_{\boldsymbol{k}-\boldsymbol{\Delta\theta}+\boldsymbol{g}}}, \quad \boldsymbol{R}_1 = (-q_y, q_x)l_B^2. \tag{139}$$

Here, $N_{\boldsymbol{k}}$ is the normalization factor appearing in Eq. 64, while $\mathcal{N}_{\boldsymbol{k}}$ normalizes the Bloch state $\psi^{\text{SkL}}_{\boldsymbol{k},\ell}(\boldsymbol{r})$. Eq. 138 is derived using the fact that the product of the two LLL factors in Eq. 136 is proportional to the $\ell = 0$ component of a skyrmion-lattice Bloch state [7]:

$$\phi^{\text{LLL}}_{\boldsymbol{k}-\boldsymbol{\Delta\theta}-\boldsymbol{q}}(\boldsymbol{r})\phi^{\text{LLL}*}_{-\boldsymbol{q}}(\boldsymbol{r}) = \tilde{\mathcal{M}}_{\boldsymbol{k}-\boldsymbol{\Delta\theta}}(\boldsymbol{q})\psi^{\text{SkL}}_{\boldsymbol{k}-\boldsymbol{\Delta\theta},0}(\boldsymbol{r}). \tag{140}$$

In Eq. 136, $\det_{\boldsymbol{k}\in\mathrm{BZ1},p}[\phi^{\mathrm{LLL}}_{\boldsymbol{k}-\boldsymbol{\Delta\theta}-\boldsymbol{q}}(\boldsymbol{r}_p)]$ is the fully filled LLL $N_\phi$-electron wavefunction on an $N_\phi$-flux torus. The conventional FCI construction consists of replacing this filled LLL factor by the corresponding fractional quantum Hall wavefunction [13], while retaining the common factor $\phi^{\mathrm{LLL}*}_{-\boldsymbol{q}}(\boldsymbol{r})$ for each electron. At $\nu=\frac{1}{3}$, this gives,

$$\Psi^{\nu=\frac{1}{3},\mathrm{FCI},(s)}_{\{\ell=0\}}(\{\boldsymbol{r}\};\boldsymbol{\Delta\theta})=\Psi^{\nu=\frac{1}{3},\mathrm{Laughlin}}_{(s)}(\{\boldsymbol{r}\};\boldsymbol{\Delta\theta}+\boldsymbol{q})\prod_{p=1}^{\frac{N_\phi}{3}}\phi^{\mathrm{LLL}*}_{-\boldsymbol{q}}(\boldsymbol{r}_p). \tag{141}$$

Because the Laughlin factor has a third-order zero when two particles coincide, this wavefunction is an exact zero-energy state of the layer-isotropic Trugman–Kivelson interaction. It can therefore also be obtained directly by band-projected exact diagonalization within the skyrmion-lattice Bloch band $\psi^{\mathrm{SkL}}_{\boldsymbol{k},\ell}$.

The same construction at $\nu=\frac{2}{3}$ is obtained by replacing the filled LLL factor by the particle-hole-conjugate $\nu=\frac{2}{3}$ state:

$$\Psi^{\nu=\frac{2}{3},\mathrm{FCI},(s)}_{\{\ell=0\}}(\{\boldsymbol{r}\};\boldsymbol{\Delta\theta})=\Psi^{\nu=\frac{2}{3},\mathrm{PH}}_{(s)}(\{\boldsymbol{r}\};\boldsymbol{\Delta\theta}+\boldsymbol{q})\prod_{p=1}^{\frac{2N_\phi}{3}}\phi^{\mathrm{LLL}*}_{-\boldsymbol{q}}(\boldsymbol{r}_p). \tag{142}$$

Using the Slater-determinant expansion derived in Sec. II C, this state can be written as,

$$\Psi^{\nu=\frac{2}{3},\mathrm{FCI},(s)}_{\{\ell=0\}}(\{\boldsymbol{r}\};\boldsymbol{\Delta\theta})=\sum_{\{\boldsymbol{k}\},\sum\boldsymbol{k}=\boldsymbol{k}^{\mathrm{tot}}_{(s)}}C^{(s)*}_{\{\boldsymbol{k}\}}(\boldsymbol{\Delta\theta}+\boldsymbol{q})\mathrm{sgn}(\{\boldsymbol{k}\})\det_{\substack{\boldsymbol{k}_i\notin\{\boldsymbol{k}\},p\\ I_1<I_2\ldots}}[\phi^{\mathrm{LLL}}_{\boldsymbol{k}_i-\boldsymbol{\Delta\theta}-\boldsymbol{q}}(\boldsymbol{r}_p)\phi^{\mathrm{LLL}*}_{-\boldsymbol{q}}(\boldsymbol{r}_p)], \tag{143}$$

and

$$\begin{aligned}\Psi^{\nu=\frac{2}{3},\mathrm{FCI},(s)}_{\{\ell\}}(\{\boldsymbol{r}\};\boldsymbol{\Delta\theta})&=\prod_p(-2i\gamma\partial_{\bar z_p})^{\ell_p}\Psi^{\nu=\frac{2}{3},\mathrm{FCI},(s)}_{\{\ell=0\}}(\{\boldsymbol{r}\};\boldsymbol{\Delta\theta})\\&=\sum_{\{\boldsymbol{k}\},\sum\boldsymbol{k}=\boldsymbol{k}^{\mathrm{tot}}_{(s)}}\mathrm{sgn}(\{\boldsymbol{k}\})C^{(s)*}_{\{\boldsymbol{k}\}}(\boldsymbol{\Delta\theta}+\boldsymbol{q})\Big(\prod_{\boldsymbol{k}_i\notin\{\boldsymbol{k}\}}\tilde{\mathcal{M}}_{\boldsymbol{k}_i-\boldsymbol{\Delta\theta}}(\boldsymbol{q})\Big)\det_{\substack{\boldsymbol{k}_i\notin\{\boldsymbol{k}\},p\\ I_1<I_2\ldots}}[\psi^{\mathrm{SkL}}_{\boldsymbol{k}_i-\boldsymbol{\Delta\theta},\ell_p}(\boldsymbol{r}_p)]\end{aligned} \tag{144}$$

Compared with the skyrmion FCI expansion, Eq. 144 is considerably simpler to evaluate numerically. The skyrmion-lattice Bloch states $\psi^{\mathrm{SkL}}_{\boldsymbol{k},\ell}$ are mutually orthogonal for distinct $\boldsymbol{k}$, such that the Slater determinants associated with distinct momentum sets are automatically orthogonal,

$$\int\prod_p d\boldsymbol{r}_p\det_{\substack{\boldsymbol{k}_i\notin\{\boldsymbol{k}\}_1,p\\ I_1<I_2\ldots}}[\psi^{\mathrm{SkL}}_{\boldsymbol{k}_i-\boldsymbol{\Delta\theta},\ell_p}(\boldsymbol{r}_p)]^*\det_{\substack{\boldsymbol{k}_i\notin\{\boldsymbol{k}\}_2,p\\ I_1<I_2\ldots}}[\psi^{\mathrm{SkL}}_{\boldsymbol{k}_i-\boldsymbol{\Delta\theta},\ell_p}(\boldsymbol{r}_p)]=0,\quad\{\boldsymbol{k}\}_1\neq\{\boldsymbol{k}\}_2. \tag{145}$$

We finally discuss the overlap between the three topological sectors. The coefficients $C^{(s)}_{\{\boldsymbol{k}\}}(\boldsymbol{\Delta\theta}+\boldsymbol{q})$ are obtained from the LLL exact diagonalization using the form factors given in Sec. II. For the $3\times3$ geometry used in this work, the three $\nu=\frac{1}{3}$ Laughlin states lie in the same total-momentum sector, $\boldsymbol{k}^{\mathrm{tot}}_{(s)}=\boldsymbol{0}$, and the corresponding coefficient vectors obtained from exact diagonalization are orthonormal,

$$\sum_{\{\boldsymbol{k}\},\sum\boldsymbol{k}=\boldsymbol{0}}C^{(s_1)*}_{\{\boldsymbol{k}\}}(\boldsymbol{\Delta\theta}+\boldsymbol{q})C^{(s_2)}_{\{\boldsymbol{k}\}}(\boldsymbol{\Delta\theta}+\boldsymbol{q})=\delta_{s_1s_2}. \tag{146}$$

In the conventional FCI wavefunction, however, each Slater determinant acquires the additional configuration-dependent factor $\prod_{\boldsymbol{k}_i\notin\{\boldsymbol{k}\}}\tilde{\mathcal{M}}_{\boldsymbol{k}_i-\boldsymbol{\Delta\theta}}(\boldsymbol{q})$. Consequently, the corresponding conventional FCI states are generally nonorthogonal. For quantities such as the energy, we construct an orthonormal basis for the three-state manifold using the procedure described in Sec. IV B 4, and report the average energy over the three orthonormalized states.

### D. Additional Variational Parameters

In the main text, we introduce a variational parameter $\xi$ that smooths the short-distance structure of the wavefunction. This is implemented by acting on the many-body wavefunction with $\exp\Big[\frac{\xi^2}{4}\sum_p\nabla_p^2\Big]$, which suppresses its

large-momentum components. Since this operator factorizes into single-particle operators, it can be applied independently to each orbital within every Slater determinant.

$$\begin{aligned}&\exp\left(\sum_p \frac{\xi^2\nabla_p^2}{4}\right)\Psi^{\text{frac-SkV},(s)}_{\{\ell\},1}(\{\boldsymbol{r}\},\boldsymbol{\Delta\theta})\\&=\sum_{\{\boldsymbol{\eta}\}}\left(\mathcal{R}(\{\boldsymbol{\eta}\},\boldsymbol{\Delta\theta})\prod_{a<b}\sigma^{*2}(\frac{\eta_a-\eta_b}{L_B}|\{\frac{L_\mu}{L_B}\})F_2^{(s)}(H,\bar{H},\boldsymbol{\Delta\theta})\right)\det_{\boldsymbol{k}\in\text{BZ2},p}[\exp\left(\frac{\xi^2\nabla_p^2}{4}\right)\psi^{\text{SkV}}_{\boldsymbol{k},\ell_p}(\boldsymbol{r}_p,\{\boldsymbol{\eta}\},\boldsymbol{\Delta\theta})].\end{aligned} \tag{147}$$

Using the plane-wave expansion in Eq. 87, the action of the smoothing operator is simple. For a plane-wave component with momentum $\boldsymbol{Q}$,

$$\exp\left(\frac{\xi^2\nabla^2}{4}\right)e^{i\boldsymbol{Q}\cdot\boldsymbol{r}}=e^{-\xi^2|\boldsymbol{Q}|^2/4}e^{i\boldsymbol{Q}\cdot\boldsymbol{r}}. \tag{148}$$

Therefore, each plane-wave component of $\psi^{\text{SkV}}_{\boldsymbol{k},\ell}$ acquires a Gaussian suppression factor:

$$\begin{aligned}&\exp\left(\frac{\xi^2\nabla^2}{4}\right)\psi^{\text{SkV}}_{\boldsymbol{k},\ell}(\boldsymbol{r},\{\boldsymbol{\eta}\},\boldsymbol{\Delta\theta})\\&=\sum_{\substack{\boldsymbol{k}_2\in\text{BZ2}\\ \boldsymbol{g}=n_1\boldsymbol{b}_{1v}+n_2\boldsymbol{b}_{2v}}} W_{\boldsymbol{k}_2}\mathcal{M}_{\boldsymbol{k},\boldsymbol{k}_2}e^{i\pi n_1n_2}e^{-\frac{l_{Bv}^2}{4}(g\bar{g}+2(k-k_2-\Delta\theta)\bar{g})}e^{i(\boldsymbol{k}-\boldsymbol{k}_2-\boldsymbol{\Delta\theta}+\boldsymbol{g})\cdot(\boldsymbol{r}-\boldsymbol{R}_0)}\frac{s_{\boldsymbol{k}-\boldsymbol{k}_2-\boldsymbol{\Delta\theta}+\boldsymbol{g},\ell}}{N_{\boldsymbol{k}-\boldsymbol{k}_2-\boldsymbol{\Delta\theta}+\boldsymbol{g}}}e^{-\frac{\xi^2|\boldsymbol{k}-\boldsymbol{k}_2-\boldsymbol{\Delta\theta}+\boldsymbol{g}|^2}{4}}.\end{aligned} \tag{149}$$

The smoothed skyrmion FCI state therefore remains a finite superposition of Slater determinants, with only the single-particle orbitals modified. All overlap and observable calculations described in Sec. IV B 4 can consequently be applied without further modification.

The conventional FCI state can be smoothed in the same way. Starting from Eq. 144, one applies $\exp(\xi^2\nabla^2/4)$ to each skyrmion-lattice orbital $\psi^{\text{SkL}}_{\boldsymbol{k},\ell}$ of Eq. 139; equivalently, each of its plane-wave components is multiplied by the corresponding factor $e^{-\xi^2|\boldsymbol{Q}|^2/4}$.

### E. Extended Data

In this subsection, we provide extended data on various quantities calculated with the FCI wavefunctions. In Fig. 2, we plot the many-body Berry curvature over the boundary-condition-twist manifold $\boldsymbol{\Delta\theta}$ for the skyrmion FCI and conventional FCI wavefunctions. With or without the smoothing parameter $\xi$, the many-body Berry curvature is nearly uniform, with $\mathcal{F}(\theta_1,\theta_2)\approx 2/(2\pi)$, and its integral over $(\theta_1,\theta_2)$ gives a total many-body Chern number $C=2$ for the three-state manifold. Thus, all of the trial wavefunctions have Hall conductivity $\sigma_H=2e^2/(3h)$.

In Fig. 3, we compare the energetics of the two skyrmion FCI trial wavefunctions, $\Psi^{\text{frac-SkV},(s)}_{\{\ell\},1}$ and $\Psi^{\text{frac-SkV},(s)}_{\{\ell\},2}$. We include both the kinetic term, Eq. 122, and the interaction term, Eq. 124. In Fig. 3(a), we show the energy difference without smoothing, $\xi=0$. The energy difference is approximately 1 meV per electron. The energy difference also remains small compared with the energy difference between the skyrmion FCI and conventional FCI states throughout the relevant parameter regime (shown in the main text). We therefore present results in the main text only for $\Psi^{\text{frac-SkV},(s)}_{\{\ell\},1}$, in which the vacancies form a bosonic $\nu=\frac{1}{2}$ Laughlin state. In Fig. 3 (b), we show the energy difference after applying the smoothing operator $\exp\left(\xi^2\sum_p\nabla_p^2/4\right)$. We find that the energy difference remains small for nonzero $\xi$.

In Fig. 4, we compare the layer-resolved charge densities of the skyrmion FCI and conventional FCI states. For the skyrmion FCI, the $\ell=0$ charge density no longer contains zeros [Fig. 4(a)]. This is because each skyrmion vacancy configuration removes a subset of the zeros present in the $\ell=0$ component of the $\nu=1$ skyrmion-lattice state. Since the skyrmion FCI wavefunction is a coherent superposition over all allowed skyrmion vacancy configurations, the zeros of $\rho_{\ell=0}(\boldsymbol{r})$ are washed out. In contrast, all electrons in the conventional FCI state, Eq. 144, retain the same common skyrmion-lattice factor $\phi^{\text{LLL}*}_{-\boldsymbol{q}}(\boldsymbol{r})$. Consequently, $\rho_{\ell=0}(\boldsymbol{r})$ retains zeros at the zeros of $\phi^{\text{LLL}*}_{-\boldsymbol{q}}(\boldsymbol{r})$, whose positions are determined by Eq. 71. However, when the smoothing operator $\exp\left(\frac{\xi^2\sum_p\nabla_p^2}{4}\right)$ acts on the conventional FCI state, it will also remove the zeros of its $\ell=0$ charge density.

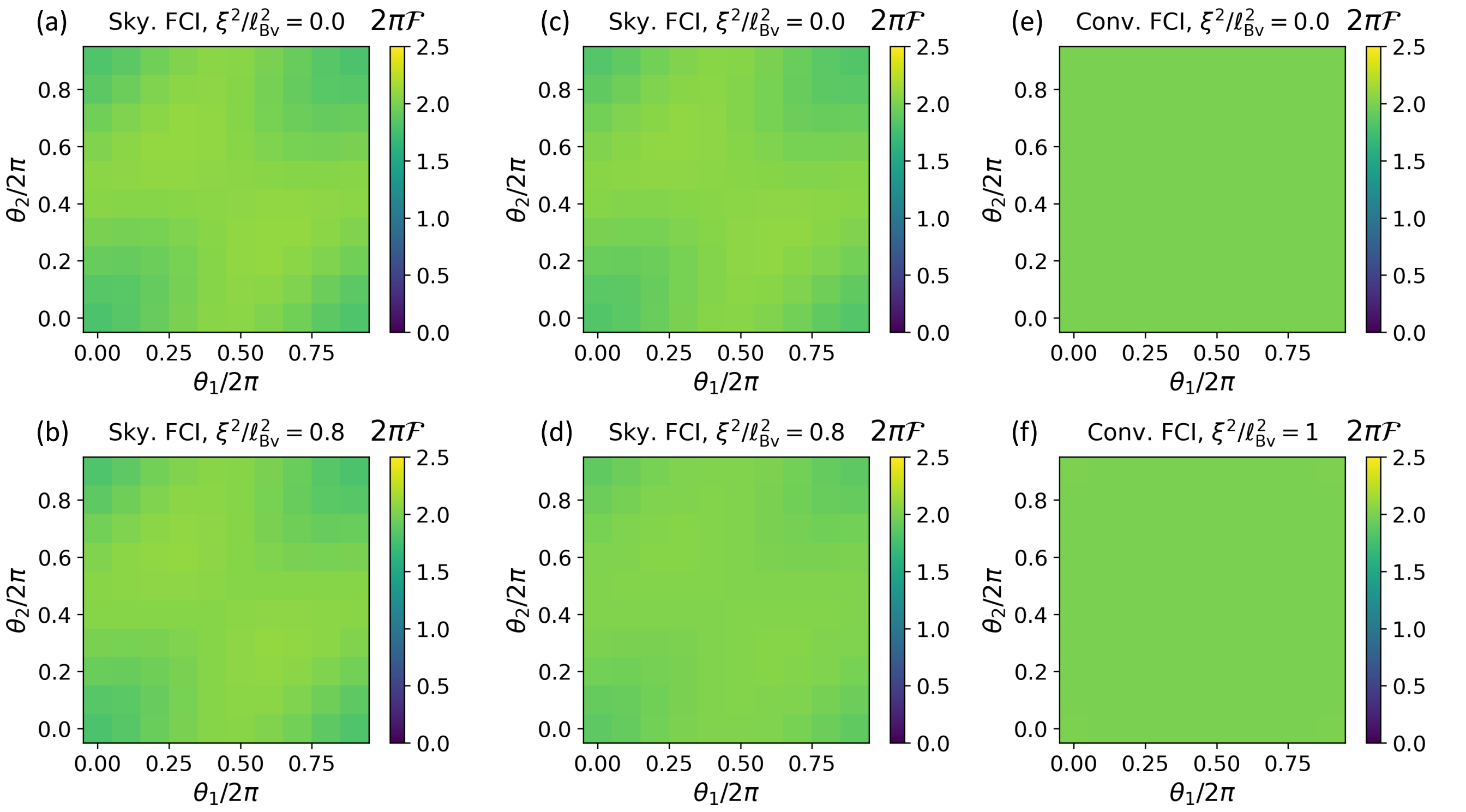


FIG. 2. Many-body Berry curvature $\mathcal{F}(\theta_1, \theta_2)$ for the trial wavefunctions **(a)** $\Psi^{\text{frac-SkV},(s)}_{\{\ell\},1}$, **(b)** $\exp\left(\sum_p \xi^2 \nabla_p^2/4\right) \Psi^{\text{frac-SkV},(s)}_{\{\ell\},1}$ with $\xi^2 = 0.8 l_{Bv}^2$, **(c)** $\Psi^{\text{frac-SkV},(s)}_{\{\ell\},2}$, **(d)** $\exp\left(\sum_p \xi^2 \nabla_p^2/4\right) \Psi^{\text{frac-SkV},(s)}_{\{\ell\},2}$ with $\xi^2 = 0.8 l_{Bv}^2$, **(e)** $\Psi^{\nu=\frac{2}{3},\text{FCI},(s)}_{\{\ell\}}$, and **(f)** $\exp\left(\sum_p \xi^2 \nabla_p^2/4\right) \Psi^{\nu=\frac{2}{3},\text{FCI},(s)}_{\{\ell\}}$ with $\xi^2 = 1.0 l_{Bv}^2$. All calculations are performed at skyrmion-lattice period $|\boldsymbol{a}_1| = |\boldsymbol{a}_2| = 11\,\text{nm}$, $N_L = 5$, corresponding to the value used in Fig. 3 of the main text. The values of $\xi$ shown here are chosen to minimize the total energy (see Fig. 3). For the kinetic term, we use the dispersion $\mathcal{E}(\boldsymbol{k})$ corresponding to $u_D = 45\,\text{meV}$. The interaction is the dual-gate-screened Coulomb interaction defined in Eq. 129.

Finally, we consider the energetics of FCI wavefunction when a periodic scalar potential is included. The scalar potential takes the form

$$H_{\rm PE} = \sum_{i=1}^{3} \sum_{\boldsymbol{k}} U_0 e^{i\phi} c^\dagger_{\boldsymbol{k}-\boldsymbol{\Delta\theta}} c_{\boldsymbol{k}-\boldsymbol{\Delta\theta}+\boldsymbol{g}_i} \langle s_{\boldsymbol{k}-\boldsymbol{\Delta\theta}} | s_{\boldsymbol{k}-\boldsymbol{\Delta\theta}+\boldsymbol{g}_i} \rangle + h.c. \tag{150}$$

Here, $\boldsymbol{g}_1 = \boldsymbol{b}_1$, $\boldsymbol{g}_2 = \boldsymbol{b}_2$, and $\boldsymbol{g}_3 = -\boldsymbol{b}_1 - \boldsymbol{b}_2$ are the first-shell reciprocal-lattice vectors. The scalar potential is fixed to have the same periodicity as the skyrmion lattice.

As shown in Fig. 3 of the main text, at their respective optimized values of $\xi$, the conventional FCI wavefunction has a higher energy than the skyrmion FCI wavefunction in the absence of the periodic potential. However, the charge density of the skyrmion FCI state is smoother than that of the conventional FCI state, as shown in Fig. 5 (a) and (b). Consequently, the conventional FCI state couples more strongly to the periodic potential, which can reverse the energetic ordering of the two states. In Fig. 5 (d), we show the energy per particle of the skyrmion FCI wavefunction and convention FCI wavefunction in the presence of a periodic potential $H_{\rm PE}$. A crossover occurs at $U_0 \approx 15$meV, above which the conventional FCI is favored.

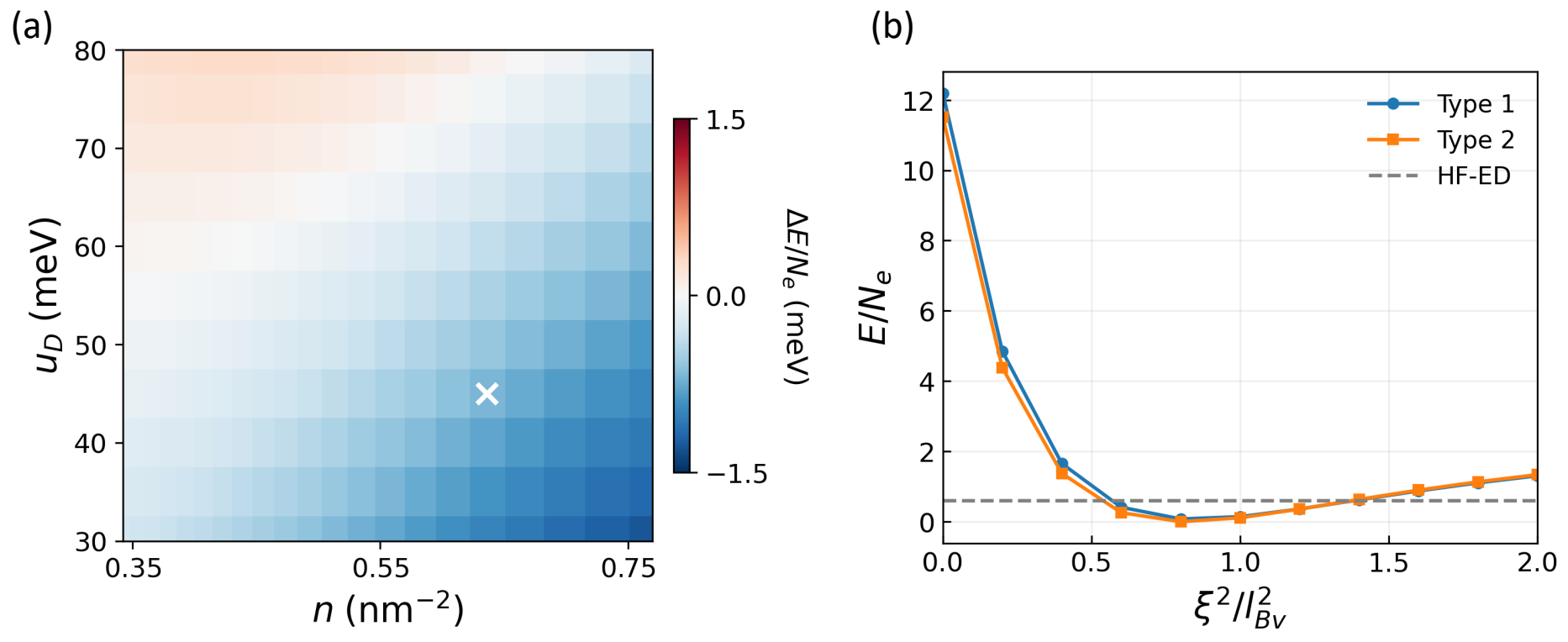


FIG. 3. **(a)** Difference in the total energy per particle between $\Psi^{\text{frac-SkV},(s)}_{\{\ell\},1}$ and $\Psi^{\text{frac-SkV},(s)}_{\{\ell\},2}$, defined as $\Delta E \equiv E(\Psi^{\text{frac-SkV},(s)}_{\{\ell\},2}) - E(\Psi^{\text{frac-SkV},(s)}_{\{\ell\},1})$. The white cross marks the parameter point further examined in Fig. 3 of the main text and in panel **(b)**. **(b)** Energy per particle for the two skyrmion FCI constructions after applying the smoothing operator $\exp\left(\xi^2 \sum_p \nabla_p^2/4\right)$, together with the Hartree-Fock-band-projected exact-diagonalization results. The calculations are performed for $N_L = 5$, $u_D = 45\,\text{meV}$, and $|\boldsymbol{a}_\mu| = 11\,\text{nm}$. The interaction is the dual-gate-screened Coulomb interaction defined in Eq. 129. Energies are shown relative to the minimum.

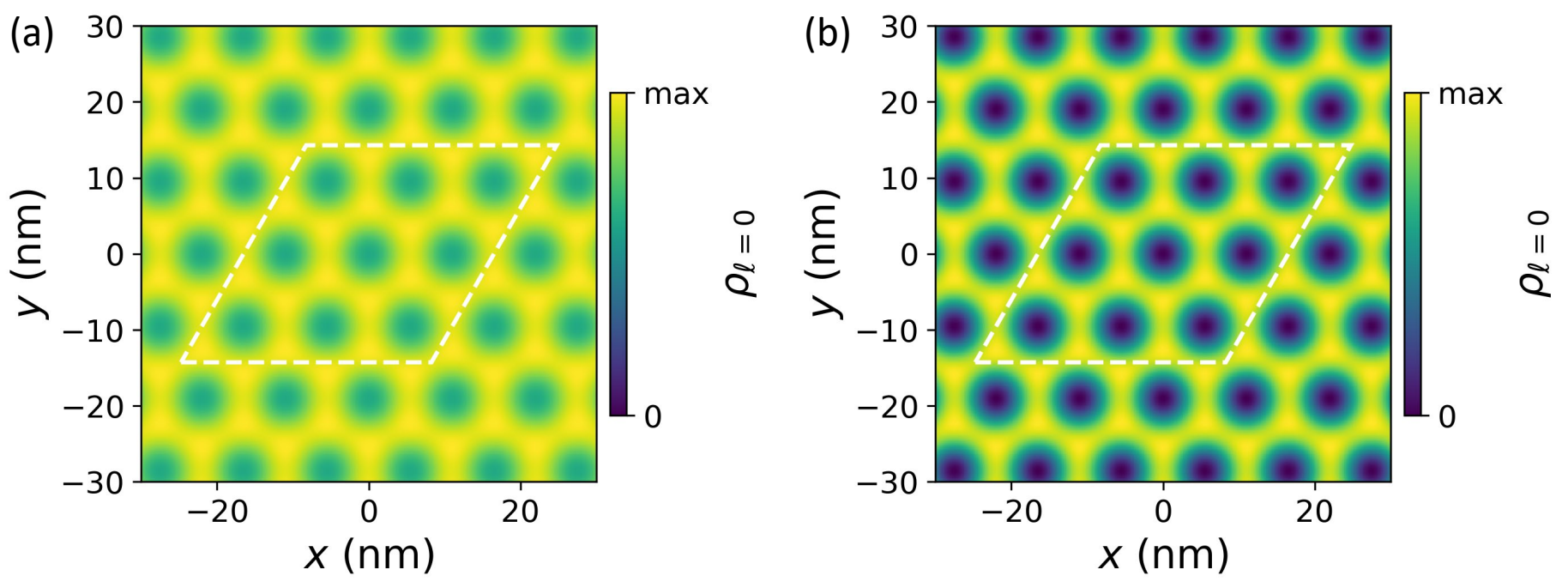


FIG. 4. Layer-resolved charge density of **(a)** the skyrmion FCI state $\Psi^{\text{frac-SkV},(s)}_{\{\ell\},1}$ and **(b)** the conventional FCI state $\Psi^{\nu=\frac{2}{3},\text{FCI},(s)}_{\{\ell\}}$. The wavefunctions are constructed for $N_L = 5$ and skyrmion-lattice period $|\boldsymbol{a}_\mu| = 11\,\text{nm}$ on a $3 \times 3$ torus. The boundary of the torus is indicated by the white dashed line.

## V. BAND WEIGHT OF THE SKYRMION VACANCY STATE AND MULTI-BANDNESS OF SKYRMION FCI STATE

### A. band weight

In this section, we will discuss how to calculate the weight of a given skyrmion vacancy state within the $\nu = 1$ skyrmion-lattice band. As discussed in the main text, this characterizes the effects of band mixing.

Let us define the fermion creation operator $f^\dagger_{\boldsymbol{k}-\boldsymbol{\Delta\theta}}$ for the skyrmion-lattice Bloch state $\psi^{\text{SkL}}_{\boldsymbol{k}-\boldsymbol{\Delta\theta},\ell}(\boldsymbol{r})$ given in Eq. 139. The momentum $\boldsymbol{q}$ entering Eq. 139 reflects a freedom in the choice of origin. For $\xi_{00} = 0$, we choose $\boldsymbol{q} = (\boldsymbol{b}_1 + \boldsymbol{b}_2)/2$, so that in the absence of vacancies the skyrmion vacancy state reduces to the corresponding $\nu = 1$ skyrmion-lattice state. For a Slater determinant $|D\rangle$ associated with a particular vacancy configuration $\{\boldsymbol{\eta}\}$ (Sec. IV B 4), we define the band weight as

$$\text{band weight} \equiv \frac{\sum_{\boldsymbol{k}\in\text{BZ1}} \langle D| f^\dagger_{\boldsymbol{k}-\boldsymbol{\Delta\theta}} f_{\boldsymbol{k}-\boldsymbol{\Delta\theta}} |D\rangle}{\sum_{\boldsymbol{k}=n_1\boldsymbol{T}_1+n_2\boldsymbol{T}_2} \langle D| c^\dagger_{\boldsymbol{k}-\boldsymbol{\Delta\theta}} c_{\boldsymbol{k}-\boldsymbol{\Delta\theta}} |D\rangle}. \tag{151}$$

Since the single-particle orbitals forming $|D\rangle$ have been orthonormalized, $|D\rangle$ is normalized and the denominator is simply the total particle number, $N_\phi - N_{\text{vac}}$. The numerator measures the number of particles contained within the

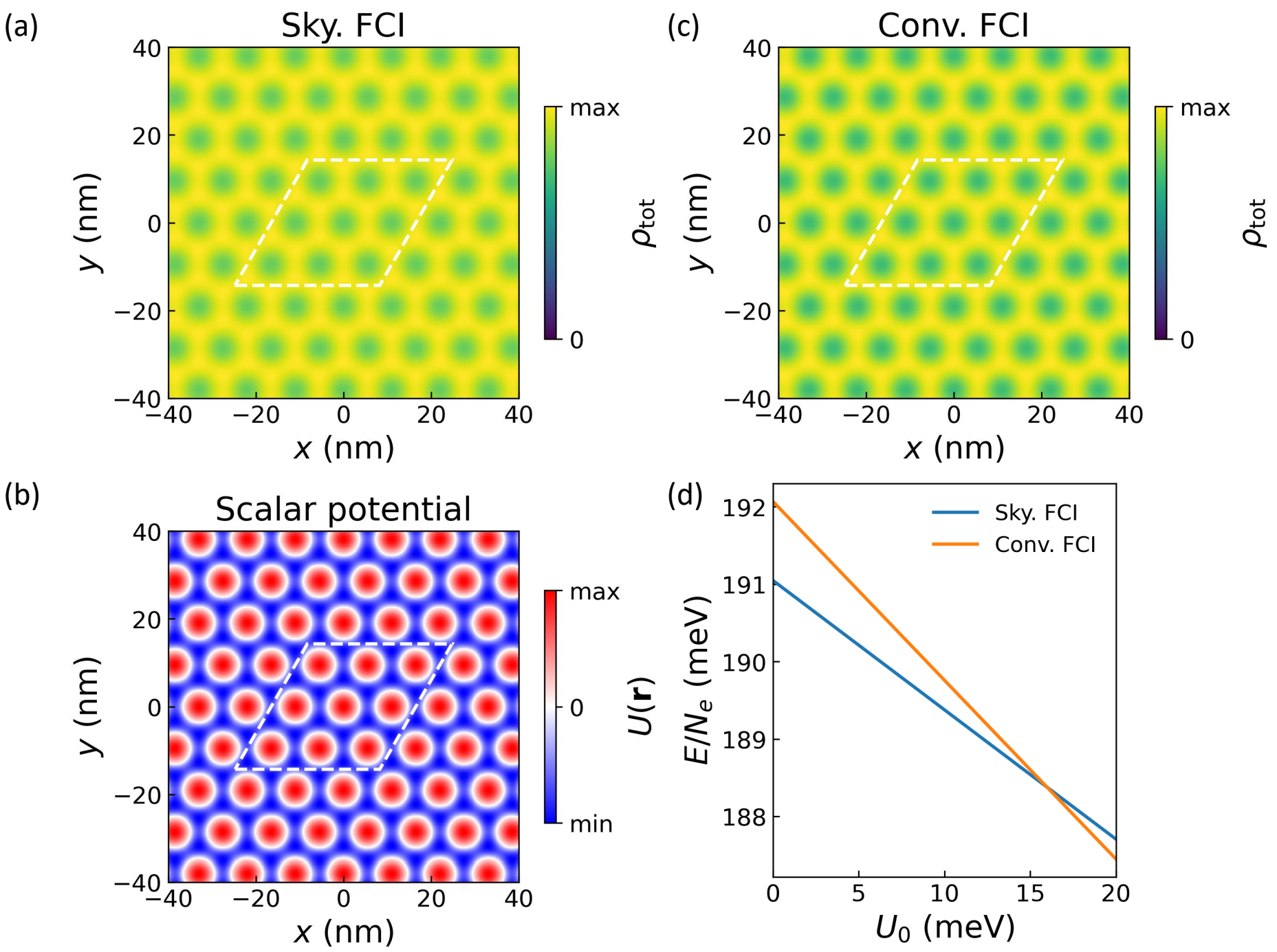


FIG. 5. Total charge density of **(a)** the skyrmion FCI state $\exp\left(\frac{\xi^2\sum_p \nabla_p^2}{4}\right)\Psi^{\text{frac-SkV},(s)}_{\{\ell\},1}$ at $\xi^2 = 0.8 l_{Bv}^2$ and **(b)** the conventional FCI state $\exp\left(\frac{\xi^2\sum_p \nabla_p^2}{4}\right)\Psi^{\nu=\frac{2}{3},\text{FCI},(s)}_{\{\ell\}}$ at $\xi^2 = l_{Bv}^2$. The wavefunctions are constructed on a $3\times 3$ torus with $N_L = 5$ and skyrmion-lattice period $|\boldsymbol{a}_\mu| = 11\,\text{nm}$. The boundary of the torus is indicated by the white dashed line. **(c)** Real-space profile of the periodic scalar potential defined in Eq. 150. We take $\phi = 0$, which minimizes the potential energy of both the skyrmion and conventional FCI states for positive $U_0$. **(d)** Energy per particle of the conventional and skyrmion FCI wavefunctions in the presence of the periodic scalar potential. We use $\xi^2 = 0.8 l_{Bv}^2$ for the skyrmion FCI state and $\xi^2 = l_{Bv}^2$ for the conventional FCI state.

$\nu = 1$ skyrmion-lattice band. Thus, the band weight gives the fraction of the skyrmion vacancy state that remains within this band.

The band weight above is defined for a fixed vacancy configuration. The results shown in the main text are obtained by averaging over random vacancy configurations. In Fig. 6, we include additional data of the band weight for different system sizes. From the plot, we conclude that the results have already converged with respect to the system size. For the largest system considered, a $6\times 6$ system, we average over 1000 random configurations.

### B. Multi-band nature of the skyrmion FCI state

The band weight discussed above is defined for a particular skyrmion-vacancy state. We now introduce another diagnostic that directly describes the multi-band nature of the skyrmion FCI wavefunction.

Since the skyrmion FCI wavefunction is Bloch periodic [Eq. 100], its one-body density matrix is block diagonal in Bloch momentum:

$$\langle \Psi_1^{\text{frac-SkV},(s)}(\boldsymbol{\Delta\theta})| c^\dagger_{\boldsymbol{k}_1-\boldsymbol{\Delta\theta}} c_{\boldsymbol{k}_2-\boldsymbol{\Delta\theta}} |\Psi_1^{\text{frac-SkV},(s)}(\boldsymbol{\Delta\theta})\rangle \propto \delta^{\text{BZ1}}_{\boldsymbol{k}_1,\boldsymbol{k}_2}. \tag{152}$$

Here, $\delta^{\text{BZ1}}_{\boldsymbol{k}_1,\boldsymbol{k}_2}$ is nonzero only when $\boldsymbol{k}_1 = \boldsymbol{k}_2$ modulo the reciprocal lattice generated by $\boldsymbol{b}_\mu$. We can therefore define, for each $\boldsymbol{k} \in \text{BZ1}$, the one-body density matrix [7]

$$\rho^{(s)}_{\boldsymbol{g}_1,\boldsymbol{g}_2}(\boldsymbol{k},\boldsymbol{\Delta\theta}) = \frac{\langle \Psi_1^{\text{frac-SkV},(s)}(\boldsymbol{\Delta\theta})| c^\dagger_{\boldsymbol{k}+\boldsymbol{g}_2-\boldsymbol{\Delta\theta}} c_{\boldsymbol{k}+\boldsymbol{g}_1-\boldsymbol{\Delta\theta}} |\Psi_1^{\text{frac-SkV},(s)}(\boldsymbol{\Delta\theta})\rangle}{\langle \Psi_1^{\text{frac-SkV},(s)}(\boldsymbol{\Delta\theta})|\Psi_1^{\text{frac-SkV},(s)}(\boldsymbol{\Delta\theta})\rangle}, \tag{153}$$

[7] Notice the index order $\boldsymbol{g}_1, \boldsymbol{g}_2$, which we choose to match the usual convention in the Hartree–Fock literature.

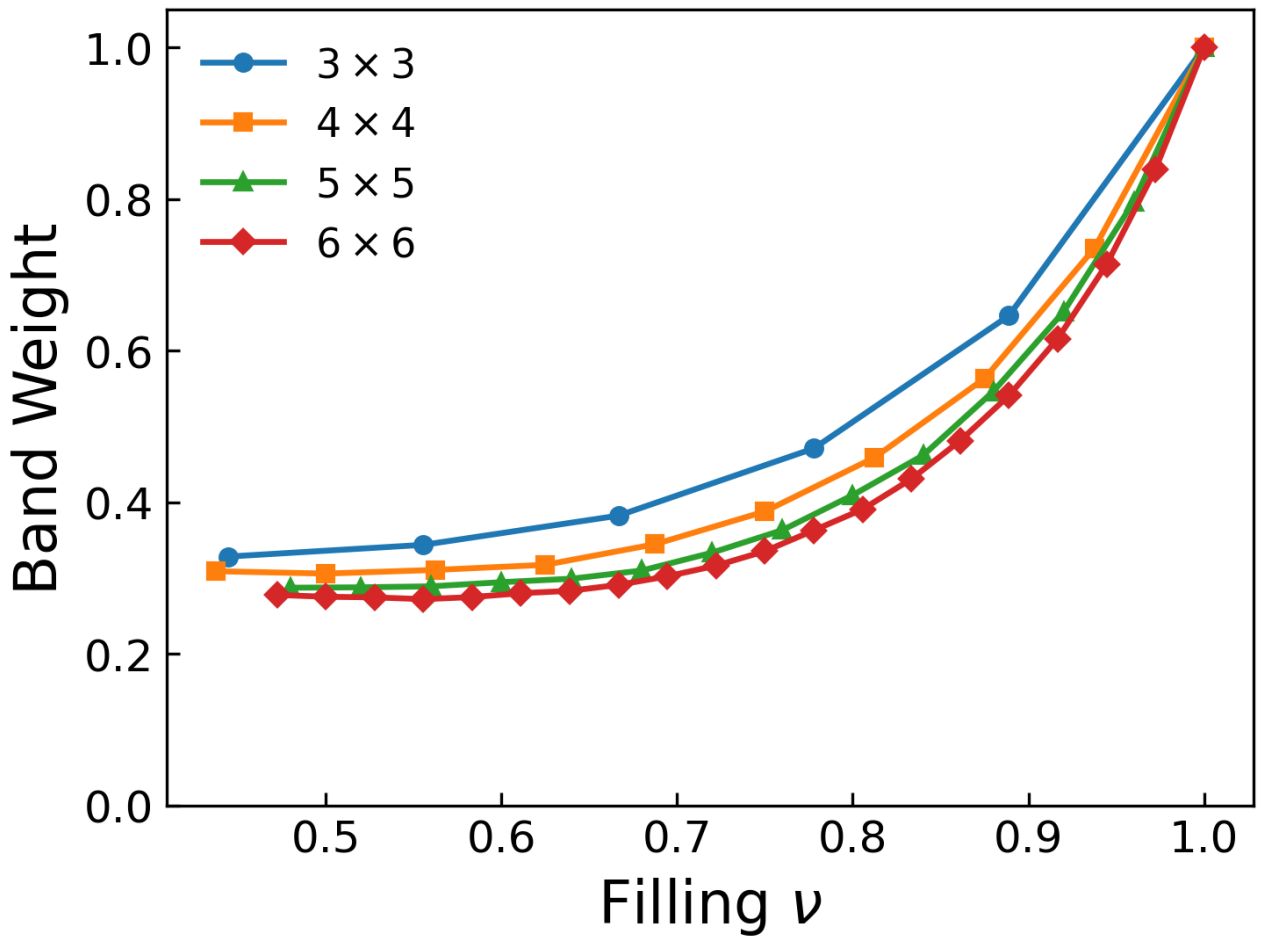


FIG. 6. The band weight, defined as in Eq.151, as a function of filling factor $\nu$, for different system size. The band weight is averaged over random vacancy configurations. The calculations are done for $\Psi^{\text{frac-SkV},(s)}_{\{\ell\},1}$ skyrmion FCI wavefunction with skyrmion lattice period $|a_\mu| = 11$nm, $N_L = 5$. This is the same set of parameters as shown in the main text.

where $\boldsymbol{g}_i$ is a reciprocal-lattice vector generated by $\boldsymbol{b}_\mu$.

The sum of the eigenvalues of $\rho^{(s)}(\boldsymbol{k}, \boldsymbol{\Delta\theta})$ over all $\boldsymbol{k}$ equals the total particle number. If the many-body state lies entirely within a single Bloch band, then $\rho^{(s)}(\boldsymbol{k}, \boldsymbol{\Delta\theta})$ has at most one nonzero eigenvalue.

In Fig. 7, we plot the three largest eigenvalues of the one-body density matrix, averaged over the orthonormalized many-body states. For the skyrmion FCI state, we find more than one nonzero eigenvalue both without smoothing [Fig. 7(a)] and with smoothing [Fig. 7(b)], directly demonstrating its multi-band nature. In contrast, the density matrix of the conventional FCI state has only one nonzero eigenvalue at each $\boldsymbol{k}$, both without smoothing [Fig. 7(c)] and with smoothing [Fig. 7(d)]. This is expected because the conventional FCI state lies entirely within the single Bloch band spanned by

$$\exp\left(\frac{\xi^2\nabla^2}{4}\right)\psi^{\text{SkL}}_{\boldsymbol{k}-\boldsymbol{\Delta\theta},\ell}(\boldsymbol{r}). \tag{154}$$

## VI. HARTREE-FOCK AND EXACT DIAGONALIZATION

In the main text, we compare the energies of the skyrmion FCI and conventional FCI trial states with exact-diagonalization results obtained after projection onto the lowest Hartree-Fock band. In Sec. VI A, we briefly describe the Hartree-Fock calculation used to obtain the band for this projection. In Sec. VI B we discuss the details of the exact diagonalization in this basis.

### A. Hartree-Fock

We use the same Hartree-Fock setup as Ref. [7]; further details of the procedure are given in its Supplementary Material. The Hamiltonian used in the Hartree-Fock calculation contains the kinetic term in Eq. 122 and the interaction term in Eq. 124. As in Sec. IV B 4, $\mathcal{E}(\boldsymbol{k})$ is taken to be the dispersion of the first conduction band of $h_{\text{RMG}}(\boldsymbol{K}+\boldsymbol{k})$, which depends on the displacement field $u_D$. The interaction $V(\boldsymbol{q})$ is taken to be the dual-gate-screened Coulomb interaction specified in Sec. IV B 4. We do not include a periodic potential in this calculation.[8] We take the lattice

[8] In an hBN/RMG moiré superlattice, a periodic potential can arise from either the filled moiré valence bands of RMG or the hBN/RMG tunneling.

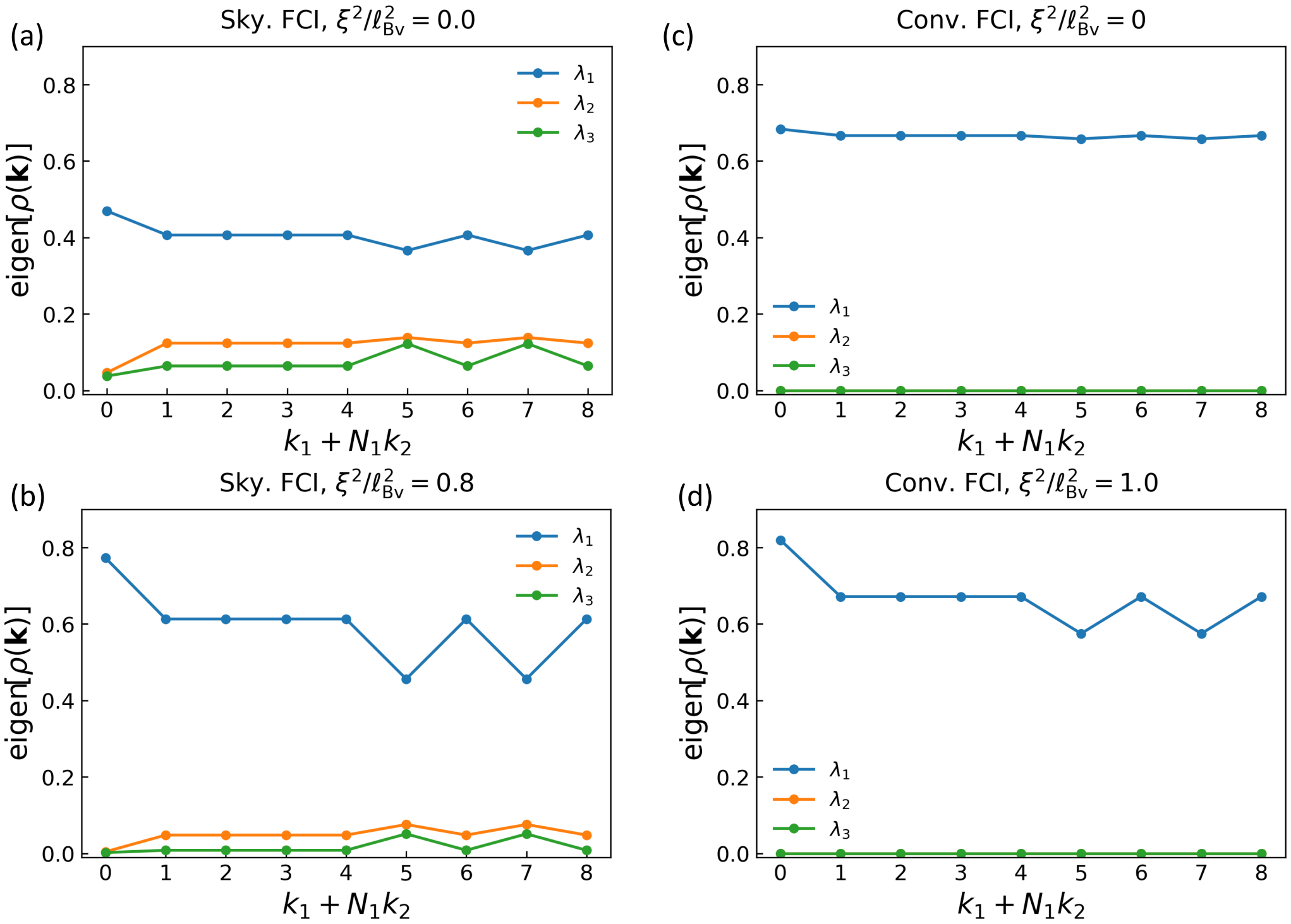


FIG. 7. Three largest eigenvalues of the one-body density matrix $\rho(\boldsymbol{k}, \boldsymbol{\Delta\theta} = \boldsymbol{0})$. The density matrix is averaged over the three orthonormalized many-body states. The plots correspond to **(a)** $\Psi^{\text{frac-SkV},(s)}_{\{\ell\},1}$; **(b)** $\exp\left(\sum_p \xi^2 \nabla_p^2/4\right) \Psi^{\text{frac-SkV},(s)}_{\{\ell\},1}$ with $\xi^2 = 0.8 l_{Bv}^2$; **(c)** $\Psi^{\nu=\frac{2}{3},\text{FCI},(s)}_{\{\ell\}}$; and **(d)** $\exp\left(\sum_p \xi^2 \nabla_p^2/4\right) \Psi^{\nu=\frac{2}{3},\text{FCI},(s)}_{\{\ell\}}$ with $\xi^2 = 1.0 l_{Bv}^2$. All calculations are performed on a 3-by-3 torus, at skyrmion-lattice period $|\boldsymbol{a}_1| = |\boldsymbol{a}_2| = 11\,\text{nm}$ and $N_L = 5$, corresponding to the parameters used in Fig. 3 of the main text. The x-axis is the momentum index, defined via $\boldsymbol{k} = k_1 \boldsymbol{T}_1 + k_2 \boldsymbol{T}_2$.

vectors to be[9]

$$\boldsymbol{a}_1 = a_m (1, 0)^T, \qquad \boldsymbol{a}_2 = a_m \left(\frac{1}{2}, \frac{\sqrt{3}}{2}\right)^T, \tag{155}$$

where $a_m$ is the lattice period. For an $N_\phi$-flux torus, the corresponding system area is

$$A = N_\phi \frac{\sqrt{3} a_m^2}{2}. \tag{156}$$

The Hartree-Fock calculation is performed on a $3 \times 3$ system, corresponding to $N_\phi = 9$, with periodic boundary condition $\boldsymbol{\Delta\theta} = \boldsymbol{0}$.

### B. Exact Diagonalization

We next perform exact diagonalization after projecting the Hamiltonian onto the lowest Hartree-Fock band. We denote the creation operator for the $n$-th Hartree-Fock band by $\tilde{f}^\dagger_{\boldsymbol{k}-\boldsymbol{\Delta\theta},n}$, with

$$\tilde{f}^\dagger_{\boldsymbol{k}-\boldsymbol{\Delta\theta},n} = \sum_{\boldsymbol{g}=n_1\boldsymbol{b}_1+n_2\boldsymbol{b}_2} U_{\boldsymbol{g},n}(\boldsymbol{k}, \boldsymbol{\Delta\theta}) c^\dagger_{\boldsymbol{k}+\boldsymbol{g}-\boldsymbol{\Delta\theta}}, \qquad \boldsymbol{k} \in \text{BZ1}. \tag{157}$$

[9] In an hBN/RMG moiré superlattice, the orientation of the moiré lattice vectors depends on the hBN/RMG twist angle $\theta$. We ignore this complication here.

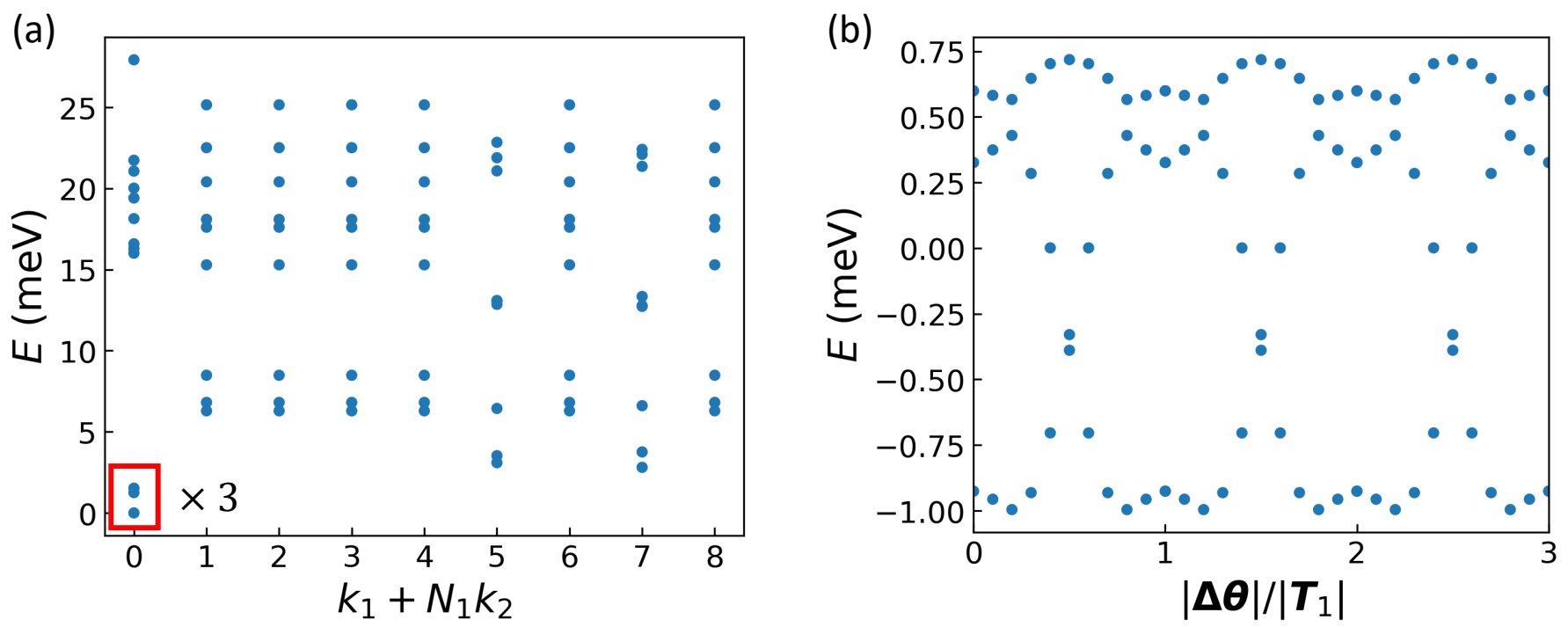


FIG. 8. **(a)** Many-body energy spectrum obtained by exact diagonalization after projecting the Hamiltonian onto the lowest Hartree–Fock band. The red box encloses the threefold nearly degenerate ground-state manifold in the zero-total-momentum sector. The energies are shown relative to the minimum energy, and the boundary-condition twist is fixed at $\boldsymbol{\Delta\theta} = \mathbf{0}$. **(b)** Spectral flow of the three nearly degenerate ground states under flux threading. At each boundary-condition twist, the energies are shown relative to their average. The calculation is performed on a $3\times3$ torus with lattice period $|\boldsymbol{a}_\mu| = 11\,\mathrm{nm}$, displacement field $u_D = 45\,\mathrm{meV}$, and electron number $N_e = 6$. These are the same parameters as those used in the corresponding main-text figure.

Here, $U_{\boldsymbol{g},n}(\boldsymbol{k}, \boldsymbol{\Delta\theta})$ is obtained by diagonalizing the Hartree-Fock Hamiltonian. In the following, we retain only the lowest Hartree-Fock band, $n = 1$.

The kinetic Hamiltonian projected onto this band is

$$H_{\mathrm{KE}}^{\mathrm{proj}} = \sum_{\boldsymbol{k}\in\mathrm{BZ1}} \tilde{f}^{\dagger}_{\boldsymbol{k}-\boldsymbol{\Delta\theta},1}\tilde{f}_{\boldsymbol{k}-\boldsymbol{\Delta\theta},1} \times \sum_{\boldsymbol{g}=n_1\boldsymbol{b}_1+n_2\boldsymbol{b}_2} U^*_{\boldsymbol{g},1}(\boldsymbol{k}, \boldsymbol{\Delta\theta})\mathcal{E}(\boldsymbol{k}+\boldsymbol{g}-\boldsymbol{\Delta\theta})U_{\boldsymbol{g},1}(\boldsymbol{k}, \boldsymbol{\Delta\theta}). \tag{158}$$

The interaction Hamiltonian projected onto the lowest Hartree-Fock band is

$$H_{\mathrm{int}}^{\mathrm{proj}} = \frac{1}{2A}\sum_{\boldsymbol{k}_1,..,\boldsymbol{k}_4\in\mathrm{BZ1}} V_{1234}\tilde{f}^{\dagger}_{\boldsymbol{k}_1-\boldsymbol{\Delta\theta},1}\tilde{f}^{\dagger}_{\boldsymbol{k}_2-\boldsymbol{\Delta\theta},1}\tilde{f}_{\boldsymbol{k}_4-\boldsymbol{\Delta\theta},1}\tilde{f}_{\boldsymbol{k}_3-\boldsymbol{\Delta\theta},1}, \tag{159}$$

where

$$V_{1234} = \sum_{\boldsymbol{q}=n_1\boldsymbol{T}_1+n_2\boldsymbol{T}_2} V(\boldsymbol{q})\left(\sum_{\boldsymbol{g}_1,\boldsymbol{g}_3} U^*_{\boldsymbol{g}_1,1}(\boldsymbol{k}_1, \boldsymbol{\Delta\theta})U_{\boldsymbol{g}_3,1}(\boldsymbol{k}_3, \boldsymbol{\Delta\theta})\langle s_{\boldsymbol{k}_1+\boldsymbol{g}_1-\boldsymbol{\Delta\theta}}|s_{\boldsymbol{k}_3+\boldsymbol{g}_3-\boldsymbol{\Delta\theta}}\rangle\delta_{\boldsymbol{k}_3+\boldsymbol{g}_3+\boldsymbol{q},\boldsymbol{k}_1+\boldsymbol{g}_1}\right) \left(\sum_{\boldsymbol{g}_2,\boldsymbol{g}_4} U^*_{\boldsymbol{g}_2,1}(\boldsymbol{k}_2, \boldsymbol{\Delta\theta})U_{\boldsymbol{g}_4,1}(\boldsymbol{k}_4, \boldsymbol{\Delta\theta})\langle s_{\boldsymbol{k}_2+\boldsymbol{g}_2-\boldsymbol{\Delta\theta}}|s_{\boldsymbol{k}_4+\boldsymbol{g}_4-\boldsymbol{\Delta\theta}}\rangle\delta_{\boldsymbol{k}_4+\boldsymbol{g}_4-\boldsymbol{q},\boldsymbol{k}_2+\boldsymbol{g}_2}\right). \tag{160}$$

We diagonalize $H_{\mathrm{KE}}^{\mathrm{proj}} + H_{\mathrm{int}}^{\mathrm{proj}}$ in the many-body Hilbert space of the lowest Hartree-Fock band. The energy reported for comparison in the main text Fig. 3 is averaged over the three nearly degenerate ground states in the zero total momentum sector. We refer the reader to, for example, the Supplementary Material of Ref. [14] for a review of Bloch-band-projected exact diagonalization.

In Fig. 8, we present many-body spectrum data obtained from our HF-ED calculation. Figure 8(a) shows the many-body exact-diagonalization spectrum. The three nearly degenerate ground states, enclosed by the red box, lie in the zero-total-momentum sector. Under flux threading, these states exhibit spectral flow and are permuted among themselves, as shown in Fig. 8(b). These results are consistent with a $\nu = \frac{2}{3}$ FCI ground-state manifold.



\end{document}